\documentclass[bibyear]{aa}  

\usepackage{graphicx}
\usepackage{natbib}  
\usepackage{float}
\usepackage{color}
\usepackage{txfonts}
\usepackage{gensymb}
\usepackage{amsmath}

\def\src {PSR~B0950+08}
\def\fdiff {$\delta \nu_\mathrm{DISS}$}
\def\FAP {$F_\mathrm{AP}$}

\begin{document}

\title{Dual-frequency single-pulse study of PSR B0950+08}

   \author{A.~V.~Bilous\inst{\ref{astron}}
   \and J.~M.~Grie{\ss}meier\inst{\ref{lpc2e},\ref{nancay}}
   \and T. Pennucci\inst{\ref{nrao},\ref{eotvos}}
   \and Z. Wu\inst{\ref{error404}}
   \and L.~Bondonneau\inst{\ref{lpc2e},\ref{cnrs}} 
   \and V.~Kondratiev\inst{\ref{astron},\ref{lebedev}}
   \and J. van Leeuwen\inst{\ref{astron},\ref{uva}}
   \and Y.~Maan\inst{\ref{astron},\ref{NCRA}}
   \and L.~Connor\inst{\ref{caltech},\ref{uva}}
   \and L.~C.~Oostrum\inst{\ref{astron},\ref{uva},\ref{escience}}
   \and E.~Petroff\inst{\ref{uva},\ref{mu},\ref{mcgill},\ref{veni}}
   \and J.~P.~W.~Verbiest\inst{\ref{error404},\ref{mpfir}}
   \and D.~Vohl\inst{\ref{astron},\ref{uva}}
   \and J.~W.~McKee\inst{\ref{cita}}
   \and G.~Shaifullah\inst{\ref{milano}}
   \and G.~Theureau\inst{\ref{lpc2e},\ref{nancay},\ref{cnrs2}}
   \and O.~M.~Ulyanov\inst{\ref{irau}}
   \and B.~Cecconi\inst{\ref{cnrs}}
   \and A.~H.~Coolen\inst{\ref{astron}}
   \and S.~Corbel\inst{\ref{saclay},\ref{nancay}}
   \and S.~Damstra\inst{\ref{astron}}
   \and H.~D\'{e}nes\inst{\ref{astron}}
   \and J.~N.~Girard\inst{\ref{cnrs}}
   \and B.~Hut\inst{\ref{astron}}
   \and M.~Ivashina\inst{\ref{chalmers}}
   \and O.~O.~Konovalenko\inst{\ref{irau}}
   \and A.~Kutkin\inst{\ref{astron},\ref{lebedev}}
   \and G.~M.~Loose\inst{\ref{astron}}
   \and H.~Mulder\inst{\ref{astron}}
   \and M.~Ruiter\inst{\ref{astron}}
   \and R.~Smits\inst{\ref{astron}}
   \and P.~L.~Tokarsky\inst{\ref{irau}}
   \and N.~J.~Vermaas\inst{\ref{astron}}
   \and V.~V.~Zakharenko\inst{\ref{irau}}
   \and P.~Zarka\inst{\ref{cnrs}}
   \and J.~Ziemke\inst{\ref{astron},\ref{oslocit}}			
          }
\institute{
    ASTRON, the Netherlands Institute for Radio Astronomy, Postbus 2, 7990 AA Dwingeloo, The Netherlands\label{astron}
    \and LPC2E - Universit\'{e} d'Orl\'{e}ans / CNRS, 45071 Orl\'{e}ans cedex 2, France\label{lpc2e}
    \and Station de Radioastronomie de Nan\c{c}ay, Observatoire de Paris, PSL Research University, CNRS, Univ. Orl\'{e}ans, OSUC, 18330 Nan\c{c}ay, France\label{nancay}
    \and National Radio Astronomy Observatory, 520 Edgemont Road, Charlottesville, VA 22903, USA\label{nrao}
    \and Institute of Physics, E\"{o}tv\"{o}s Lor\'{a}nd University, P\'{a}zm\'{a}ny P. s. 1/A, 1117 Budapest, Hungary\label{eotvos}
    \and Fakult\"{a}t f\"{u}r Physik, Universit\"{a}t Bielefeld, Postfach 100131, 33501 Bielefeld, Germany\label{error404}
    \and LESIA \& USN, Observatoire de Paris, CNRS, PSL, SU/UP/UO, 92195 Meudon, France\label{cnrs}
    \and Astro Space Center of Lebedev Physical Institute, Profsoyuznaya Str. 84/32, 117997 Moscow, Russia\label{lebedev}
    \and Anton Pannekoek Institute, University of Amsterdam, Postbus 94249, 1090 GE Amsterdam, The Netherlands\label{uva}
    \and National Centre for Radio Astrophysics, Tata Institute of Fundamental Research, Pune 411007, Maharashtra, India\label{NCRA}
    \and Cahill Center for Astronomy, California Institute of Technology, Pasadena, CA, USA\label{caltech}
    \and Netherlands eScience Center, Science Park 140, 1098 XG, Amsterdam, The Netherlands\label{escience}
    \and Department of Physics, McGill University, 3600 rue University, Montr\'{e}al, QC H3A 2T8, Canada\label{mu}
    \and McGill Space Institute, McGill University, 3550 rue University, Montr\'{e}al, QC H3A 2A7, Canada\label{mcgill}
    \and Veni Fellow\label{veni}
    \and Max-Planck-Institut f\"{u}r Radioastronomie, Auf dem H\"{u}gel 69, 53121 Bonn, Germany\label{mpfir}
    \and Canadian Institute for Theoretical Astrophysics, University of Toronto, 60 Saint George Street, Toronto, ON M5S 3H8, Canada\label{cita}
    \and Dipartimento di Fisica ``G. Occhialini'', Universit\'a degli Studi di Milano-Bicocca, Piazza della Scienza 3, 20126 Milano, Italy\label{milano}
    \and Laboratoire Univers et Th\'{e}ories, Observatoire de Paris, Universit\'{e} PSL, CNRS, Universit\'{e} de Paris, 92190 Meudon, France\label{cnrs2}
    \and Institute of Radio Astronomy of the National Academy of Sciences of Ukraine, 4, Mystetstv St., Kharkiv, 61002, Ukraine\label{irau}
    \and AIM, CEA, CNRS, Universit\'{e} de Paris, Université Paris-Saclay, F-91191 Gif-sur-Yvette, France\label{saclay}  
    \and Dept. of Electrical Engineering, Chalmers University of Technology, Gothenburg, Sweden\label{chalmers}
    \and University of Oslo Center for Information Technology, P.O. Box 1059, 0316 Oslo, Norway\label{oslocit}    
             }

\abstract{\src\ is a bright nonrecycled pulsar whose single-pulse fluence variability  
is reportedly large. Based on observations at two widely separated frequencies, 
55\,MHz (NenuFAR) and 1.4\,GHz (Westerbork Synthesis Radio Telescope), we review 
the properties of these single pulses. We conclude that they are  more similar 
to ordinary pulses of radio emission than to a special kind of short and bright 
giant pulses, observed from only a handful of pulsars. We argue that a temporal variation 
of the properties of the interstellar medium along the line of sight to this nearby pulsar,
namely the fluctuating size of the decorrelation bandwidth of diffractive scintillation 
makes an important contribution to the observed single-pulse fluence variability. We further present 
interesting structures in the low-frequency single-pulse spectra that resemble
the ``sad trombones'' seen in fast radio bursts (FRBs); although for \src\ the 
upward frequency drift is also routinely present. We explain these spectral 
features with radius-to-frequency mapping, similar to the model developed by 
Wang et al. (2019) for FRBs. Finally, we speculate that $\mu$s-scale fluence variability 
of the general pulsar population remains poorly known, and that its further study may bring 
important clues about the nature of FRBs.
  \vspace{3ex}
}
   

\titlerunning{}
\maketitle

\section{Introduction}

\src\ \citep[CP 0950,][]{Pilkington1968} is one of the three pulsars detected 
in the follow-up survey of compact radio sources that was conducted after the 
first pulsar, CP 1919, was serendipitously discovered. The relatively large flux 
density and close proximity to Earth made \src\ a subject of numerous studies, 
resulting in a wealth of observational facts, which were accumulated over decades. However, 
as is common in pulsar radio astronomy, the interpretation of this information is 
still subject to debate, as there is no consensus on the magnetospheric 
geometry, the location of radio emission, or the emission mechanism of \src\ 
\citep[e.g.,][]{Hankins1981}.

The fluences (the spectral flux densities integrated over time) of its 
single pulses were quickly found to exhibit a great deal of variability \citep{Smith1973}.
Recently, a series of studies \citep[e.g.,][]{Cairns2004,Singal2012,Smirnova2008,Tsai2016,Kuiack2020} 
have claimed that at least some of these pulses are in fact giant pulses (GPs), 
a special and rare kind of pulses  recorded only from a handful of pulsars. 
GPs are narrow (ns-$\mu$s), very bright, have a power-law fluence distribution, 
occur in restricted phase windows, and generally coincide with high-energy emission 
components \citep{Knight2006a,Bilous2015}. 

In this work, we analyze single-pulse data from nonsimultaneous radio observations 
at two widely separated frequency ranges. We review the pulse properties and 
compare them both to ``classical'' GPs, and to the single pulses of the general pulsar 
population. Knowing the fluence distribution and frequency structure of single 
pulses of pulsar radio emission is important not only for constraining the elusive
emission mechanism, but also for future pulsars searches, and for studies of 
fast radio bursts (FRBs).

\section{Observations and data processing}

This work utilizes observations recorded with Apertif on the Westerbork telescope 
\citep{al19}, as well as a number of dedicated  low-frequency observations conducted 
with the NenuFAR telescope\footnote{\url{https://nenufar.obs-nancay.fr/en/astronomer}} 
\citep{Bondonneau2021,Zarka2020}. Hereafter, we refer to the datasets by the 
Institute of Electrical and Electronics Engineers (IEEE) names of the corresponding 
frequency bands\footnote{\url{http://standards.ieee.org/findstds/standard/521-2002.html}}, 
VHF for the NenuFAR \mbox{20--83}\,MHz band (although frequencies 
below 30\,MHz belong to the HF IEEE band), and L-band for the 
1250--1550\,MHz of Apertif.

The L-band observations are comprised of seven 5-min sessions from November 2019--August 
2020. These were primarily intended as short test observations that verified the
Apertif system performance and real-time single-pulse detection pipeline 
\citep{Sclocco2016,Connor2018} of the ALERT FRB survey \citep{Maan2017}. The Apertif 
data acquisition followed the same setup as outlined in, for example, \citet{Oostrum2020}.
Filterbank files were recorded with a time resolution of 81.92\,$\mu$s and channels 
195\,kHz wide. From these, single-pulse archives were prepared with 
\texttt{dspsr}\footnote{\url{http://dspsr.sourceforge.net/}} \citep{vanStraten2011}
using ephemerides from the ATNF pulsar 
catalogue\footnote{\url{http://www.atnf.csiro.au/people/pulsar/psrcat/}} 
\citep{Manchester2005}. The number of phase bins (3080) was chosen to roughly match
the filterbank time resolution. Radio frequency interference (RFI) was manually 
excised with the PSRCHIVE tool \texttt{psrzap} \citep{Hotan2004}, which resulted in 
zero-weighting of some of the subbands and single-pulse subintegrations. 
Archives were subsequently downsampled in frequency to ten subbands. In total, 18177 
pulses were collected. 

In the VHF band, four observations (each lasting for one hour) were carried out 
in January and April 2020. Coherent compensation of dispersion and Faraday rotation 
was applied in real time to the signal within the 8--83\,MHz frequency range, using 
the single-pulse mode of LUPPI (the Low frequency Ultimate Pulsar Processing 
Instrumentation) \citep{Bondonneau2021}. The dispersion measure (DM) used for the 
compensation was obtained during previous observations and was subsequently refined 
for the correct dispersion removal in the whole band extent. In this work, we  
focus on the uncalibrated total intensity, as we  currently lack a well-tested 
calibration scheme for NenuFAR data. In  VHF band the time resolution was more coarse 
than in L-band, resulting in only 386 phase bins. In total, 53518 pulses were examined.

\begin{figure*}
\centering
 \includegraphics[width=0.5\textwidth]{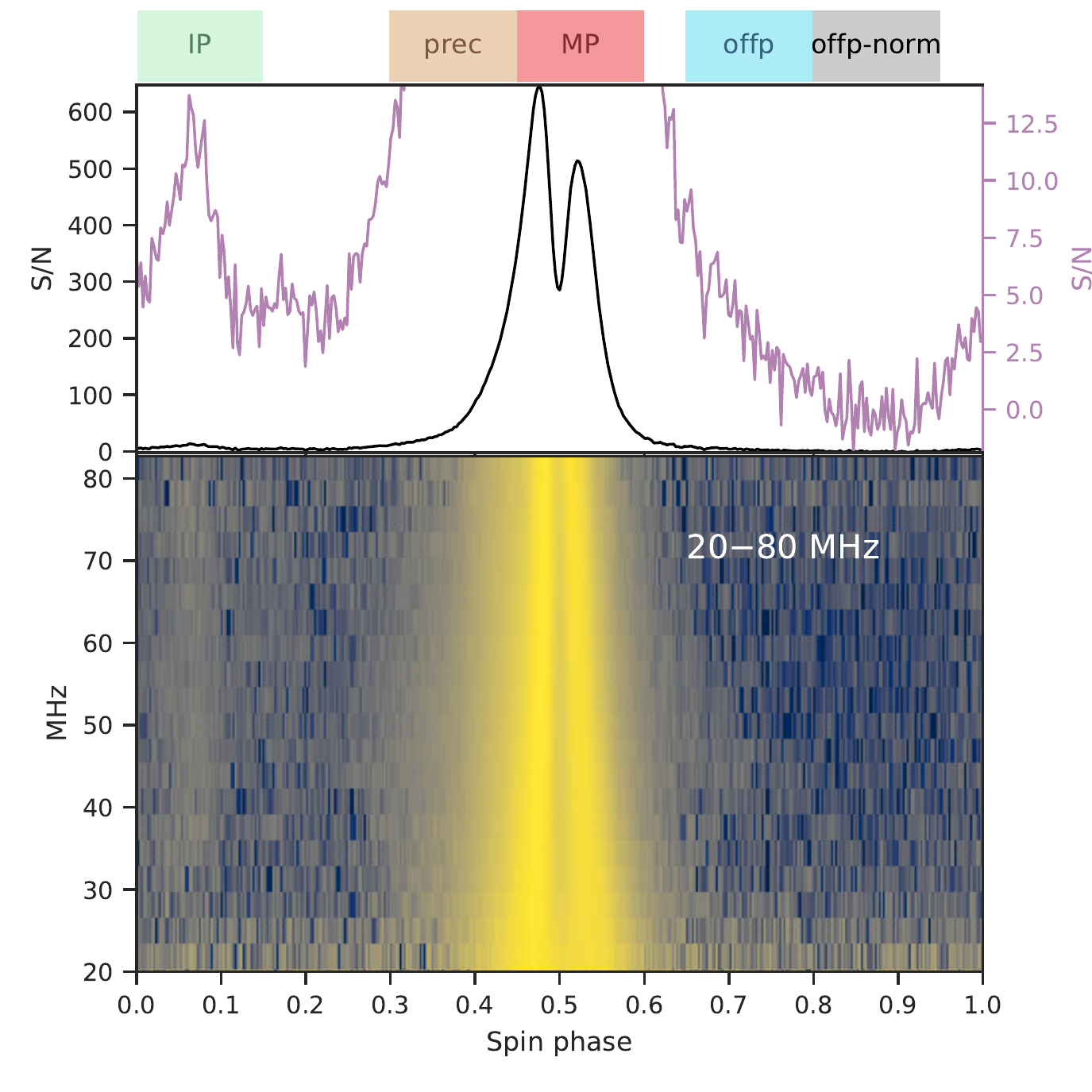}\includegraphics[width=0.5\textwidth]{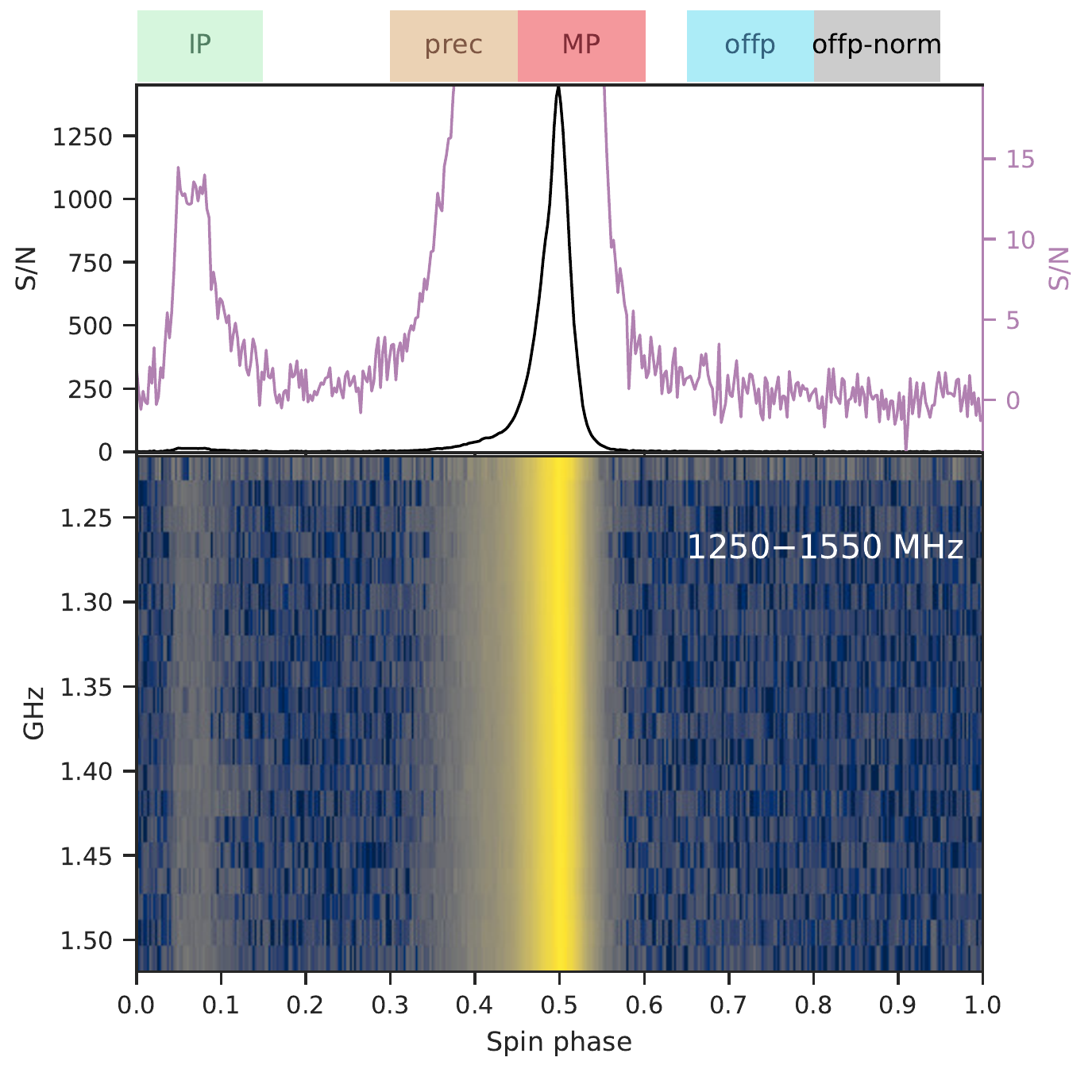}
 \caption{Average profiles (upper row) and spectra (lower row) of \src\ in 
 VHF band (left) and L-band (right). Here, the average profiles are normalized 
 by the mean and standard deviation of the signal in specifically selected
 ``off-pulse-norm'' window of the frequency-integrated data. The spectra are 
 normalized by the profile maximum in each frequency subband and the logarithm of 
 the absolute value of the signal is plotted in order to highlight faint spectral 
 features. For the average profiles, the light purple curves show the zoom-in to 
 the low signal-to-noise (S/N). The colored boxes at the top mark the borders of  
 selected phase windows. These cover the interpulse (IP), precursor, main pulse 
 (MP), off-pulse (``offp'') for the noise-induced fluence variability calculations, 
 and another off-pulse window for baseline subtraction (``off-pulse-norm'').}
\label{fig:AP}
\end{figure*}

All single-pulse data were next corrected for the dispersive delay due to propagation 
through the ionized interstellar medium (ISM).  Determining the dispersion measure (DM) 
is not straightforward, since the L-band observations are not sensitive to small DM 
variations, and prominent profile evolution in the VHF band confounds the measurement 
of the absolute DM \citep{Hassall2012}.  We used a hybrid method to make a model for 
profile evolution, which was then used to measure the DM for each VHF observation.  
The earliest VHF observation was aligned against a four-component Gaussian model in 
which the amplitudes and widths were all allowed to vary as a function of a frequency, 
but their positions were fixed \citep{Pennucci2015}; this alignment permits a 
frequency-dependent separation between the two observed profile components while 
maintaining the usual assumption of a fixed-in-frequency fiducial point, such as the 
mid-point between the observed components.  We then model the profile evolution in this 
aligned VHF observation following \citet{Pennucci2019}, in which a principal component 
analysis is used to reproduce the frequency dependent shape, resulting in a higher 
fidelity, noise-free template than when using a Gaussian component decomposition. The 
DM from each VHF observation was then determined by effectively measuring the $\nu^{-2}$ 
dependence of the delays of the profiles relative to this template \citep{Pennucci2014}.  
The DMs from the April VHF observations are $\sim 1\sigma$ consistent, and they are 
different from the January VHF DM by $\delta\mathrm{DM}=5\times 10^{-4}$\,pc~cm$^{-3}$, 
which amounts to a linear trend in DM of $-1.8\times 10^{-3}$\,pc~cm$^{-3}$~yr$^{-1}$. 
A trend of such magnitude for a pulsar with $\mathrm{DM}\approx 3$\,pc~cm$^{-3}$ is not 
uncommon \citep{Hobbs2004, Bilous2016}. 

All L-band data were dedispersed using the April 2020 VHF DM value of 2.97052\,pc~cm$^{-3}$. 
Assuming that the DM follows the linear trend calculated above throughout all eight months 
of L-band observations, the maximum pulse delay across the L-band would be $\sim2$\,$\mu$s, 
much smaller than our time resolution.
To each L-band observation a phase offset was applied that placed the peak of the 
average profile at phase 0.5. In VHF, the 0.5 phase was assigned to the dip between 
the two strongest components of the main profile.

\section{Average profile}

The average radio profile of \src\ consists of two components, a main pulse (MP) 
and an interpulse (IP), separated by roughly 0.4 in spin phase (Fig.~\ref{fig:AP}). 
In the VHF band, the MP is composed of two separate components which move further 
apart toward lower frequencies. The leading edge of the MP exhibits a flat 
shoulder (hereafter the ``precursor``). The pulses originating in this precursor 
are known to have larger fluence variability (with respect to the average fluence 
in the same phase region) than the pulses from the MP \citep{Cairns2004,Smirnova2012}. 

Faint emission is, notably, also present in the bridge between IP and MP, similar 
to the previous reports \citep[e.g.,][at 430\,MHz]{Hankins1981}. The bridge emission 
becomes relatively stronger at lower frequencies.

For the subsequent analysis we have selected five phase windows of equal width 
(0.15 in spin phase), which roughly cover the IP component (starting at phase 0.0),
precursor (starting at 0.3), MP (0.45), and two noise windows, off-pulse (``offp'', 
0.65) and off-pulse normalization (``offp-norm'', starting at 0.8). For each pulsar 
rotation we subtracted the mean value calculated in the off-pulse normalization window.
We note that in the VHF band faint emission is present throughout virtually the full
pulsar period: single-pulse fluences may thus be underestimated, and the noise 
contribution overestimated.

The choice of window locations was affected by our wish to have windows of the same 
size, to ensure the noise contribution had the same magnitude for fluences calculated 
in all windows. In the VHF band, the precursor window contains some emission from the 
MP component. 

\begin{figure*}
\centering
\includegraphics[width=1.0\textwidth]{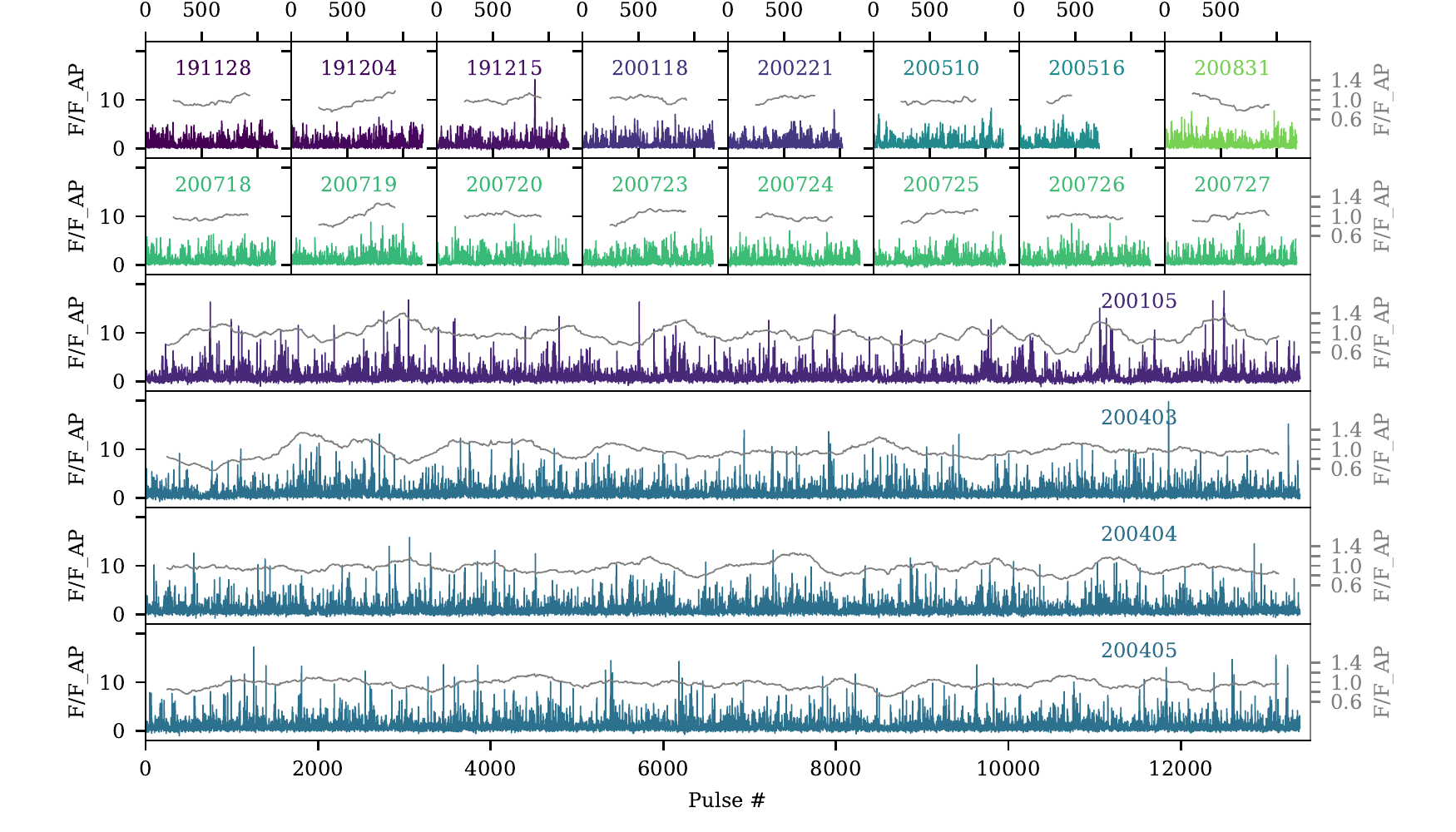}
\caption{Single-pulse fluences integrated within MP, precursor, and IP regions 
combined. The top two rows show L-band observations; the others correspond to the VHF 
sessions. Color encodes the epoch of observation, varying smoothly between darker blue 
(first session, November 2019) to lighter green (last session, August 2020). Grey 
lines mark the rolling 500-pulse average, with corresponding fluence scale displayed 
on the right. The observing date is noted for each session, in \textit{yymmdd} format.}
\label{fig:Etemp}
\end{figure*}

We have used uncalibrated data for this study, so we normalize the single-pulse data 
by the period-integrated fluence \FAP\  of the average profile (AP), in each observing 
session separately. Because of the small session duration, the source elevation did 
not change over the observation; and the fluences averaged in 2-min rolling windows 
do not exhibit variation on the scales larger than $\sim 10$\,min for the longer VHF 
sessions (Fig.~\ref{fig:Etemp}). Shorter-term variations in both bands can be 
attributed to scintillation (see Sec.~\ref{subsec:scintl}).

\begin{figure*}
\centering
\includegraphics[width=0.95\textwidth]{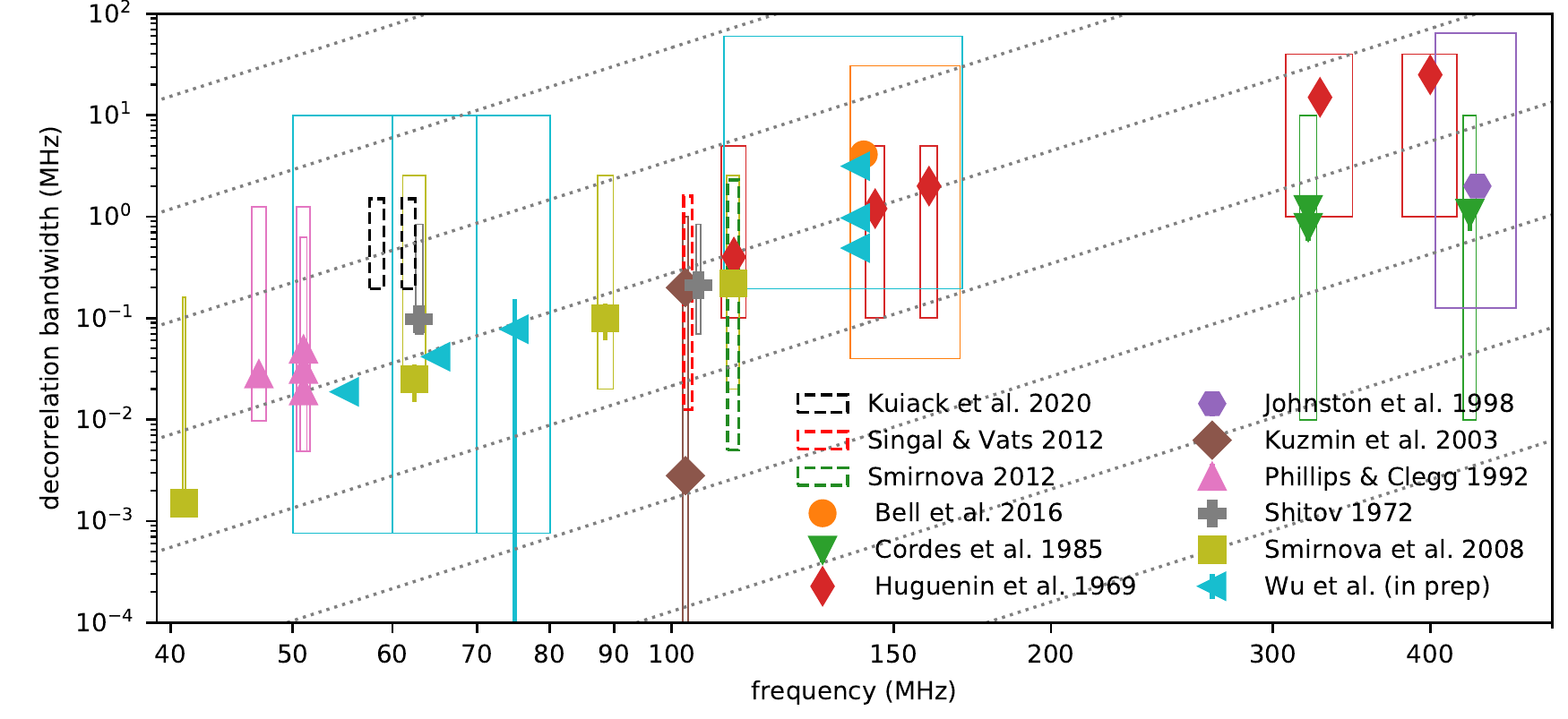}
\caption{Measurements of the decorrelation bandwidth of diffractive scintillation
versus observing frequency for \src.  For Wu et al. (in prep.), the markers at 
145\,MHz indicate the median as well as 2.5 and 97.5 percentiles of a series of \fdiff\ 
measurements. The width and height of the same-color rectangle around each marker 
shows observing bandwidth as well as  minimum and maximum spectral resolution 
(corresponding to channel and band widths, respectively), 
respectively. Dashed rectangles show, in the same manner, the observing setup 
for three observing campaigns which did not make direct measurements of \fdiff, 
but reported abrupt increases in \src's intensity. Dotted diagonal lines 
illustrate $\nu^{4}$ dependence. }
\label{fig:dec_bw}
\end{figure*}

\section{Single-pulse properties}

\subsection{Long-term single-pulse fluence variability and scintillation}
\label{subsec:scintl}

\subsubsection{Previous studies}
Numerous observing campaigns conducted at frequencies below 200\,MHz have reported 
high variability of the flux density exhibited by \src\ 
\citep{Gupta1993,Singal2012,Smirnova2012,Bell2016}, with occasional peaks several 
times higher than the average flux density. At least partly, this variation is 
caused by scintillation in the ISM, which thus must be taken into account when 
analysing single-pulse fluence distributions. 

For the observations below 200\,MHz, the existing measurements of the decorrelation 
bandwidth of the diffractive scintillation (Fig.~\ref{fig:dec_bw}) indicate that 
scintillation is in its strong regime \citep{Rickett1990}.  In this regime, the 
magnitude of the observed scintillation-induced variation of the signal depends 
on the number of scintles averaged within the observing bandwidth (BW) and 
integration time ($T_\mathrm{intg}$). Such variation is commonly characterized by 
the modulation index, defined as the ratio of the standard deviation of the signal 
to its mean: 
\begin{equation}
m = \sigma_\mathrm{F} / \langle F \rangle.  
\end{equation}
Following \citet{Cordes1991}, the modulation index for diffractive scintillation 
can be expressed as follows:
\begin{equation}
 m^{\mathrm{obs}}_{\mathrm{DISS}} \approx \left[\left(1+\kappa N_t \right)\left(1+\kappa N_\nu\right)\right]^{-1/2},
\end{equation}
where $\kappa$ is an empirical factor on the order of 0.1--0.2, while $N_\nu$ and 
$N_t$ are the number of averaged scintles with characteristic dimensions \fdiff\ 
in frequency and $\delta t_\mathrm{DISS}$ in time:
\begin{equation}
N_t = \dfrac{T_\mathrm{intg}}{\delta t_\mathrm{DISS}}, 
N_\nu = \dfrac{\mathrm{BW}}{\delta \nu_\mathrm{DISS}}. 
\end{equation}
If $N_\nu N_t \gg 1$, the signal strength is distributed according to a Gaussian 
law with the following survival function (SF, probability of measuring a value 
larger than a threshold $a$):
\begin{equation}
P(F> a \langle F \rangle) = \dfrac{1}{2} - \dfrac{1}{2} \mathrm{erf}\left( \dfrac{ a -1}{m\sqrt{2}}  \right),
\end{equation}
where $\mathrm{erf}$ is the Gauss error function.
If the number of scintles averaged is small ($N_\nu N_t \sim 1$), then the 
signal strength follows an exponential distribution with the following SF 
\citep{Cordes1991}:
\begin{equation}
\label{eq:scint}
P(F> a \langle F \rangle) = e^{-a}.
\end{equation}

Although  outliers are present, the bulk of the \fdiff\ measurements included 
in Fig.~\ref{fig:dec_bw} suggest that between 45 and 200\,MHz, \fdiff\ is 
described by the following relation:
\begin{equation}
\label{eq:fdiss}
\nu_\mathrm{DISS}(\nu) \approx 200\,\mathrm{kHz}\left(\dfrac{\nu}{100\,\mathrm{MHz}}\right)^{4} .
\end{equation}
The measured \fdiff\ is smaller than predicted by Eq.~\ref{eq:fdiss} at 40\,MHz 
and around 400\,MHz. At the higher frequencies the values obtained in different 
works disagree by an order of magnitude \citep{Cordes1985, Johnston1998,Huguenin1969}. 
Some of this disparity may be caused by differences in observing setup, such as 
the session duration (maximum integration time), observing bandwidth, and 
frequency resolution: \fdiff\ is much less likely to be measured properly in cases 
where it is close to the band- or channel width. Similar reasoning applies to 
the scintillation timescale. At the same time, \citet{Kuzmin2003} reported on two 
\fdiff\ scales  simultaneously present in their data at the frequencies close 
100\,MHz. The smaller scale is roughly in agreement with \citet{Cordes1985} and 
\citet{Johnston1998}, adjusted for frequency evolution.

There is also an evidence of occasionally large decorrelation bandwidth. Recent 
wideband low-frequency observing campaigns showed that \fdiff\ at 145\,MHz sometimes 
becomes as large as a few MHz, several times larger than the median value. 
\citet{Bell2016} reported \fdiff\ of 4\,MHz in one of their eight observing sessions 
with the Murchison Widefield Array. Wu et al.\ (in prep.) analyzed a series of 303 
observations at 145\,MHz, spread over six years, using six German LOFAR stations, and
found that in 2.5\% of cases the \fdiff\ of \src\  was larger than 3.1\,MHz, with two 
observations yielding $5.5\pm1$\,MHz. It is worth noting that a large, five-fold 
variation in \fdiff\ has been detected from another nearby pulsar, PSR J0030+0451. 
This variation  was attributed to anisotropies in the density or the magnetic field 
structure of the local ISM \citep{Nicastro2001}. 

Direct measurements of the scintillation timescale $\delta t_\mathrm{DISS}$ at 
145\,MHz yield values of 15--40 min, with a larger $\delta t_\mathrm{DISS}$ being 
generally accompanied by a larger \fdiff\ (Wu et al.\ in prep.). \citet{Bell2016} 
measured $\delta t_\mathrm{DISS} = 28.8$\,min during their event of increased source 
brightness. These measurements correspond to 10--25 min at 100\,MHz using the 
$\nu^{1.2}$ scaling from \citet{Lorimer2005}, and agree well with measured 
$\delta t_\mathrm{DISS} = 400\pm 100$\,s at 60\,MHz from \citet{Smirnova2008}.  

To our knowledge, the work of \citet{Bell2016} contains the only published 
simultaneous measurements of flux density variations together with scintillation 
parameters. The latter were measured during an increased brightness event when 
the flux integrated for 112\,s within 30-MHz bandwidth increased by a factor 
of 6 with respect to the flux measured during other sessions. Visually, the 
dynamic spectrum on their Fig.~4 contained three scintles. Using Eq.~\ref{eq:scint}, 
for this low-scintle regime, the probability of a flux increase larger than six is 
$e^{-6}\approx0.0025$. Considering that the large \fdiff\ event is also quite rare, 
this level of variability would be quite hard to detect with only a few 
observing sessions.

The same holds true for \citet{Smirnova2012} and \citet{Singal2012}. Extrapolating 
the possible range of scintillation parameters at 150\,MHz 
to their frequencies of 103\,MHz, we see that for the former work, $N_\nu\sim 1 - 20$, 
$N_t <1$ while for the latter, $N_\nu\sim 1 - 20$, $N_t \sim 1-3 $. For the lowest 
possible number of scintles, the reported flux density variations of 6 and 8 are too strong to 
be described by Eq.~\ref{eq:scint} based on the number of observing sessions. On 
the other hand, the shape of the flux density curve versus time is similar to the 
low-scintle regime, with strong peaks and long troughs (Fig.~4 in \citealt{Cordes1991}). 
Perhaps refractive interstellar scintillation is providing extra variability. 
Accounting for that is difficult, since for {\src} the long-term flux modulation indices and refractive 
scintillation timescales are known to deviate from the textbook predictions \citep{Gupta1993}, 
requiring more complex models of ISM inhomogeneties \citep{Goldman2021}.

Diffractive scintillation may be the source of the bright 195.3-kHz spectral features 
of one-second snapshots during the periods of high single-pulse activity in \citet{Kuiack2020}:
\fdiff(60\,MHz)\,$\approx$\,0.2\,MHz corresponds to \fdiff(145\,MHz)\,$\approx$\, 6.8\,MHz.
This value is higher, but close to the largest \fdiff\ values that have been directly measured
at these frequencies. We note that the exact 
frequency extent of the correlation in \citet{Kuiack2020} is not known very well because of the coarse 
resolution. A dedicated long-term study of the dynamic spectra of \src\  below 
100\,MHz is needed to test the scintillation hypothesis;  it is, however, indirectly 
corroborated by the temporal correlation of the bright features on the timescales 
of few minutes, similar to the measured $\delta t_\mathrm{DISS} = 400\pm 100$\,s 
at 60\,MHz \citep{Smirnova2008}.

\begin{figure*}
\centering
\includegraphics[width=0.5\textwidth]{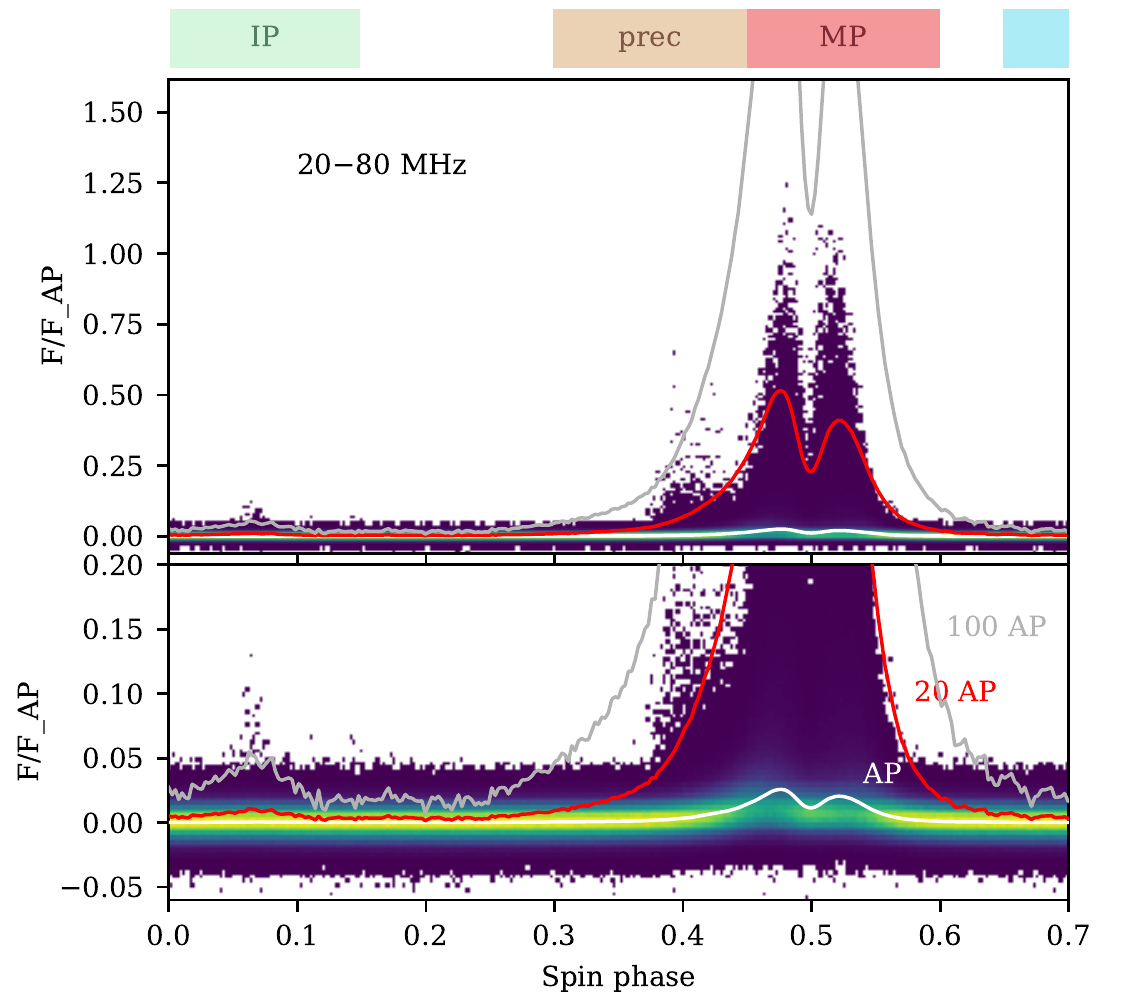}\includegraphics[width=0.5\textwidth]{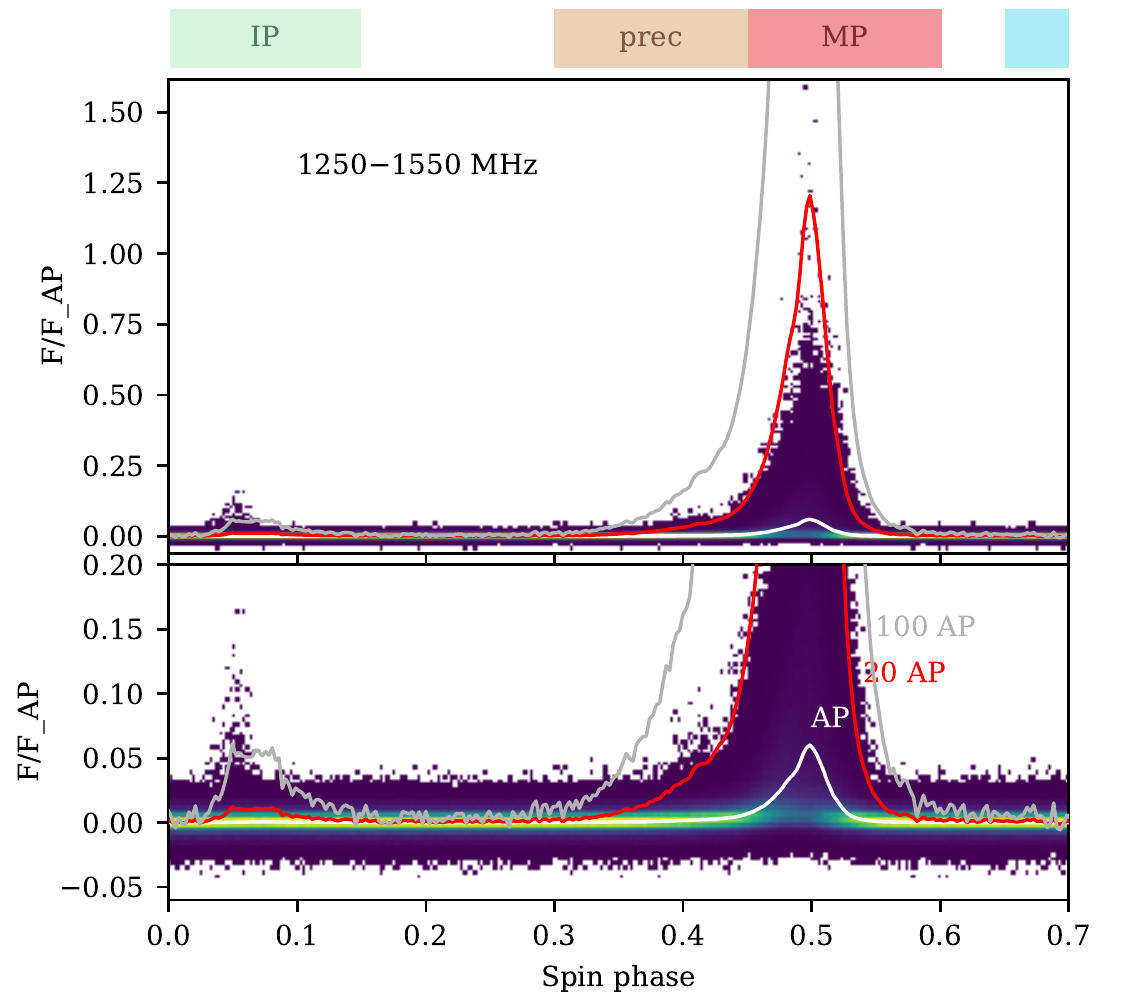}
\caption{Longitude-resolved fluence distributions per $\approx650$\,$\mu$s, 
normalized by the average period-integrated fluence for each observing session. 
On 2D histogram, lighter colors correspond to larger number of counts in the given 
(phase $\times$ fluence) bin, with the darkest color corresponding to one count. The 
average phase-resolved fluence and its $20\times$ and $100\times$ multiplicatives are shown by 
white, red and gray lines, respectively. Spin phases beyond 0.7 are not shown.}
\label{fig:Fluence_2D}
\end{figure*}

\subsubsection{This work}
The VHF observations were carried out with a moderate frequency resolution, of 195\,kHz. 
Although scintillation was visible in the dynamic spectrum at the top of the band, 
the frequency resolution was too coarse to measure the width of individual scintles 
with the standard autocorrelation function analysis. 

In this work we investigate the fluences integrated over the 20--83\,MHz frequency 
range. Even in the case of unusually large \fdiff\ of $\approx 0.2$\,MHz at 60\,MHz, 
our band would contain dozens of scintles. Very roughly -- neglecting the frequency 
dependence of the scintle size within the band -- the modulation index for 
session-averaged fluence is $<0.15$ and for single-pulse fluence $<0.18$.

In L-band, \fdiff\ is much larger than our observing bandwidth and the scintillation 
is in its weak regime \citep{Rickett1990,Lorimer2005}. The expected modulation index 
is 0.1--0.3 for single-pulse and session-integrated fluences.

\begin{figure}
\centering
\includegraphics[width=0.25\textwidth]{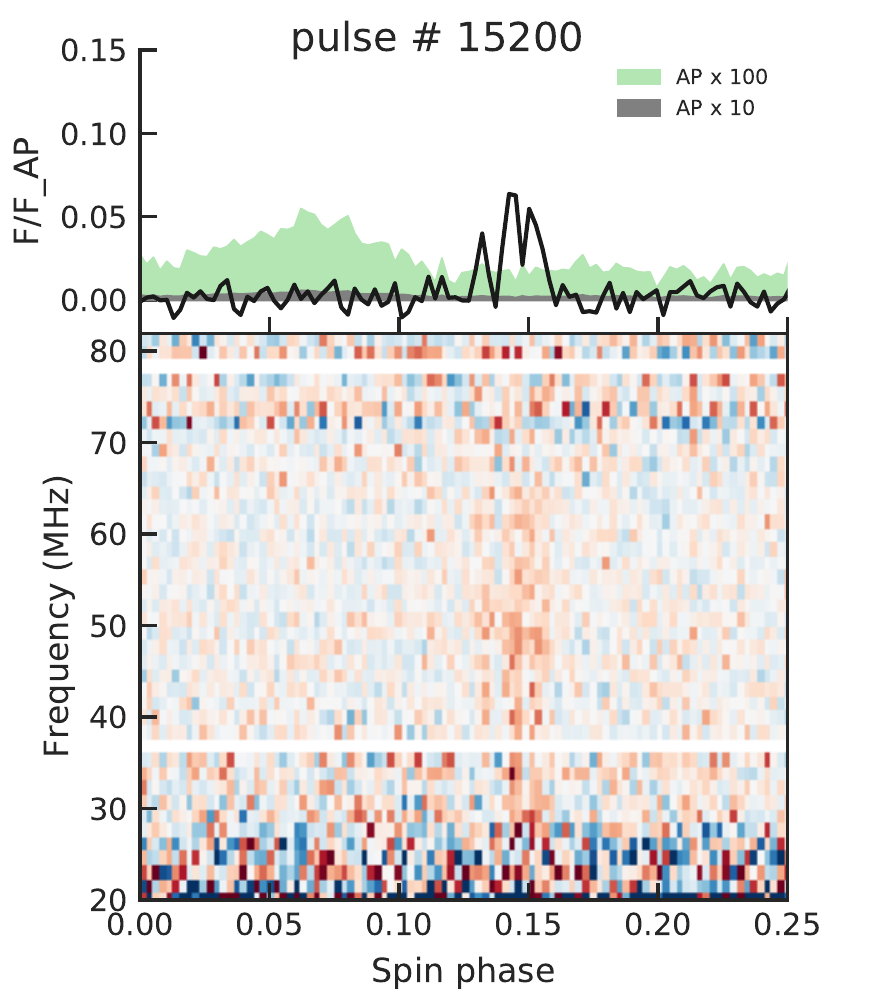}\includegraphics[width=0.25\textwidth]{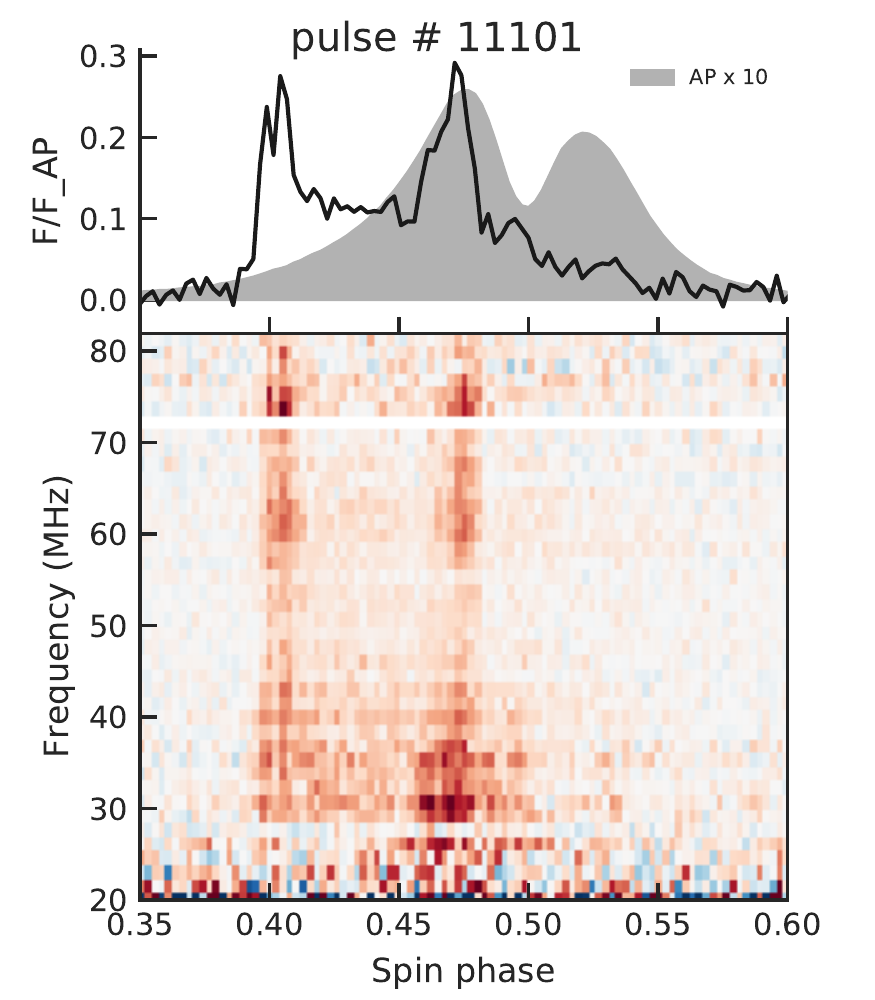}
\caption{Strongest pulse from the bridge phase region (\textit{left}) and an example 
of a strong pulse with components in both the precursor and 
 MP phase regions (\textit{right}). \textit{Upper plot}: 
band-integrated signal, with shaded regions marking $10\times$ and (left only) $100\times$ the average 
pulse strength. \textit{Lower plot}: pulse spectrum with each frequency subband 
normalized by the period-integrated fluence of the average pulse in the same 
subband. The colorbar span is set by the maximum value on the spectrum of this 
particular pulse. Here and for the other VHF pulses the maximum value is calculated 
for frequencies above 30\,MHz.}
\label{fig:bridge_pulse}
\end{figure}

\subsection{Phase of occurrence}

\begin{figure*}
\centering
\includegraphics[width=0.5\textwidth]{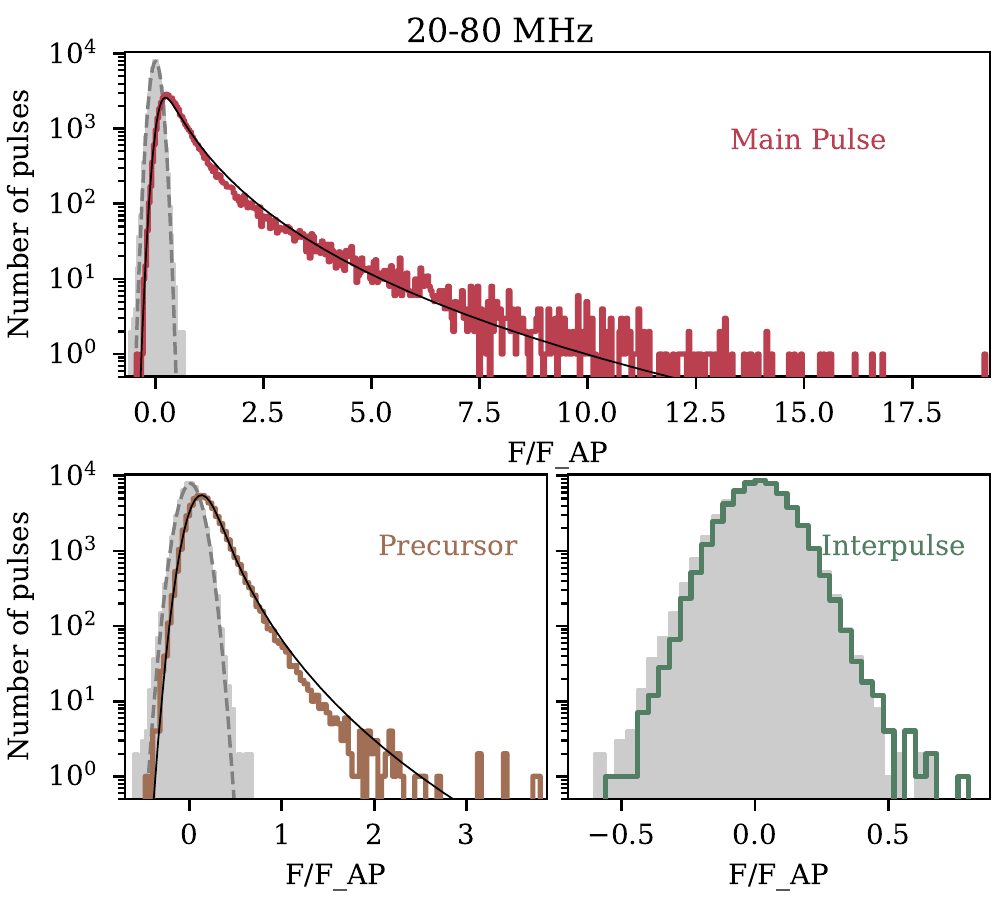}\includegraphics[width=0.5\textwidth]{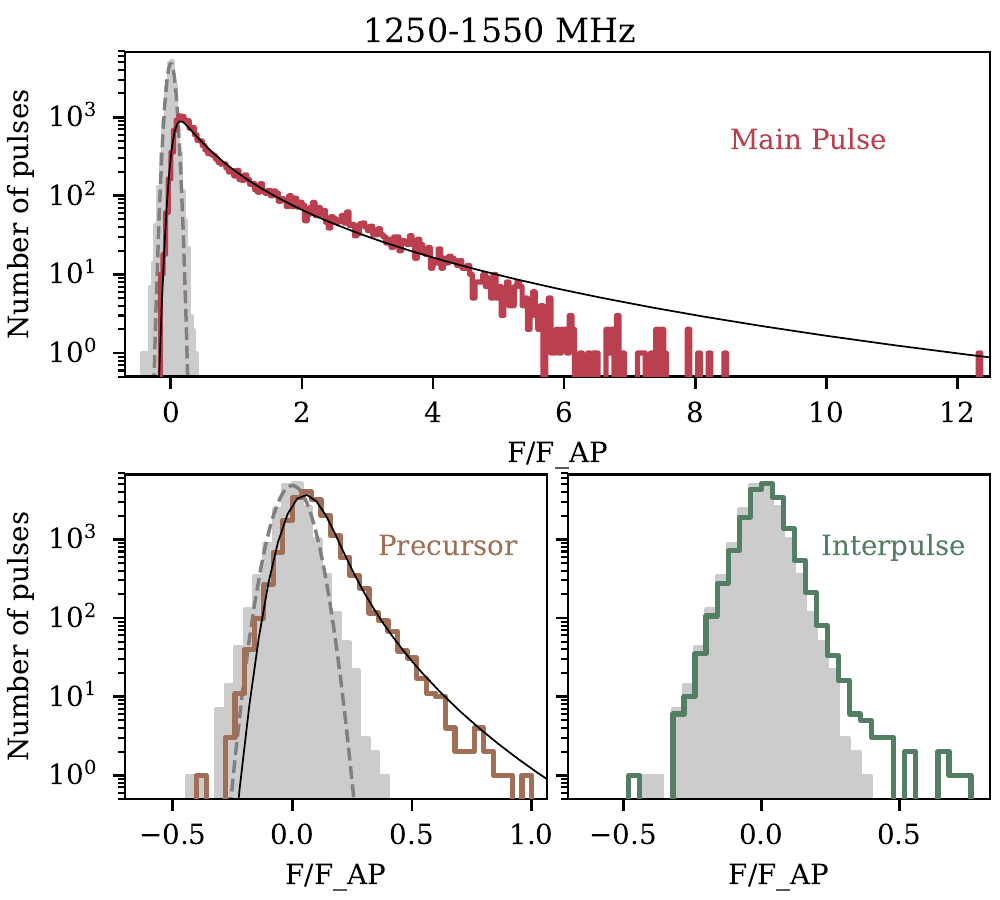}
\caption{Fluence histograms for three selected phase windows (color lines). 
Fluences are normalized by the average profile fluence integrated over whole 
period during the respective observing session. The distribution of fluences 
in the off-pulse window is shown with gray shade. For off-pulse window, gray 
dashed line indicates normal distribution based on mean and standard deviation
of fluences integrated over this window. For main pulse and precursor window, 
the best-fit convolution of lognormal and Gaussian noise distributions is 
shown with black line (see text for details).}
\label{fig:Fdistr_all_win}
\end{figure*}

Figure~\ref{fig:Fluence_2D} shows the distribution of the single-pulse fluences
integrated over $2.6\times10^{-3}$ of spin phase (equivalent to the time resolution 
of the VHF data), and normalized by the period-integrated mean fluence \FAP\ of
each respective session. This method of plotting highlights potential candidates 
for GPs: narrow pulses with fluences comparable to or exceeding \FAP\ which come 
from specific phase regions, such as the  leading or trailing edges of the profile 
components.

The brightest single pulses in our observations occur at the peak phases  of the 
respective average profile components. The pulse-to-pulse fluence variability 
increases at lower frequencies in the precursor and MP regions and tentatively decreases 
in the IP region. In L-band, the phase-resolved fluence distribution is somewhat 
skewed toward the leading edge of IP and the trailing edge of MP. Such behavior 
is not observed in the VHF band.

We have visually inspected all periods with signal exceeding some set threshold outside 
on-pulse windows (0.045\FAP\ for the VHF and 0.035\FAP\ for L-band). In this manner, 
two distinct single pulses were found in the bridge region of two VHF observations.  
Both pulses came from the spin phase of $\approx 0.15$, well outside the phase 
region for a discernible IP profile. The stronger pulse had a peak fluence of 0.064\FAP\, 
the other was 30\% fainter. In both cases these are over a hundred times larger than 
the local average emission fluence. Both pulses had a width of about 7\,ms, comparable 
to IP pulses and similar several-component broadband spectra (Fig.~\ref{fig:bridge_pulse}). 

The bridge pulses found in the VHF band bear a curious resemblance to the unusually 
strong pulses found by \citet{Cairns2004} at the leading edge of precursor and the 
trailing edge of MP component, at 430\,MHz. Within 7072 pulsar rotations, the authors 
detected a few such pulses, one of which (corresponding to our spin phase 0.35) was 
more than 100 times brighter than the average profile at that phase. The pulses had 
width of a few ms, similar to our bridge pulses.

So far, little can be inferred about the origin of bridge pulses and whether they 
represent a different pulse population. \citet{Cairns2004} speculate they may be 
related to the micro GPs seen in the Vela pulsar \citep{Kramer2002}. Although the 
bridge pulses easily outshine the average emission in their respective phases, they 
are much dimmer than regular MP or IP single pulses. They are, furthermore, prone 
to being overlooked by  the standard practice of only investigating single pulses in 
the on-pulse windows. We urge more sensitive, dedicated studies  be carried out, to 
explore these bridge pulses further.

Our VHF observations did not confirm a few single-pulse properties that were reported 
previously from observations with BSA telescope in Pushchino Observatory. 
\citet{Smirnova2012} noted that at 112\,MHz, strong pulses in the  precursor region 
were accompanied by an absence of emission during main pulse, and vice versa. This 
was interpreted as evidence supporting  models in which precursor pulses are due to 
induced scattering of the MP emission by relativistic particles of strongly 
magnetized plasma in the pulsar magnetosphere \citep{Petrova2008}. While our frequency 
span does not cover that of \citet{Smirnova2012}, we note that our strong precursor 
pulses occasionally have emission in the MP region as well (Fig.~\ref{fig:bridge_pulse}). 
Meanwhile, \citet{Kazantsev2020} reported on occurrence of strong pulses at 
frequencies close to 100\,MHz, which come predominantly from the leading part of 
the two-peaked MP component. This is not supported by our observations 
(Fig.~\ref{fig:Fluence_2D}). \citet{Shabanova2004} observed the variations in the 
detected shape of the average profile with the same instrument, resulting in 
variable relative subcomponent height in MP, which the authors concluded to be 
due to Faraday rotation.

\subsection{Single-pulse fluence distribution} 
\label{sec:cdf_fits}

\begin{table}
\begin{center} 
\caption{Average and maximum fluences integrated within selected phase windows normalized 
by the average fluence integrated over the whole spin period (\FAP). This average fluence 
includes the emission from bridge region, not covered by selected windows. 
\label{table:fnorm}}
\setlength\tabcolsep{1.5pt}
\renewcommand{\arraystretch}{1.2}
\begin{tabular}{|l|ccc|ccc|}
\hline 
& \multicolumn{3}{c}{VHF} & \multicolumn{3}{|c|}{L}\\
\cline{2-7}
$\phantom{\dfrac{\frac{1}{1}}{\frac{1}{1}}}$&
\parbox{0.8cm}{\centering $\dfrac{F_\mathrm{mean}}{F_\mathrm{AP}}$} &
\parbox{1.0cm}{\centering $\dfrac{F_\mathrm{max}}{F_\mathrm{AP}}$} &
\parbox{1.0cm}{\centering $\dfrac{F_\mathrm{max}}{F_\mathrm{mean}}$} &
\parbox{0.8cm}{\centering $\dfrac{F_\mathrm{mean}}{F_\mathrm{AP}}$} &
\parbox{1.0cm}{\centering $\dfrac{F_\mathrm{max}}{F_\mathrm{AP}}$} &
\parbox{1.0cm}{\centering $\dfrac{F_\mathrm{max}}{F_\mathrm{mean}}$}\\
\hline
Main Pulse & 0.80 & 19.8 & 25 & 0.90 & 12.2 & 14\\ 
Interpulse & 0.02 & 0.8 & 46 & 0.01 & 0.73 & 52\\
Precursor & 0.15 & 3.5 & 23 & 0.08 & 1.0 & 12\\
Offpulse & 0.01 & 0.7 & $-$ & 0.00 & 0.4 & $-$\\
\hline 
\end{tabular} 
\end{center}
\end{table}

We  measured the fluences of pulses originating from the phase regions of 
interest, by integrating the signal within MP, precursor and IP phase regions. 
It is customary to normalize the fluences by the ``mean pulse'' fluence,
although it remains a matter of choice whether to normalize by the mean 
fluence in the respective phase windows or by the mean fluence integrated 
over the whole spin period. In this work, we used period-integrated fluences 
calculated in each session separately, which provided uniform normalization 
for all three main phase regions. Individual pulse fluences can be renormalized 
to the average fluences of their respective phase regions by using 
Table~\ref{table:fnorm}.

The resulting band-integrated fluence histograms are shown on Fig.~\ref{fig:Fdistr_all_win}.
The distribution in the off-pulse window (we note that the baseline was subtracted 
in a different window labeled ``off-pulse-norm'') resembles a Gaussian and 
occupies fluences between $\pm 0.5 F_\mathrm{AP}$. In both bands there is a small 
level of pulsar signal in the off-pulse window, so the mean fluence is somewhat 
larger than 0 (Table~\ref{table:fnorm}). The distribution of fluences from the 
IP window is dominated by noise and the maximum fluences are not larger 
than $0.8F_\mathrm{AP}$. However, they are about 50 times larger than the average 
fluence in the IP window, indicating the largest degree of variability among three 
emission windows. This degree must be taken with caution, since at this low level 
of average fluence its exact value is affected by details of baseline subtraction. 
Unlike MP, IP emission shows more variability at higher frequencies.

Although some VHF pulses in the precursor window have flux density values larger 
than $100\times$ the average flux density in the exact same phase window, their 
integrated fluences are only up to 20 times larger. The L-band data shows less 
fluence variability. For the MP phase region, the distribution in L-band appear 
to steepen at $F\gtrsim 5 F_\mathrm{AP}$, however relatively strong pulses with 
$F\gtrsim 8 F_\mathrm{AP}$ are also present. Interestingly enough, the fluence 
distribution at 430 MHz from \citet{Hankins1981} also shows steepening at similar 
fluences, and the shape of the distributions are qualitatively similar. 

Following a procedure similar to  \citet{BurkeSpolaor2012}, we fit the observed 
MP and precursor fluence distributions with lognormal distributions 
($P_\mathrm{intrinsic}$) convolved with a Gaussian model for the off-pulse region:
\begin{equation}
 P_\mathrm{observed} = P_\mathrm{intrinsic}\otimes P_\mathrm{offp}.
 \label{eq:PL_conv}
\end{equation}
$P_\mathrm{offp}$ was approximated with the probability density function of the 
normal distribution, with $\mu_\mathrm{N}$ and $\sigma_\mathrm{N}$ measured 
directly from the data:
\begin{equation}
\label{eq:norm}
\mathrm{N}(\mu_\mathrm{N}, \sigma_\mathrm{N}) \sim \frac{1}{\sigma_\mathrm{N}}\exp\left[-\frac{(F/F_\mathrm{AP}-\mu_\mathrm{N})^2}{2\sigma_\mathrm{N}^2}\right].
\end{equation}
The off-pulse flux density distribution appeared reasonably well described by the 
normal distribution, with some excess of pulses on both tails of the Gaussian curve, 
which we attribute to residual RFI. Because of this excess the formal reduced $\chi^2$, 
defined as sum of squared difference between model and data in the fluence bins with 
at least one pulse:
\begin{equation}
\label{eq:chi2}
 \chi^2 = \dfrac{1}{N_\mathrm{bin}-3}\sum \dfrac{(\mathrm{data} -\mathrm{ model})^2}{\mathrm{model}},
\end{equation}
was rather high ($>4$); we chose to proceed with a Gaussian model because of its 
simplicity.  

\begin{figure*}
\centering
\includegraphics[width=0.33\textwidth]{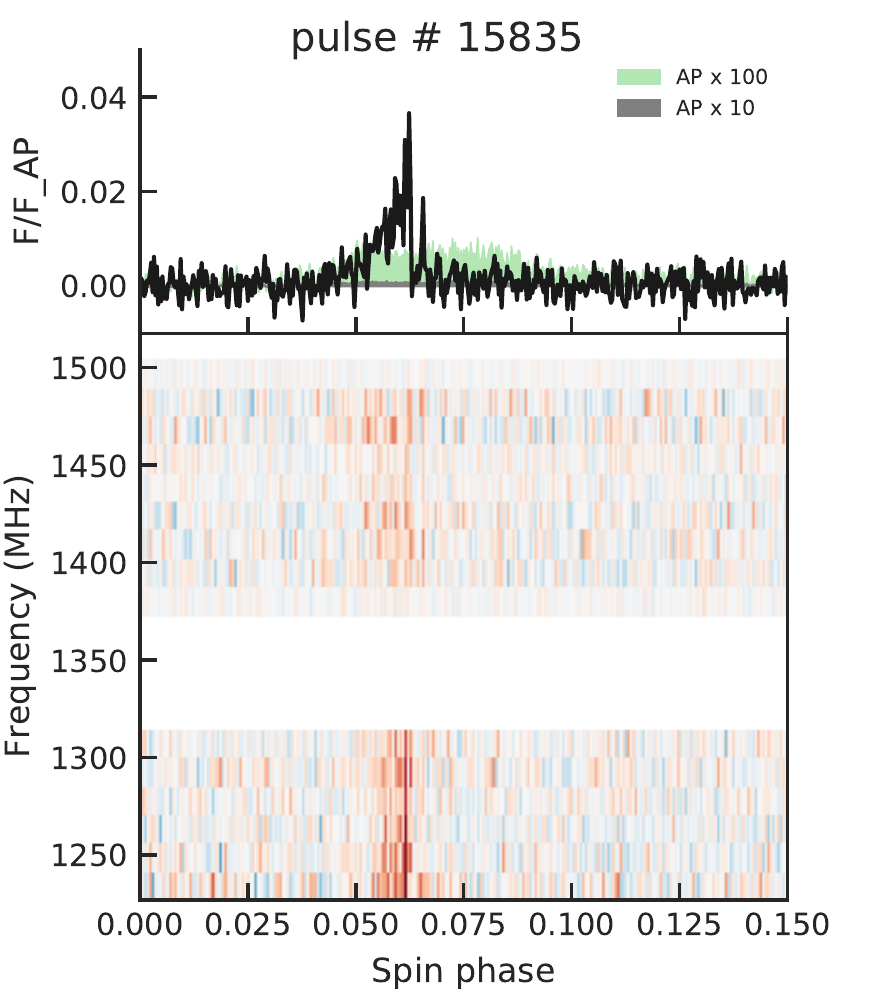}\includegraphics[width=0.33\textwidth]{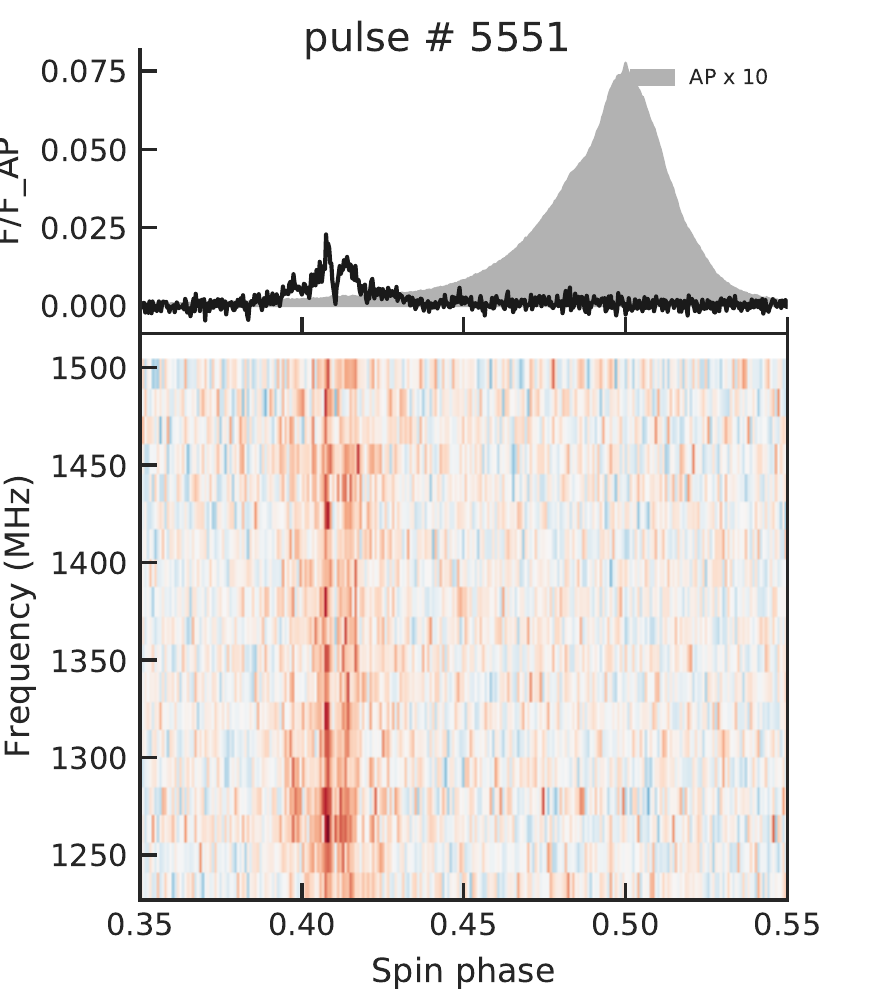}\includegraphics[width=0.33\textwidth]{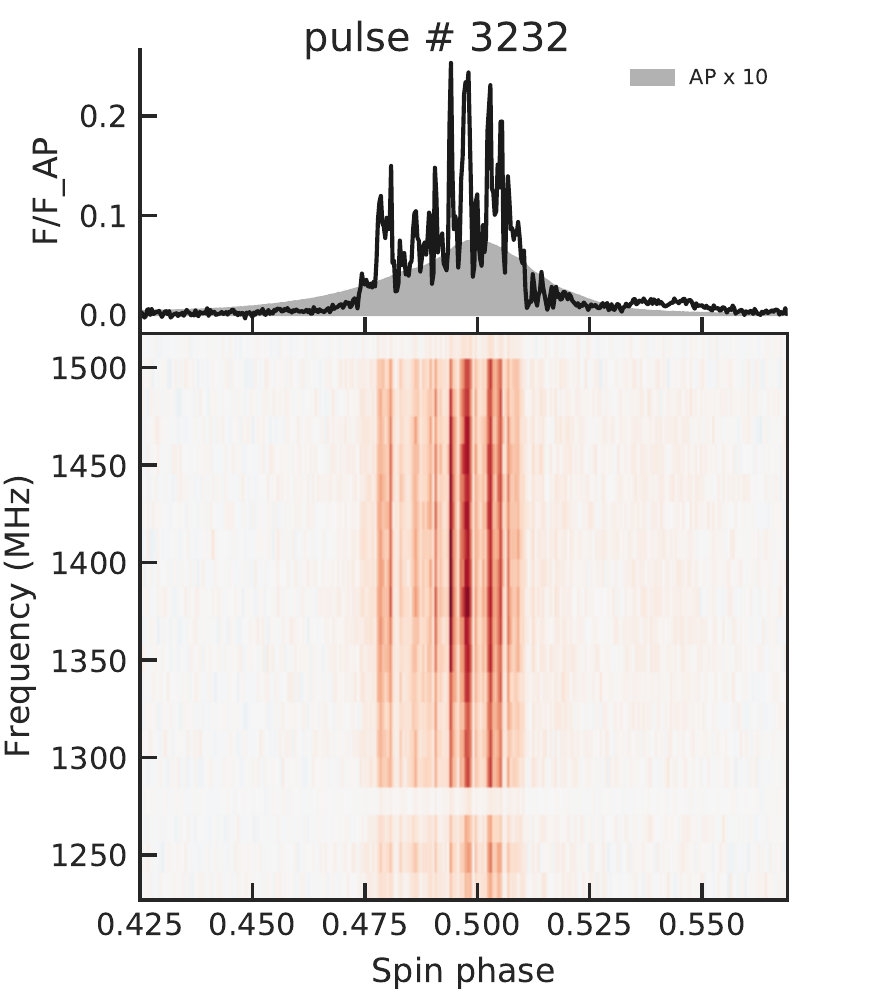}
\includegraphics[width=0.33\textwidth]{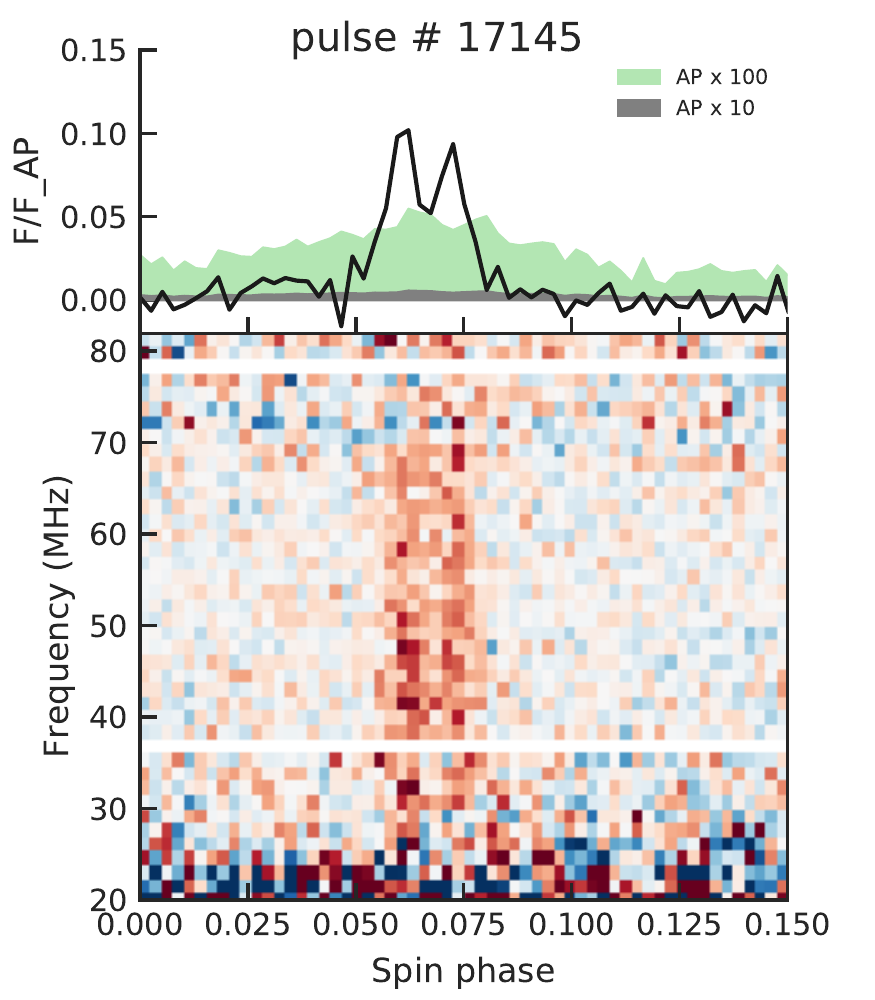}\includegraphics[width=0.33\textwidth]{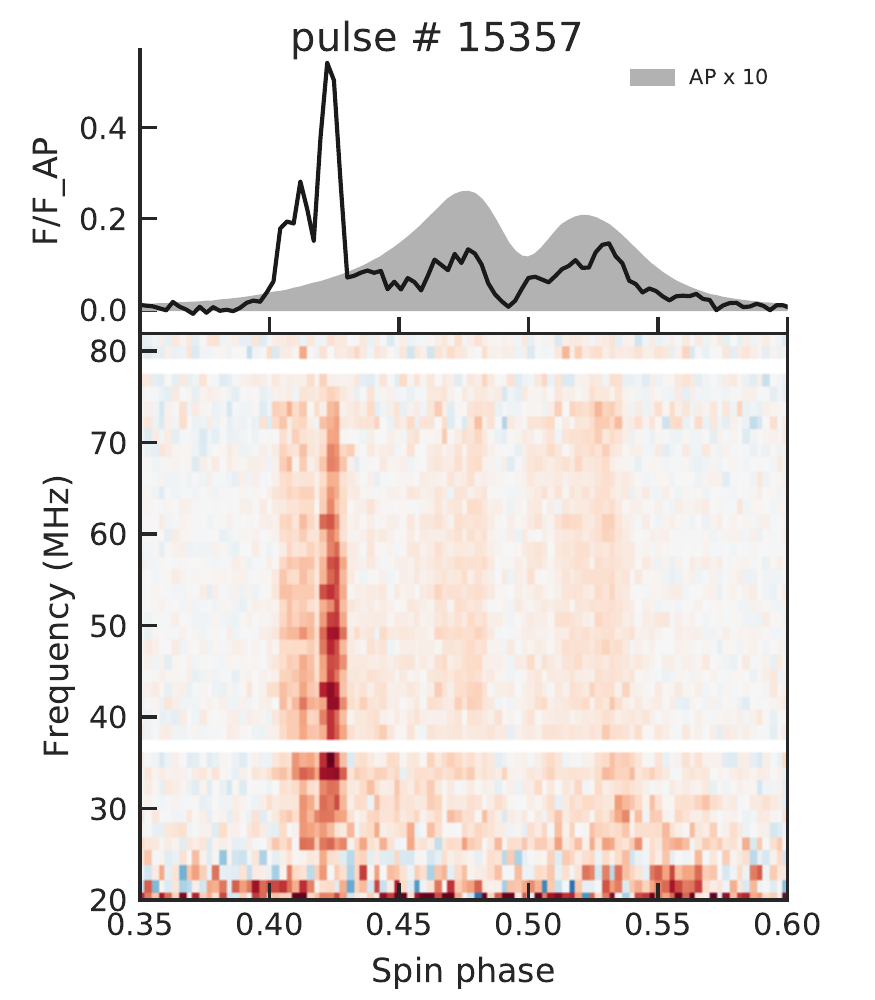}\includegraphics[width=0.33\textwidth]{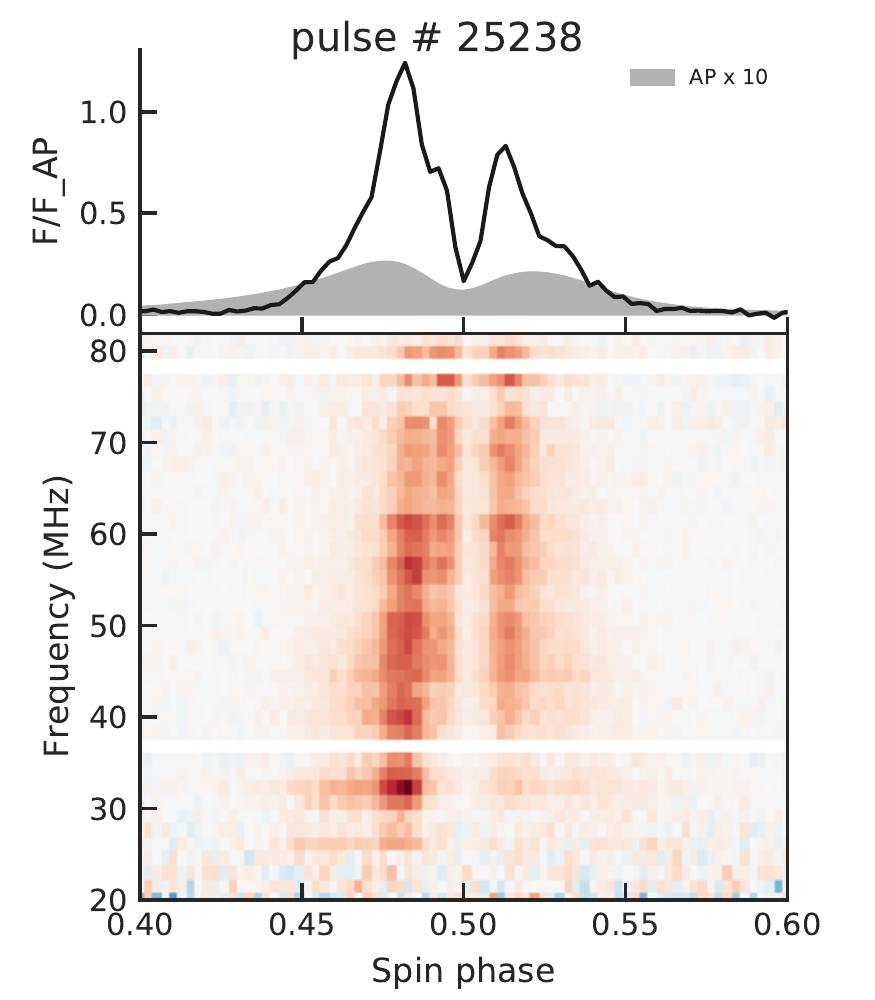}
\caption{Pulses with largest fluence in IP (left), precursor (middle) and MP 
(right column). \textit{Top row}: L-band, \textit{bottom row}: VHF. For 
the plotting conventions, see the caption to Fig.~\ref{fig:bridge_pulse}.
}
\label{fig:strongSP}
\end{figure*}

The intrinsic flux density distribution was modeled with log-normal (LN) distribution, 
with the same parametrization as in  
\citet{BurkeSpolaor2012}:
\begin{equation}
\label{eq:lognorm}
P_\mathrm{intrinsic}\equiv\mathrm{LN}(\mu_\mathrm{LN}, \sigma_\mathrm{LN}) \sim \frac{F_\mathrm{AP}}{F\sigma_\mathrm{LN} }\exp\left[ -\frac{[\log_{10}(F/F_\mathrm{AP})-\mu_\mathrm{LN}]^2}{2\sigma_\mathrm{LN}^2} \right].
\end{equation}
The convolution of $P_\mathrm{intrinsic}$ and best-fit $P_\mathrm{offp}$ 
(Eq.~\ref{eq:PL_conv}) was fit to the observed histogram $P_\mathrm{observed}$ 
by minimizing the $\chi^2$ statistics from Eq.~\ref{eq:chi2}.

\begin{table}
\begin{center} 
\caption{Estimated parameters of the Gaussian fluence distribution in the off-pulse 
window (Eq.~\ref{eq:norm}) and the best-fit values of the intrinsic lognormal 
single-pulse fluence distribution in the MP and precursor windows (Eq.~\ref{eq:lognorm}). 
Reduced $\chi^2$ values were calculated according to Eq.~\ref{eq:chi2}. \label{table:LNfit}}
\setlength\tabcolsep{1.5pt}
\renewcommand{\arraystretch}{1.2}
\begin{tabular}{|l|ccc|ccc|}
\hline 
& \multicolumn{3}{c}{VHF} & \multicolumn{3}{|c|}{L} \\
\cline{2-7}
$\phantom{\dfrac{\frac{1}{1}}{\frac{1}{1}}}$&
\parbox{0.8cm}{\centering $\mu$} &
\parbox{0.8cm}{\centering $\sigma$} &
\parbox{0.8cm}{\centering $\chi^2$} &
\parbox{0.8cm}{\centering $\mu$} &
\parbox{0.8cm}{\centering $\sigma$} &
\parbox{0.8cm}{\centering $\chi^2$}\\
\hline
Offpulse (Norm)& $\phantom{-}0.01$ & $0.11$ &  & $\phantom{-}0.00$ & $0.06$ & \\ 
Main Pulse (LogNorm)& $-0.35$ & $0.45$ & $5.3$ & $-0.28$ & $0.56$ & $\phantom{0}2.7$\\
Precursor (LogNorm)& $-0.89$ & $0.37$ & $3.0$ & $-1.28$ & $0.38$ & $23.2$\\
\hline 
\end{tabular} 
\end{center}
\end{table}

\begin{figure}
\centering
\includegraphics[width=0.5\textwidth]{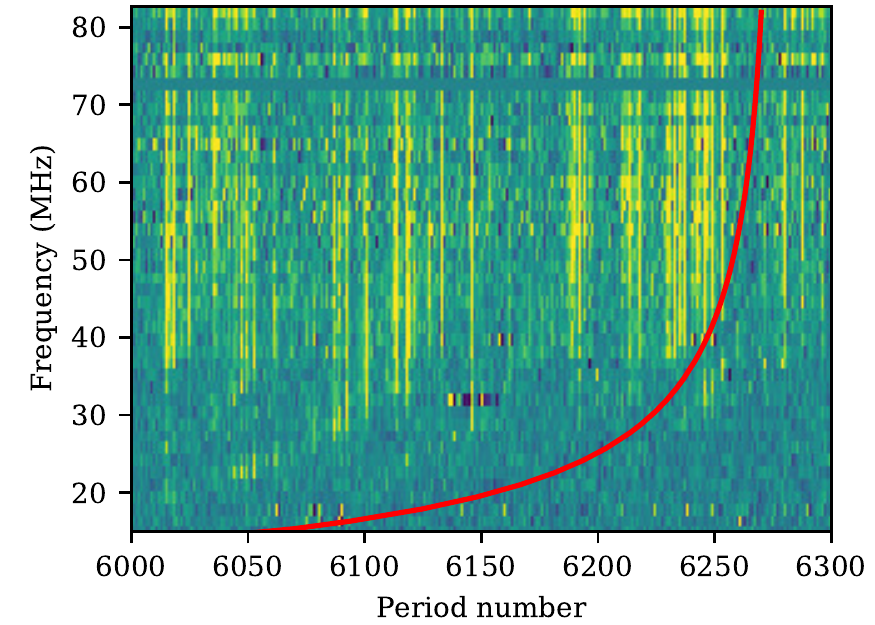}
\caption{Dynamic spectra for 300 pulse periods, averaged in MP+precursor phase 
region. Color is saturated. Red line shows dedispersed track for $\mathrm{DM}=0$.}
\label{fig:Dynspec}
\end{figure}

\begin{figure*}
\centering
\includegraphics[width=0.33\textwidth]{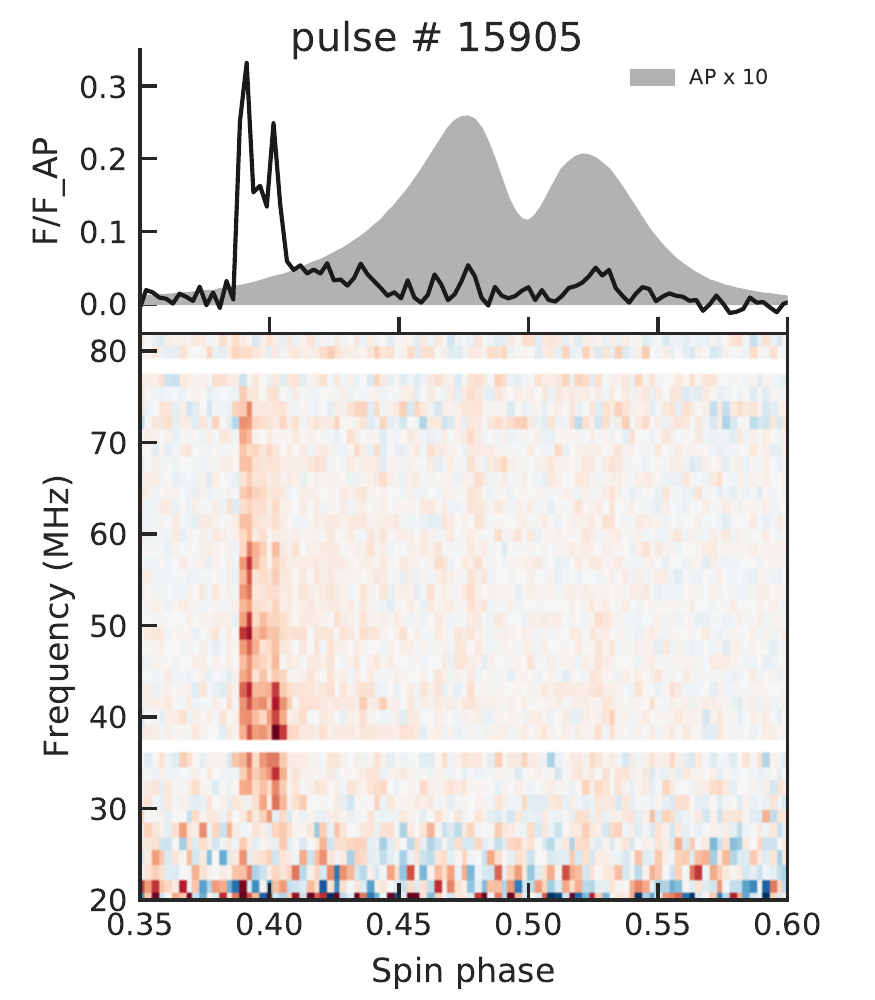}\includegraphics[width=0.33\textwidth]{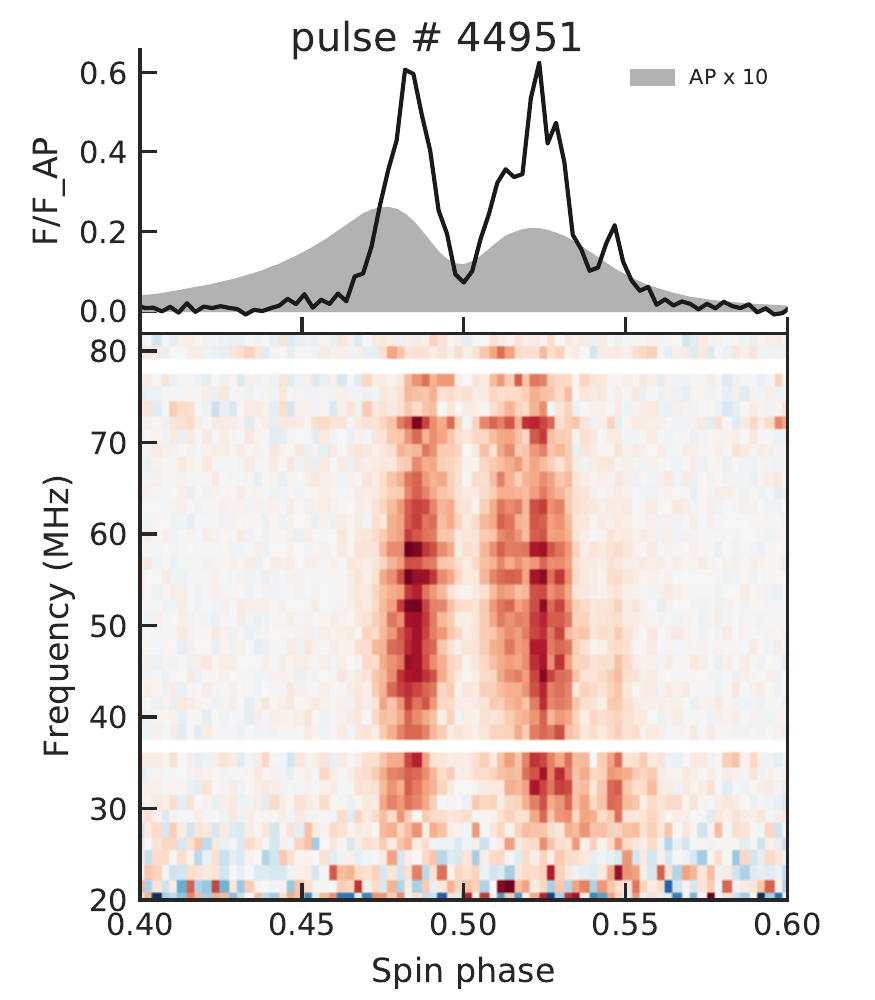}\includegraphics[width=0.33\textwidth]{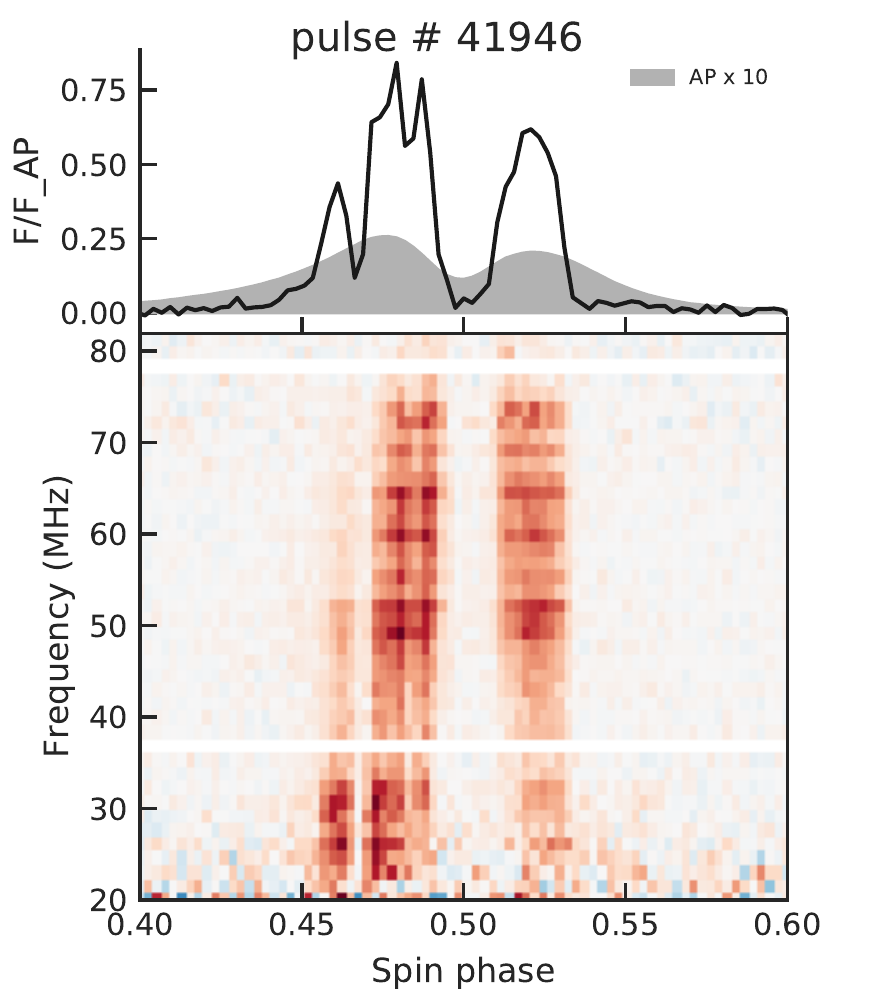}
\caption{Examples of subpulses drifting in frequency in precursor and MP. For 
the plotting conventions, see the caption to Fig.~\ref{fig:bridge_pulse}.}
\label{fig:SP_trombones}
\end{figure*}

The fit results are given in Table~\ref{table:LNfit}. The minimum reduced 
$\chi^2$ value was sensitive to the number of bins in the fluence histogram 
and was generally higher for the smaller number of bins. The much larger 
reduced $\chi^2$ value for the precursor window in L-band is mostly caused by 
the small number of bins per extent of the distribution. Increasing bin number 
results in similar best-fit parameter values, but a reduced $\chi^2$ comparable 
to the distributions from the other windows. For the MP window in L-band, 
$\chi^2$ indicates only a marginally good lognormal fit if
we follow the $\chi^2 = 2.8$ threshold accepted by \citet{BurkeSpolaor2012}. 
However, our best-fit values of $\mu_\mathrm{LN}$ and $\sigma_\mathrm{LN}$
are outliers in the distribution of corresponding quantities from \citet{BurkeSpolaor2012}. 
None of their lognormal fluence distributions have $\mu_\mathrm{LN}<0$. This could, 
in particular, stem from the different fluence normalizations we adopted: we 
normalize MP fluences by total mean \FAP, including signal from IP and precursor.
Re-fitting the distribution of MP fluences normalized by the average in MP phase 
window increases $\mu_\mathrm{LN}$, but it still remains negative. On the other 
hand, $\sigma_\mathrm{LN}$ is larger than almost all of corresponding values on 
\citet{BurkeSpolaor2012} sample, indicating an unusually large variability of the 
single-pulse fluences of {\src}. 

From the visual estimation in VHF, the fit underestimates the number of high-fluence 
pulses (we note that bins with one pulse were not used in $\chi^2$ minimization). In 
L-band there is a shortage of pulses with moderate fluences \citep[6--12\FAP, same 
as in][]{Hankins1981}, but the lognormal distribution still predicts the highest 
observed fluence correctly. It must be kept in mind that many factors may distort 
the true shape of single-pulse fluence distribution: non-Gaussian noise, contamination 
of MP fluences by precursor pulses, broadband intensity variations of the dispersed 
signal in VHF, scintillation, etc. The influence of these factors is generally 
hard to quantify, since it depends on the observing setup, preprocessing, and exact 
ways of calculating fluences. Nevertheless, some rough estimates may be derived 
from comparing fluence distributions obtained in different works at similar radio frequencies.
Detailed discussion is given in Sec.~\ref{subsec:distr_lit}, however we may state that 
unless for some reasons the fluence distributions are considerably offset from each 
other, at the same level of incidence probabilities the single-pulse fluences 
from different works may differ by  up to $\sim 25\%$.

\subsection{Pulse spectra}

A number of examples of individual pulse spectra from the strongest pulses in MP, 
precursor and IP regions are shown in Fig.~\ref{fig:strongSP}. In L-band the 
spectra do not exhibit any noticeable frequency dependence over the 300\,MHz 
bandwidth. The spectrum of the strongest, 12-\FAP\ pulse (\#3232) does not have 
any special features, compared to other MP pulses. The strongest pulses from 
precursor and IP region are much weaker than the MP pulses, but their spectra do 
not seem to be qualitatively different.

In the VHF band, the moderate time resolution did not allow us to explore the 
fine features of individual pulses on the same timescale as in L-band. Also, 
spectra of individual pulses below roughly 40\,MHz are affected by approximately
minute-long broadband fluctuations of the pulsar signal, possibly due to 
refraction or scintillation in the Earth ionosphere, or telescope issues 
\citep{Song2018,Fallows2016}. After compensating for dispersive delay, these 
intensity fluctuations manifest themselves as narrowband, bright patches that 
move upward in frequency, across several pulsar periods  (Fig.~\ref{fig:Dynspec}). 
Thus, any low-frequency bright spots on individual spectra must be interpreted 
with caution. 

Below 100\,MHz, many pulses appeared to consist of subpulses of limited 
bandwidth, which drifted upward on the leading edge of MP and downward 
on its trailing edge (Fig.~\ref{fig:SP_trombones}). Sometimes a downward 
(but no upward) drift was observed in the precursor region. This frequency 
drift is different from the artifacts caused by broadband variation of 
dispersed signal, as it happens on much smaller timescale and can go both 
ways in frequency. The strongest pulses from the IP and bridge phase regions 
(Fig.~\ref{fig:bridge_pulse} and \ref{fig:strongSP}) consist of broadband 
components with no sign of frequency drift. The small number of strong pulses 
does not, however, allow us to draw any definitive conclusions on the 
prevalence of the drift.

The positions of low-frequency subpulses do not seem to shift away from each 
other at low frequencies in a manner similar to components of the average 
profile (Fig.~\ref{fig:AP}). Interestingly, this same behavior was previously 
noticed during narrowband simultaneous low-frequency observations: 
\citet{Rickett1975} and \citet{Rickett1981} reported that micropulse positions 
are highly correlated between 111 and 318\,MHz and that the distance between 
them remains frequency-independent. 

In our observations, the spectra of individual pulse components had widths of 
$\gtrsim 30$\,MHz. We can conjecture that the spectral width is frequency-dependent 
and gradually increases with frequency. According to \citet{Rickett1975}, 
individual micropulses have wideband spectra spanning at least from 111 to 
318\,MHz and in L-band they are larger than our band width of 300\,MHz. 

\begin{figure*}
\centering
\includegraphics[width=\textwidth]{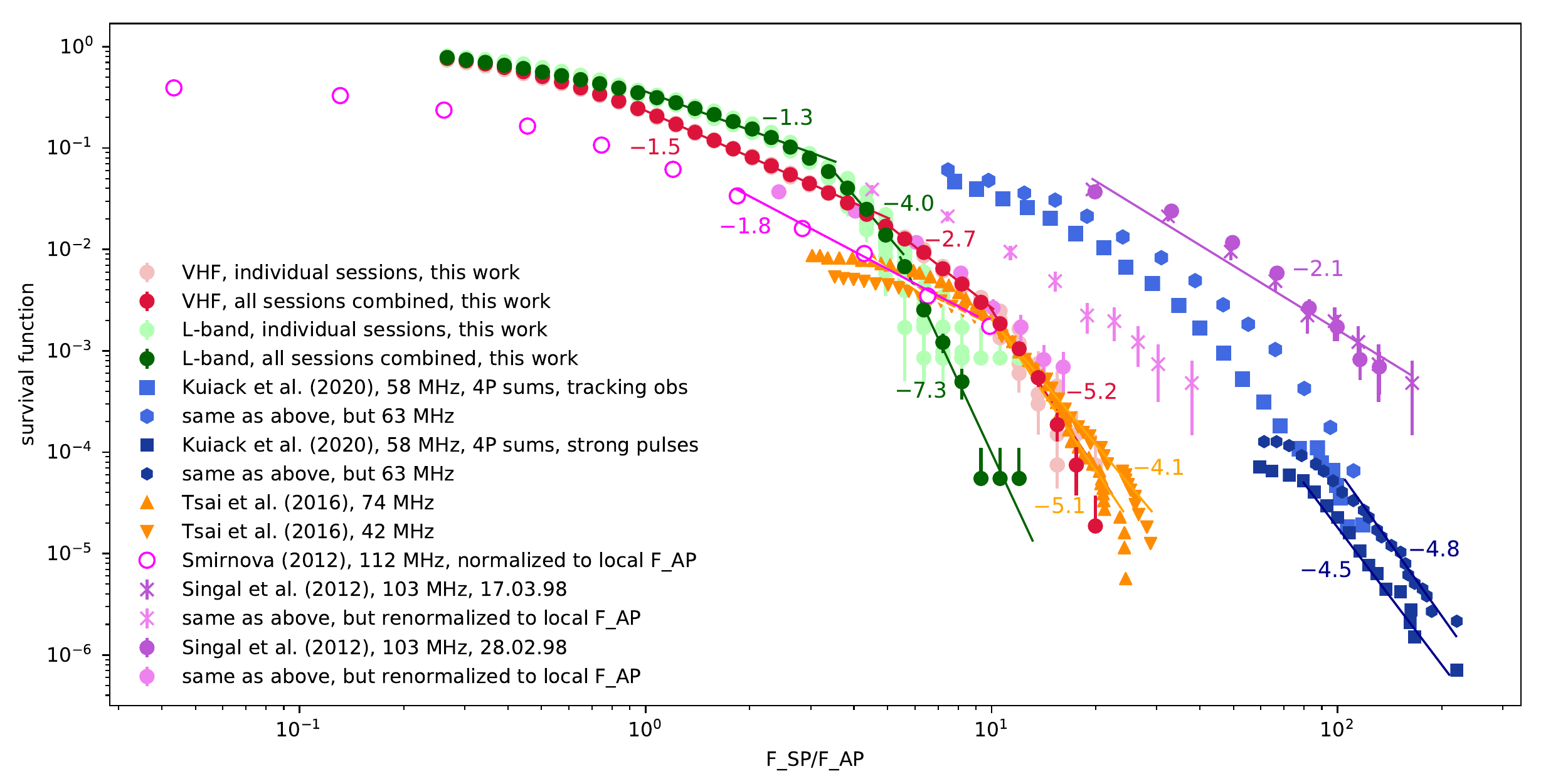}
\caption{Survival function of the \src\  single-pulse fluence distribution in 
the MP+precursor phase regions, from our observations and literature. For our 
observations, dark and light green circles mark the combined distribution in
L-band and individual sessions, respectively. Similarly, light and dark red 
circles show the distributions for the VHF fluences. One-second continuum 
integrations from \citet{Kuiack2020}, equivalent to 4-pulse fluence sums, are 
shown with blue circles and squares. Orange triangles mark single-pulse fluences 
from \citet{Tsai2016}. Purple circles and squares mark the measurements of 
\citet{Singal2012}, with lighter shade showing values renormalized to their 
local average profile fluences. Observations from \citet{Smirnova2012} are shown 
as unfilled magenta circles. Power-law indices from the literature or this work 
are given next to the corresponding distributions in matching color.}
\label{fig:B0950_cf_lit}
\end{figure*}

\section{Discussion}

\subsection{Comparison of fluence distribution to the literature}
\label{subsec:distr_lit}

In the literature, precursor single pulses were rarely explored separately. Also,
at the lowest frequencies, these  precursor single pulses may extend into the MP 
phase region (Fig.~\ref{fig:bridge_pulse}). Therefore, in order to compare 
the single-pulse fluence distributions from our observations to previous work, 
we integrated fluences within the combined MP and precursor windows, and normalized 
these by the total AP fluence (we note that the average fluence in MP+precursor is 
95\% and 98\% of the \FAP\ in VHF and L-band, respectively: normalizations by 
\FAP\ and the average flux in MP+precursor are almost identical). The survival 
functions for individual sessions did not differ significantly from each other 
(Figure~\ref{fig:B0950_cf_lit}). On a log-log scale, the SFs exhibited a smooth, 
curved shape, affected by thermal noise below $F_0 \approx 2F_\mathrm{AP}$.

We did not attempt to identify an exact functional form of the distribution, 
since for L-band the limited total observing duration resulted in only a moderate 
range of recorded  fluences; while in VHF the broadband intensity fluctuations 
may obscure the true SF shape.
For the sake of comparison with the literature, we fit a collection of broken 
power laws (PLs) to our distributions. We fit $\alpha(F_\mathrm{min}, F_\mathrm{max})$ 
such that the survival function of the fluence distribution in the fluence range 
$F_\mathrm{min}/F_\mathrm{AP} < F \leq F_\mathrm{max}/F_\mathrm{AP}$ was described by 
\begin{equation}
\label{eq:pl}
 N(F\geq F_0/F_\mathrm{AP}) = K \left(\dfrac{F_0}{F_\mathrm{AP}}\right)^\alpha.
\end{equation}

\begin{table}[h]
\begin{center} 
\caption{Power-law indices for the piece-wise SF fit (Eq.~\ref{eq:pl}) for fluences 
in MP+precursor window in the range of $[F_\mathrm{min}/F_\mathrm{AP}, F_\mathrm{max}/F_\mathrm{AP}]$. \label{table:PLfit}}
\setlength\tabcolsep{1.5pt}
\renewcommand{\arraystretch}{1.2}
\begin{tabular}{|ccc|ccc|}
\hline 
 \multicolumn{3}{|c}{VHF} & \multicolumn{3}{|c|}{L}\\
\hline
\parbox{0.8cm}{\centering $\frac{F_\mathrm{min}}{F_\mathrm{AP}}$} &
\parbox{1.0cm}{\centering $\frac{F_\mathrm{max}}{F_\mathrm{AP}}$} &
\parbox{1.0cm}{\centering $\alpha$} &
\parbox{0.8cm}{\centering $\frac{F_\mathrm{min}}{F_\mathrm{AP}}$} &
\parbox{1.0cm}{\centering $\frac{F_\mathrm{max}}{F_\mathrm{AP}}$} &
\parbox{1.0cm}{\centering $\alpha$}\\
\hline
 1.0 & 5.0  & $-1.51 \pm 0.02$ & 1.0 & 3.5 & $-1.26 \pm 0.06$\\ 
 5.0 & 10.0 & $-2.72\pm 0.09$  & 3.5 & 5.5 & $-4.05\pm 0.23$\\
 10.0 & 22.0 & $-5.22 \pm 0.25$ & 5.5 & 13.0 & $-7.31 \pm 0.15$\\
\hline 
\end{tabular} 
\end{center}
\end{table}

The literature single-pulse fluence SFs are collected in Fig.~\ref{fig:B0950_cf_lit}. 
All were constructed from pulses at 100\,MHz or below and should thus  be compared 
to our VHF observations. Some authors \citep[e.g.,][]{Tsai2016} normalized single-pulse 
fluences by \FAP\ measured in the same session. Others such as  \citet{Singal2012} 
used \FAP\ averaged over many sessions. Finally, \citet{Smirnova2008} provided 
calibrated fluences. Further details on these literature distributions are found 
in \ref{subsec:0950lit}. Interestingly, the distributions at similar radio frequencies 
tend to agree with each other, if the fluences are normalized by the local average 
fluence of the session for the work of \citet{Smirnova2008} and one session in 
\citet{Singal2012}. For another session in \citet{Singal2012}, the reported average 
flux is too low and normalized fluence distribution is offset from the bulk of other 
measurements, albeit having roughly the same slope. 

\begin{figure}
\centering
\includegraphics[width=0.5\textwidth]{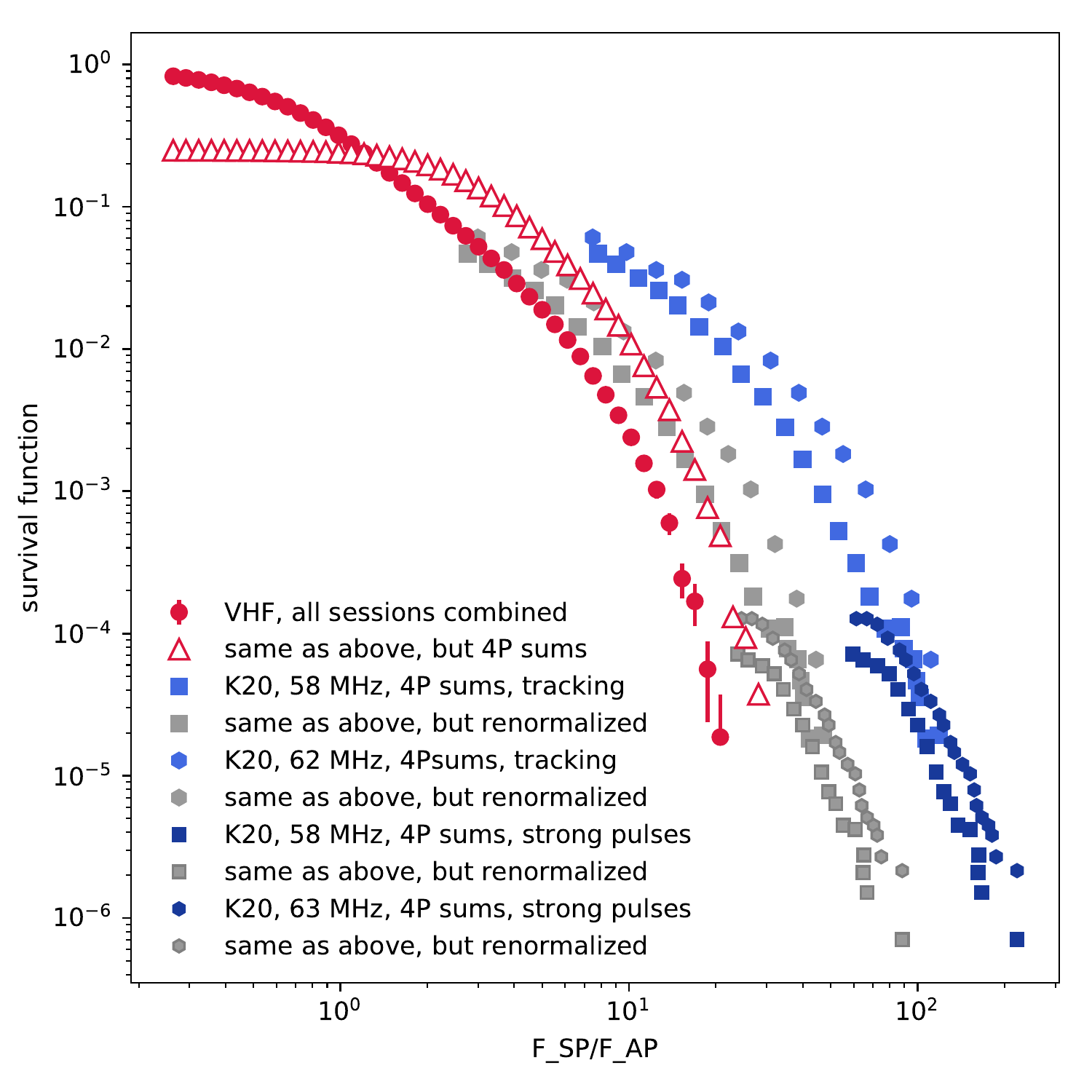}
\caption{Single-pulse MP+precursor fluence survival functions from all sessions of VHF 
observations combined (red circles) together with 4-P sums of fluences normalized by 
total number of periods (triangles). The latter SF emulates observing setup of 
\citet{Kuiack2020} who integrated continuum fluence in one-second ($4P$) intervals, but 
constructed SF by normalizing by total number of periods in their session. The 
distributions of fluences from \citet{Kuiack2020}  are shown as the shades of blue 
for the normalization provided by the authors and as shades of gray with normalization 
of 2.5 larger \FAP\ (see text for details). }
\label{fig:cdf_sums}
\end{figure}

Comparing SFs with  \citet{Kuiack2020} requires special care because their result is an 
SF of continuum fluences integrated within one second ($\approx4P$). These SF are 
constructed by counting the number of events above a set threshold and dividing it 
by the number of pulsar periods in all sessions combined, not the number of 
one-second integrations. While \citet{Kuiack2020} remark that 
their SFs  are dominated by single pulses,
we find this not to be valid for the range of probabilities explored in their work.
We simulated a distribution with the power-law probability density function equivalent 
to the SF of the observed pulses (with index $-5$, see Fig.~\ref{fig:B0950_cf_lit}),
and find that the probability that SF constructed from four-element sums 
matches the individual-element 
SF is smaller than $10^{-8}$. Given the number of pulses included in \citet{Kuiack2020}, 
their smallest possible probability is $10^{-6}$. The effect  is further illustrated 
by distribution of fluences from consecutively summed 4 pulses in our VHF band 
(Fig.~\ref{fig:cdf_sums}), normalized by total number of pulses to match 
\citet{Kuiack2020}. For all probabilities smaller than $\approx 0.2$ the distribution 
for fluences integrated within one second is shifted to the right with respect to 
single-pulse distribution. Thus, we have to compare the distributions from 
\citet{Kuiack2020} to our four-pulse distributions.

\begin{figure}
\centering
\includegraphics[width=0.5\textwidth]{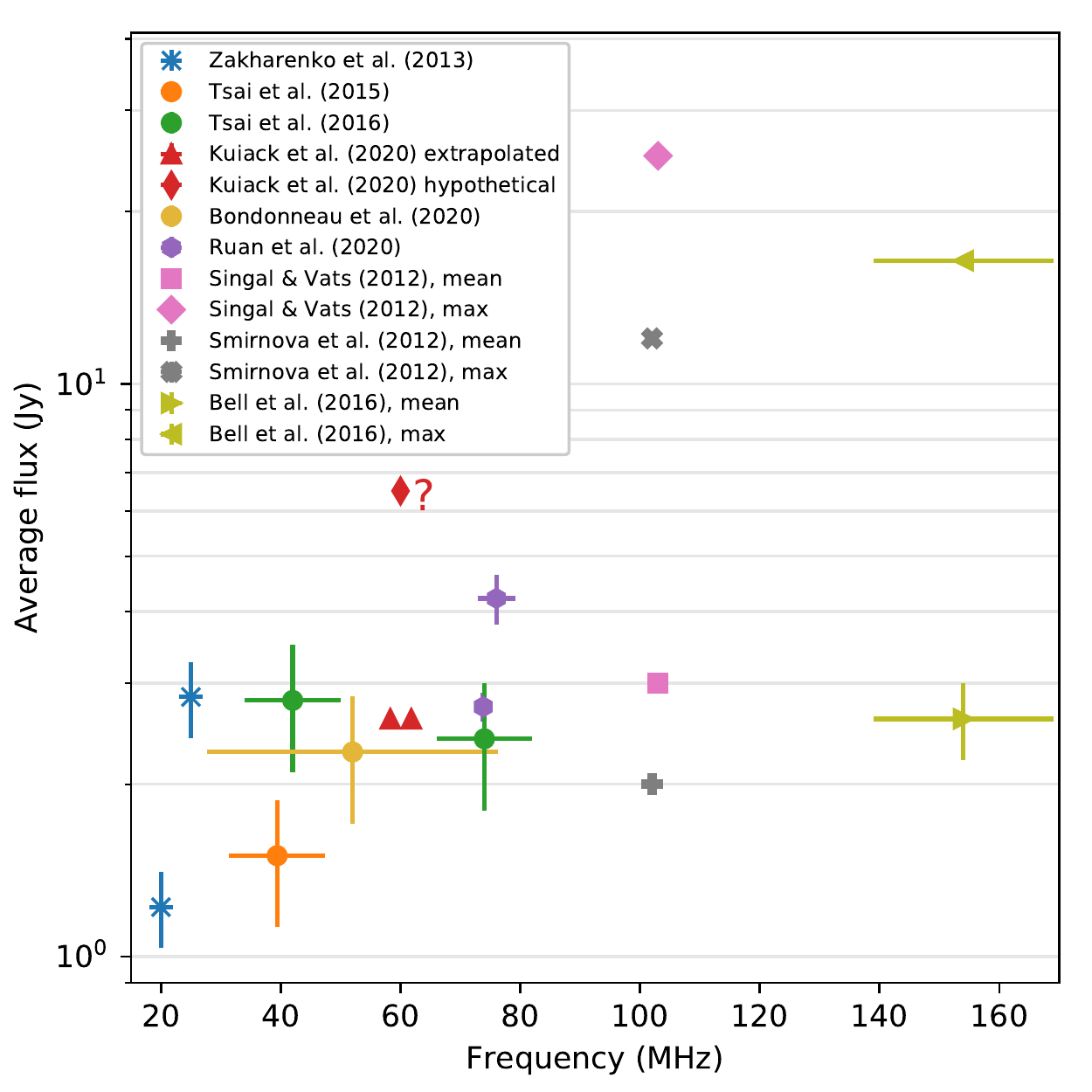}
\caption{Average flux density spectrum for \src\ constructed from the literature 
with uncertainties given by the authors
\citep{Zakharenko2013,Tsai2015,Tsai2016,Bondonneau2020,Ruan2020,Singal2012,Smirnova2008,Bell2016}. 
The horizontal errorbars indicate bandwidth.  For observations at frequencies higher 
than 100\,MHz both mean and maximum fluxes per observing campaign are shown. 
\citet{Kuiack2020} did not measure the average flux density values directly, but 
extrapolated from \citet{Tsai2016}. The question mark points to the conjectured 
flux density value that would align the single-pulse fluence distribution of 
\citet{Kuiack2020} with other work (Figs.~\ref{fig:B0950_cf_lit} and \ref{fig:cdf_sums}).}
\label{fig:flux}
\end{figure}

The SFs from \citet{Kuiack2020} have similar shape to ours, but their single-pulse 
fluences are approximately 2.5 larger. Their work  measures continuum flux, which 
includes an unpulsed component that is entirely missed by the traditional pulsar 
observing setup that we use. \citet{Ruan2020} reported on unpulsed flux of \src\ to be
$0.59\pm0.16$\,Jy at 74\,MHz, with, so far, a poorly constrained spectral index. 
While the corresponding fluence should be subtracted from the measured by 
\citet{Kuiack2020}, it  is likely to only be a minor correction.

\citet{Kuiack2020} do not themselves measure the average fluence, extrapolating it 
from the flux density measurements of \citet{Tsai2016}. For \src\, the flux density 
measurements in the  20--170\,MHz frequency range exhibit a considerable amount of 
variation (Fig.~\ref{fig:flux}). For frequencies below 100\,MHz the measurements 
were made over the course of one or few hours, or over some number of sessions. 
It is possible that the scatter of the reported flux density values is partly 
influenced by scintillation, and partly by systematic flux calibration errors. 
Individual observations at frequencies above 100\,MHz exhibit a great deal of 
variation, with flux increases by a factor of a few, and in this variation 
scintillation is likely to play a considerable role (see Sec.~\ref{subsec:scintl}).

The distributions from \citet{Kuiack2020} can be made to match our 4-P sums if 
we assume the average flux during their observing epoch was actually about 
2.5$\times$ higher than the normalization value that was used \citep[based on][]{Tsai2016}.
Considering the overall uncertainty on the average flux density, possible
intrinsic flux variability, and the influence of scintillation for a given 
observing setup, it is difficult to assess whether the true average flux  during this 
observation could have been 
6.2\,Jy, rather than the 2.6\,Jy that was assumed. 
 The scintillation timescale at their frequencies should be $\sim$5--14 min, compatible 
with the flux density variation scale from their Fig.~5 (i.e., the Fig. does not 
provide clear evidence of the contrary). A potential, unusually large \fdiff\  
would make the instantaneous modulation index to be close to 0.6--0.7, making 
the observed fluence distribution more shallow. However, over the four-hour observations 
the average flux density would have a modulation index of 0.2--0.5, meaning it is
not likely the average flux over the session as whole would be increased.
Clearly, more observations are needed to investigate this matter. These should 
include simultaneous measurements of single-pulse fluence distributions, calibrated 
average fluence and, preferably, direct scintillation measurements.

\begin{figure*}
\centering
\includegraphics[width=\textwidth]{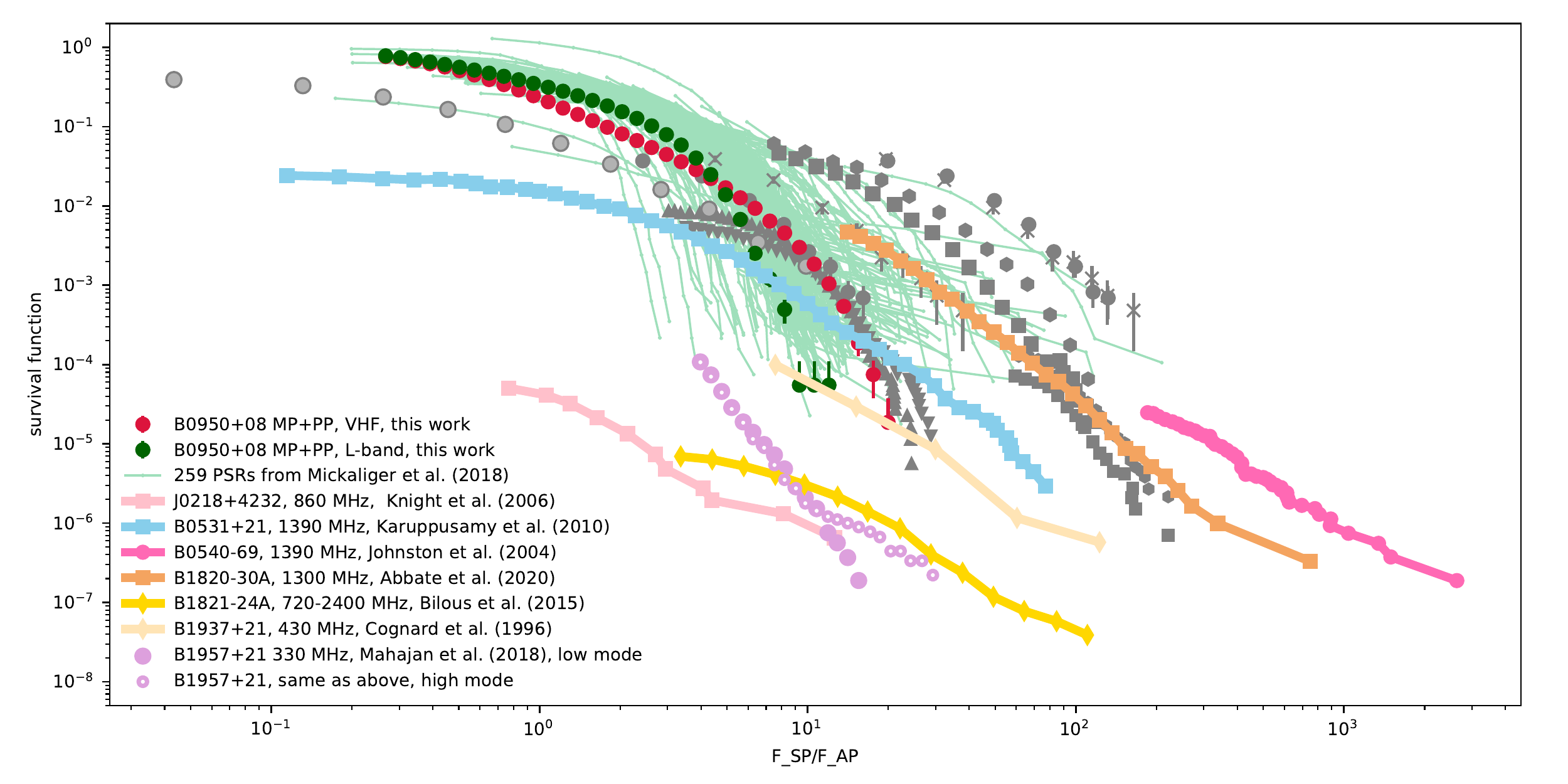}
\caption{Comparison of \src's MP+precursor fluence distribution to the sample of 
normal pulsars from \citet[][light green lines]{Mickaliger2018} and the 
collection of  ``classical'' GPs normalized to \FAP\ (colored lines). Other 
measurements of the \src\  single-pulse fluence distributions from 
Fig.~\ref{fig:B0950_cf_lit} are given in gray (using the same markers as 
in that Figure). }
\label{fig:B0950_cf_GP}
\end{figure*}

\subsection{On the possibility of \src\ emitting GPs.}

The single-pulse variability of the first few dozens of pulsars, studied in 1970s 
at frequencies above 100\,MHz, lacked  single pulses brighter than  ten times the 
average intensity \citep[see references in][]{Cognard1996}. The subsequently 
discovered strong pulses from PSR B1937+21 and the Crab pulsar painted a sharp 
contrast. Thsese high-intensity pulses seemingly followed a separate distribution, 
suggesting a distinct pulse population. Such pulses were named ``giant pulses''
and, for the time being, the ``ten times the average'' criterion discriminated 
between the two populations\footnote{A threshold 2$\times$ higher is also 
extensively used in the literature, and flux density integration  methods varies: 
average fluence in the whole on-pulse window, in a given component, in a phase
window of GPs, peak flux density and so on.}. Several other pulsars joined the 
GP-emitting set, their GPs roughly sharing the following properties: ns-$\mu$s 
width, high brightness temperature, restricted window of occurrence coinciding 
with high-energy emission, strong magnetic field at the light cylinder. This led 
to the speculation of GPs being associated with caustic radio emission produced 
close to the light cylinder \citep[see review in][]{Bilous2015}. In this paper, we 
 call such GPs ``classical GPs''. 

While there has been some progress in constraining single-pulse fluence 
distributions on a population level \citep{BurkeSpolaor2012, Mickaliger2018} 
since the 1970s, much of the field is still there to explore. At the same time 
a plethora of peculiar single-pulse variability patterns has been discovered, 
including, but not limited to giant micropulses \citep[$\mu$s-wide, 
$F>10F_\mathrm{AP}$ for longitude-resolved fluence only, PL energy distribution 
with indices similar to those of GPs, see][]{Johnston2002,Cairns2004,Raithel2015}; 
individual pulses from sparsely radio-emitting neutron stars \citep[``rotating 
radio transients'', RRATs][]{McLaughlin2006}, which are wider than GPs and have 
a lognormal fluence distribution \citep{Keane2010, Mickaliger2018};   
RRAT-like pulses from normal pulsars \citep[e.g.,][]{Esamdin2012}, ``bursting 
modes'' characterized by abrupt onset, changes in the shape of the single-pulse 
fluence distribution and the shape of the average profile 
\citep[e.g.,][]{Seymour2014,Wang2020}; prolonged periods of absence of any 
emission (nulling, resulting in very low values of average flux, see, for example, 
\citealt[][]{Gajjar2014}). At low radio frequencies of $\lesssim100$\,MHz 
single-pulse fluences seem to be more variable, resulting in regular detection 
of ms-wide pulses above the 10\FAP\ GP threshold \citep{Kuzmin2007, Ulyanov2006}. 

In this section we try to compare some properties of single pulses from \src\ 
to GPs and to  ``normal'' pulses. The task is somewhat complicated by the 
absence of clear definition of a ``giant'' or a ``not a giant'' pulse and the 
dearth of sufficient observational knowledge to distinguish between the two 
classes. It is also possible that these two classes are mere extremes and 
in fact pulsars' single-pulse properties form a continuum.

\subsubsection{Fluence distribution}

In our L-band observations of \src, only one pulse in the MP+precursor region 
exceeds the historical 10\FAP\ threshold. This pulse occurred in the middle 
of the MP window. Its shape and spectrum resembled the other, less energetic 
single pulses we observe. If we compare to the average fluence only in the 
occurrence phase window, all three phase regions exhibited single pulses 
above this factor-10 threshold. None of those bright pulses in the MP or 
precursor region, however, appeared to have properties (such as widths, pulse 
shapes and spectra) distinctly different from the fainter pulses. For the IP, 
the generally low signal-to-noise ratio (S/N) of individual pulses precludes 
drawing any similar conclusions, although individual IP pulses definitely 
show the most variability with respect to their average. At the same time, 
peak flux densities occasionally exceed $20\times$ absolute or local peak flux 
density of the average pulse, especially for the IP region. In the VHF data, 
single-pulse fluences exhibit a greater degree of variability in MP and 
precursor, but smaller in the IP, but the trend described above remains 
the same: pulses above 10\FAP\ threshold do not have properties visibly 
distinct from fainter pulses.

Let us discuss how does the single-pulse fluence variability of  \src\   compare to the 
rest of the pulsar population. As discussed in Sect.~\ref{sec:cdf_fits}, a 
lognormal distribution provides a marginally good fit to the single-pulse 
fluences from the MP and precursor regions. The best-fit coefficients
indicate an unusually large degree of variability (more shallow SF), when 
compared to other pulsars with lognormal single-pulse fluence distributions 
from the 315-pulsar L-band study by \citet{BurkeSpolaor2012}. It is worth 
noting that among 225 non-nulling pulsars, only 33\% had a log-normal 
single-pulse fluence distribution, while 48\% were classified as ``other'', 
that is neither Gaussian, lognormal nor bimodal. \citet{BurkeSpolaor2012} also
revealed that for their brightest sources (approximately 20\% of the total 
sample, regardless of distribution shape) the fluence distribution integrated 
within the on-pulse window did not exceed 6.4\FAP. Taking into account the 
distribution of single-pulse counts collected in their sample, our L-band 
observations suggest a larger degree of fluence variability.

\citet{Mickaliger2018} explored observations in the frequency range similar 
to the one of  \citet{BurkeSpolaor2012} and  analyzed a partly overlapping
source sample of a similar size, albeit with roughly an order of magnitude more 
observing time. Interestingly, 58\% of sources showed single pulses with 
fluences $F>10$\FAP, with a probability of getting by chance such a large fluence values 
ranging from $10^{-4.5}$ to $10^{-1.5}$. About 16\% of the sources exceeded 
20\FAP, with approximately the same probability range. The L-band single-pulse 
fluence distribution for \src\ that we report here falls well within the 
sample SFs from \citet{Mickaliger2018}, not showing any signs of unusual 
single-pulse fluence variability.  

The tentative inconsistency between \citet{Mickaliger2018} and \citet{BurkeSpolaor2012} 
is puzzling and might be at least partially explained with different source 
samples and observing time. Ultimately, a larger study of single-pulse fluences 
is needed both in L-band and at low frequencies. For the low frequencies, there 
has been no population study of single-pulse fluence distributions yet, so we 
do not really know the typical degree of single-pulse fluence variability. We 
hypothesize that  single-pulse fluences become increasingly more variable at 
lower frequencies, however for \src\ the SF still lies within the 
\citet{Mickaliger2018} sample.

Let us discuss how do the single-pulse  fluences for \src\ compare to ``classical'' GPs.
Figure~\ref{fig:B0950_cf_GP} shows GP fluence distributions from studies that 
measured both the average flux and the GP flux (see details in Appendix~\ref{ap:flgp}).
Three things can immediately be gleaned from Fig.~\ref{fig:B0950_cf_GP}. First of 
all, the SFs  of GPs are more shallow than those of normal pulses and they seem to
follow PL distributions for a larger fractional range of fluences (many 
show flattening at lower fluences, due to intrinsic properties, or to observing 
artifacts such as incomplete sampling near the selection threshold, or noise 
influence). Secondly, the fluence distribution for the ``classical'' GPs is on 
average much better studied due to dedicated long-term observations. However, one 
must keep in mind that most of these observations were performed \textit{after} 
bright pulses caught attention, so there is a strong selection bias present. In 
this regard, it would be very interesting to explore the high-fluence tails of 
the distributions from \citet{Mickaliger2018}. Finally, it is interesting to note 
that, when normalized by their respective \FAP\, most of the GP distributions 
continue below 10\FAP\ and that for the same SF probabilities the relative fluences 
from different pulsars span two orders of magnitude.  Obviously, if GPs were a separate 
pulse population and do not contribute much to \FAP, normalizing them by \FAP\ is to 
some extent arbitrary.

\begin{figure}
\centering
\includegraphics[width=0.5 \textwidth]{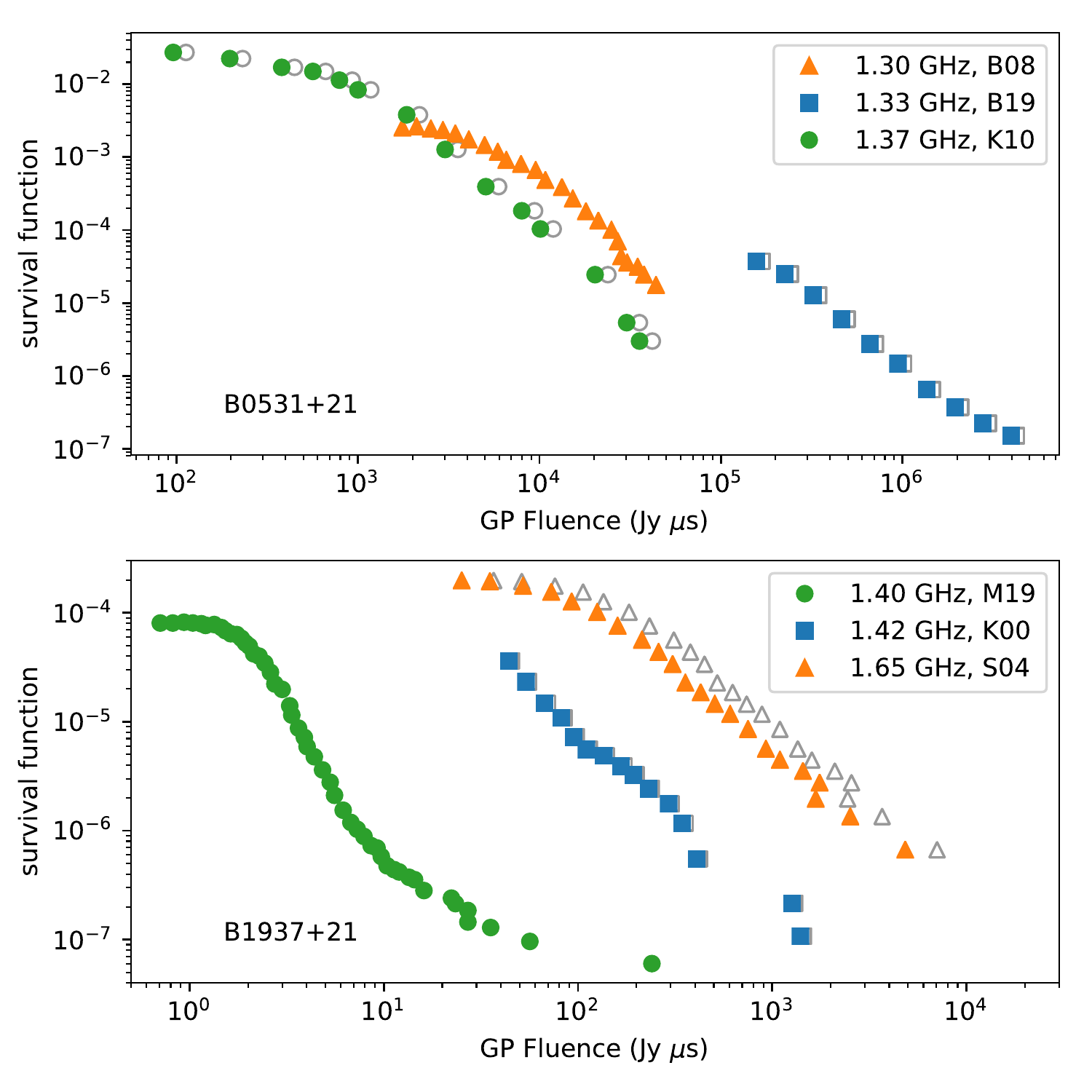}
\caption{Survival function of the fluence distributions for two most studied GP pulsars, 
the Crab pulsar PSR B0531+21 and PSR B1937+21 from the works of  
\citet[][B08]{Bhat2008}, \citet[][B19]{Bera2019}, \citet[][K10]{Karuppusamy2010}, 
\citet[][M19]{Mckee2019}, \citet[][K00]{Kinkhabwala2000}, and
\citet[][S04]{Soglasnov2004}. The gray unfilled markers show 
fluences rescaled to a common frequency (1.3\,GHz for the Crab and 1.4\,GHz for PSR~B1937+21) 
using broadband spectral fits from \citet{Bilous2016} and \citet{Kondratiev2018}.}
\label{fig:GP_other}
\end{figure}

On the log-log scale the distribution of single-pulse fluences from \src\ is steeper 
than that of GPs and generally can be fit with a single PL over a smaller range of 
fluences. However, neglecting the fluence offset between our distribution and the one 
of \citet{Kuiack2020} and combining two distributions together produces a seemingly-PL
distribution spanning an order of magnitude in pulse fluence and four orders of 
magnitude in probability, which is similar to known GP distributions, albeit with the 
steeper slope. Interestingly, PSR B1957+21 exhibits a similar 
PL slope over a similar span in magnitude in its low mode, and has a clear flattening 
of at the large-fluence end, in the high mode. It is yet unclear whether these low-mode 
single pulses are GPs in the classical sense.

It is currently unknown which form of 
distribution, lognormal or PL,  better describes the high-fluence tail of the fluence 
distribution (or the full distribution) for single pulses from \src, other ``normal''
pulsars and GPs. Careful studies of such kind are required.

Finally, we note that a fluence offset between distributions from different authors, as 
observed for \src\ (Fig.~\ref{fig:B0950_cf_lit}), is also present for the GP fluence 
distributions of the two most studied GP emitters, the Crab pulsar and PSR B1937+21. 
Most of the works on GPs from these pulsars provide only fluences of individual GPs, 
not the average profile. While the slope of the PL fit is similar at similar observing 
frequencies \citep[e.g., references in ][]{Mickaliger2012,Knight2006}, the fluences of GPs 
at similar probabilities are different by a factor of few or even a few orders of 
magnitude (Fig.~\ref{fig:GP_other}). These discrepancies may be only partly explained 
by scintillation, and we believe they may stem from unaccounted systematic errors in 
flux density calibration or from intrinsic variability. In any case such a large 
uncertainty in the rate of occurrence of a GP of specific fluence has considerable 
influence on the estimates of prospects of detecting GPs from pulsars beyond our Galaxy.

\begin{figure*}
\centering
\includegraphics[width=0.95\textwidth]{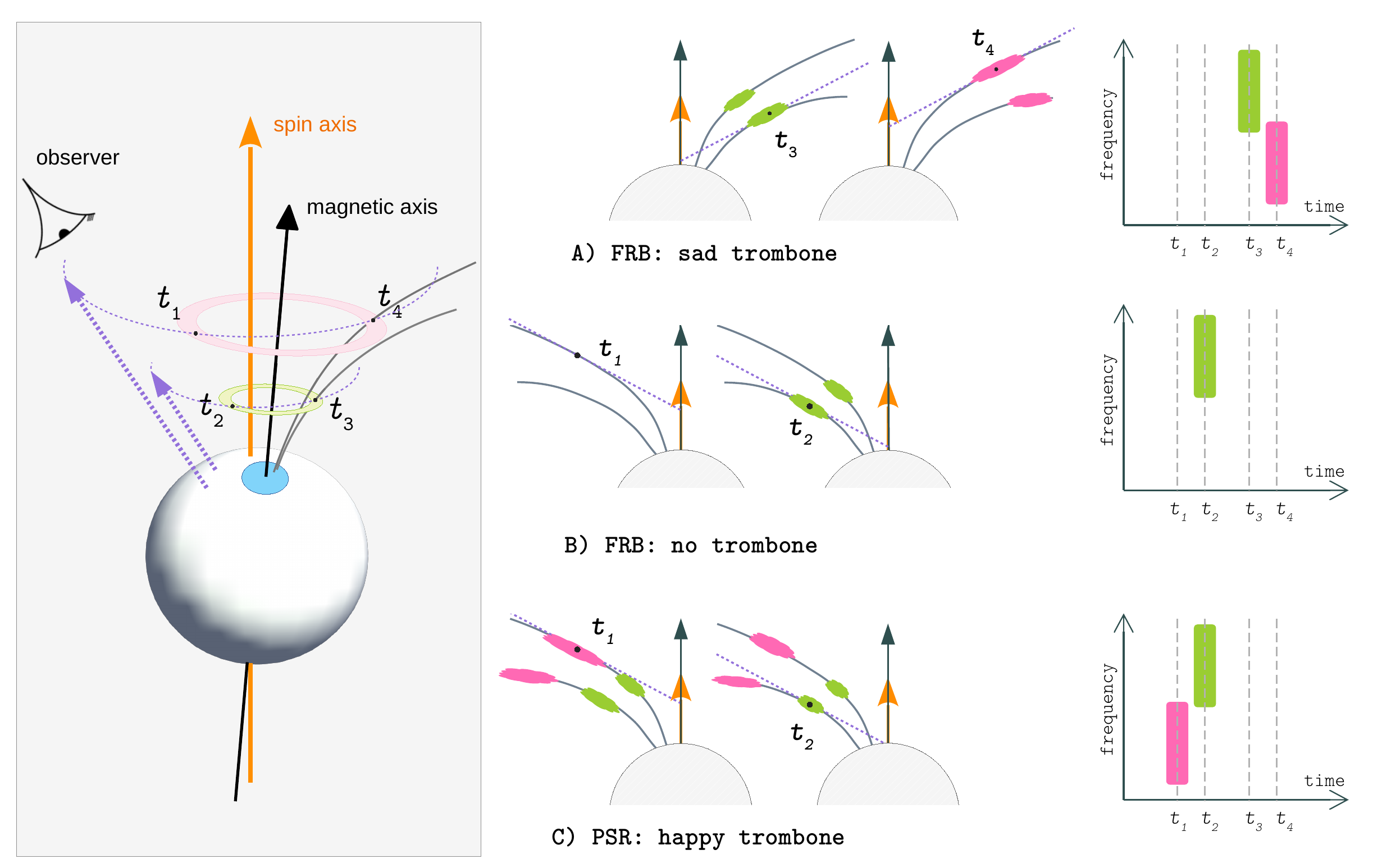}
\caption{Cartoon model of the frequency drift in subpulse components in pulsars and 
FRBs from the radius-to-frequency mapping of emission from separate plasma clumps 
propagating in neutron star magnetosphere \citep{Wang2019}. The inset on the left side 
shows neutron star with magnetic dipole axis (black arrow) inclined with respect to 
the spin axis (orange arrow). The light blue circle on the surface of the star shows 
the footpoints of the open field lines, with a couple of field lines drawn as an example. 
Purple dashed arrows indicate LOS direction. Following classical radius-to-frequency 
mapping, radio waves are emitted tangentially to the field line, with higher frequencies 
coming from smaller altitudes. The emission is postulated to come from a ring of field 
lines with the same magnetic collatitude. As pulsar rotates, the observer detects 
emission from a range altitudes at the moments of time $t_{1-4}$ (black dots). In the 
center column the 2D projection of the same setup is shown with magnetic dipole axis 
lying in the picture plane and the spin axis pointing up and away from the reader. The 
line of sight (LOS) at $t_{1-2}$ or $t_{3-4}$ is indicated by the dotted line. Pulse 
spectra are plotted in the rightmost column. In all three rows some event generates 
clumps of plasma simultaneously on two separate field lines. These clumps are moving 
along the field lines emitting radio waves with altitude-dependent radio frequency. For 
the row A) LOS fortuitously aligns with radio emission from one clump at higher radio 
frequency, and, shortly afterward, with another clump at lower frequency, resulting in 
the ``sad trombone'' spectrum. For B) the same scenario is shown for the leading part of 
the profile. In this case the clump is not there yet when LOS passes through outer 
magnetic field line, resulting in single-component ``no trombone'' spectrum. For pulsars 
(C) clumps are supposed to be much more frequent (but still distinct), so LOS crosses 
beams of radio waves from two different batches of clumps, generating upward frequency 
drift, a.k.a. ``happy trombone''. 
 }
\label{fig:FRB_cartoon}
\end{figure*}

\subsubsection{Pulse duration and brightness temperature}

Because of their high peak-flux density and narrow widths, GPs sometimes have the largest 
implied brightness temperatures among all known astrophysical sources: $T_b>5\times 10^{39}$\,K 
for PSR B1937+21 \citep{Soglasnov2004}, $10^{37}$\,K or even $10^{41}$\,K for the Crab pulsar \citep{Hankins2003,Hankins2007}. However, such measurements are anecdotal since they require 
a special observing setup and can be performed only on strong pulses at the frequencies where 
pulses are unaffected by scattering. For the same Crab pulsar,  $T_b$ is several orders of 
magnitude lower when time resolution is larger than the duration of pulse or its subcomponents 
\citep[$10^{32}$\,K,][]{Cordes2004}. 

Individual pulses of \src\ are much wider than ``classical'' GPs, and typically even have 
widths comparable to that  of the average profile components. At the same time, observing 
with sufficient time resolution reveals that individual pulses are often composed of a 
collection of separate $\mu$s-wide subpulses \citep[micropulses, also dubbed ``microstructure'', 
e.g.,][]{Hankins1971}. Micropulses have been recorded from a substantial fraction of strong 
nearby pulsars and are speculated to be a common trait of pulsar radio emission  \citep{Lange1998,Popov2002,Kuzmin2003,De2016}. Their origin is still as unknown as the GP 
formation mechanism; although some studies  link  these phenomena  \citep{Petrova2004a, Petrova2004b}. 
Most existing studies of the microstructure focus on the quasi-periodicity of micropulses 
\citep[e.g.,][]{Cordes1990}, and, as of now, the other properties of micropulses (e.g., 
fluence distribution, polarization) are not known well enough to compare them thoroughly 
to ``classical'' GPs.

In our study pulsar emission was recorded with rather coarse time resolution, which 
precluded studying \src's microstructure, especially in the VHF band. We still observe 
sharp single-pulse components both in VHF and L-band (Figs.~\ref{fig:strongSP}, 
\ref{fig:SP_trombones}). The brightest such component in L-band peaked at 0.3\FAP.
Assuming average flux in L-band to be 100 mJy \citep{Jankowski2018}, this indicates 
the peak flux density of 91\,Jy and $T_b \sim 10^{27}$\,K for the effective width 
equal to 81.92\,$\mu$s time resolution. This is much smaller than the typical $T_b$ 
of GPs, however it is currently unclear what the smallest timescale present in the 
\src's  microstructure is, and how bright individual subpulses can be. \citet{Hankins1978} 
reported a strong (1\,kJy) sub-$\mu$s unresolved subpulse at 430\,MHz, with implied 
$T_b > 3\times 10^{31}$\,K. At the same time \citet{Popov2002}, observing for an hour 
at 1.6\,GHz with time resolution of only 62.5\,ns did not find any subpulses shorter 
than few $\mu$s. 

\subsubsection{Constraints on the place of origin}

Observed single pulses from \src\ contribute directly to its average emission.
Thus, one can use the average profile to constrain the emission altitudes with the 
help of radius-to-frequency mapping and the rotating vector model 
\citep{Radhakrishnan1969,Kijak2003}. An apparent average-profile bifurcation below 
400\,MHz is a common feature, interpreted as emission coming from the diverging 
magnetic field lines. Within the core/cone emission region model 
\citep{Rankin1990,Rankin1993}, the presence of interpulse emission means that we 
observe either two poles of an orthogonal rotator or a very wide cone of an almost 
aligned one. Both models appear to have difficulties explaining the observed data 
\citep{Hankins1981}. Fitting a rotating vector model to polarization data, and 
modeling magnetospheric X-ray emission, preferred an orthogonal rotator 
\citep{Everett2001,Becker2004}, but a series of works employing more sophisticated 
beam shapes spoke in favor of an aligned one \citep[e.g.,][]{Narayan1983,Gil1983}. 
Thus, standard techniques do not provide an unambiguous picture of the magnetic 
field configuration and the emission heights in {\src}, since no model explains 
all observed characteristics.

While the region of the GP origin is still unknown, GP phase windows are hinted 
to be associated with average-profile components thought to be made of caustic 
radio waves emitted close to the light cylinder (LC). The high  magnetic field 
strengths on the LC for the ``classical'' GP pulsars may be important for their 
emission mechanism \citep[e.g.,][]{Lyubarsky2019}. Caustic radio components coincide 
with X-ray and gamma-ray emission. For \src, nonthermal magnetospheric X-rays are 
indeed observed, however the peaks of X-ray profile are substantially offset from 
the radio peaks, speaking against caustic origin of the radio components 
\citep{Becker2004}. In $\gamma$-rays \src\ has not been detected yet -- it is absent 
from the most recent Fermi Large Area Telescope Fourth Source Catalog 
\citep{Abdollahi2020}, although no targeted search has been published either. Finally, 
its magnetic field on the light cylinder, calculated using the standard dipole magnetic
field model, is several orders of magnitude smaller than those of classical GP pulsars.

The frequency evolution  of the profile components in the VHF band can, however, be 
qualitatively explained by geometrical models which involve diverging magnetic field 
lines (see Sec.~\ref{sec:frb}), thus speaking against an LC origin. It is interesting 
to note that no direct comparison can be made between VHF pulses observed in  {\src},
and the classical GPs at the same frequencies. To our knowledge, only the Crab pulsar 
has individual GPs recorded below 200\,MHz \citep{Popov2006,Eftekhari2016,Karuppusamy2012,vanLeeuwen2020}, 
however scattering in the ISM precludes any studies of pulse structure at these frequencies. 
The rest of ``classical'' GP pulsars show similar or larger levels of scattering and 
are generally fainter than Crab GPs.

\begin{figure*}
\centering
 \includegraphics[width=\textwidth]{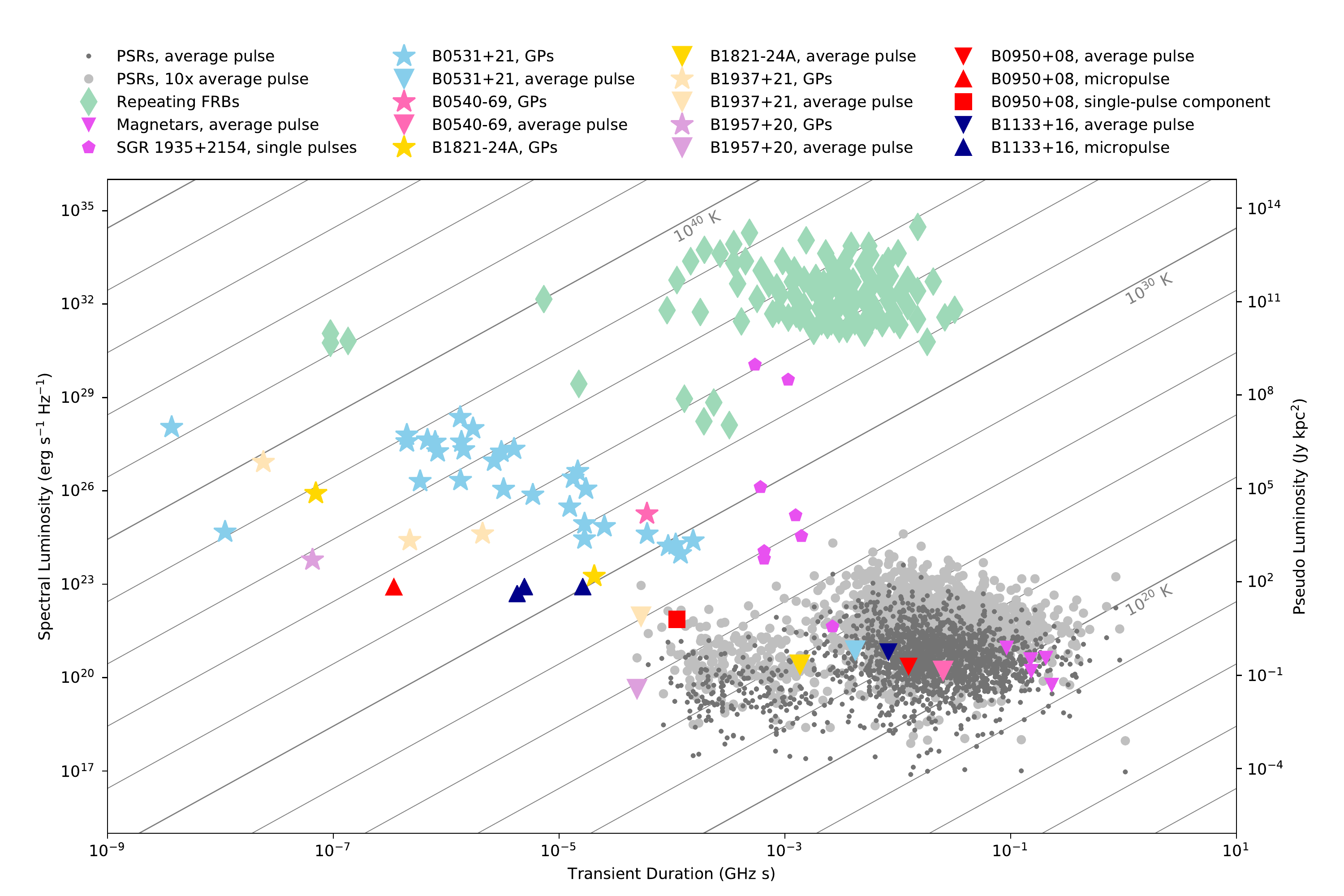}
 \caption{ Time-luminosity phase-space plot for radio pulses from neutron stars and 
 repeating FRBs showing the product of observing frequency and pulse width ($\nu W$ 
 in GHz\,s) against equivalent isotropic spectral luminosity ($4\pi S_\mathrm{peak}d^2$, 
 left) and the pseudoluminosity ($S_\mathrm{peak}d^2$, right). The brightness 
 temperature $T_b = (1/2k_b) \times (S_\mathrm{peak}d^2)/(\nu W)^2$ is indicated 
 by gray lines. Black dots show pulsar spectral luminosity calculated from continuum 
 flux density at 1400\,MHz, with pulse duration assumed to be full width at half 
 maximum of the average profile at the same frequency \citep[ATNF pulsar catalog 
 v.1.64][]{Manchester2005} or $0.04P$ if no such value was given in the catalog. 
 Lighter dots mark common single-pulse fluence variability at $10\times$ the average pulse 
 luminosity. Light green diamonds show FRBs from repeating sources  with known 
 distances (FRB 20121102A, FRB 20180916B, FRB 20190711A, FRB 20200120E see 
 \citet{Nimmo2021}, and references therein). Purple pentagon mark individual bursts 
 from magnetar SGR~1935+2154 taken from compilation by \citet{Nimmo2021}, and the 
 downward triangles indicate the average pulse luminosities from the neutron 
 stars with surface magnetic field larger than $5\times10^{13}$\,G in the ATNF 
 catalog. Stars show representative GPs or their components from the studies 
 where both peak flux density and widths were measured, with the latter not biased 
 by scattering (but still possibly below time resolution). Light blue color marks 
 the Crab pulsar \citep{Hankins2003,Hankins2007,Karuppusamy2010,Popov2009,Bera2019,Bhat2008}, 
 beige corresponds to PSR B1937+21 \citep{Kinkhabwala2000,Soglasnov2004,Mckee2019}, 
 yellow to PSR~B1821-24A \citep{Knight2006,Bilous2015}, pink  to PSR~B0540-69 
 \citep{Geyer2021}, and light violet to PSR~B1957+21 \citep{Main2017}. 
 Finally,  upward red triangle shows unresolved microstructure component for 
 \src\ from \citet{Hankins1978}.  Blue upward triangles mark strong microcomponents 
 from a pulse from another ordinary nearby PSR~B1133+16 \citep{Bartel1978,Bartel1982}. 
 The brightest unresolved component for our L-band observation of \src\ is shown 
 with the red square (see text for details). Downward triangles of corresponding 
 colors show the average pulses from the respective pulsars. }
\label{fig:lum_vs_w}
\end{figure*}

\subsection{What can \src\ tell us about FRBs}
\label{sec:frb}

Fast radio bursts (FRBs) are sub-ms radio pulses of  extragalactic origin 
(\citealt{Lorimer2007}; see \citealt{Petroff2019} and \citealt{Petroff2021} for the recent reviews).
As the precise nature of FRBs is yet to be determined, dozens of theories abound to 
explain their properties \citep[cf. the theory catalog from][]{Platts2019}. 
Magnetospheres of neutron stars are popular candidate formation localizations of 
FRBs. Non-cataclysmic, repeating FRBs are for example being attributed to super-GPs 
from young neutron stars or MSPs \citep{Cordes2016, Connor2016}. Recently, an 
extremely bright (but still few orders of magnitude fainter than the bulk of 
extragalactic FRBs) radio burst was detected from a Galactic magnetar SGR J1935+2154 
\citep{Bochenek2020}, accompanied by two much fainter bursts \citep{Kirsten2020}.

\begin{figure*}
\centering
 \includegraphics[width=\textwidth]{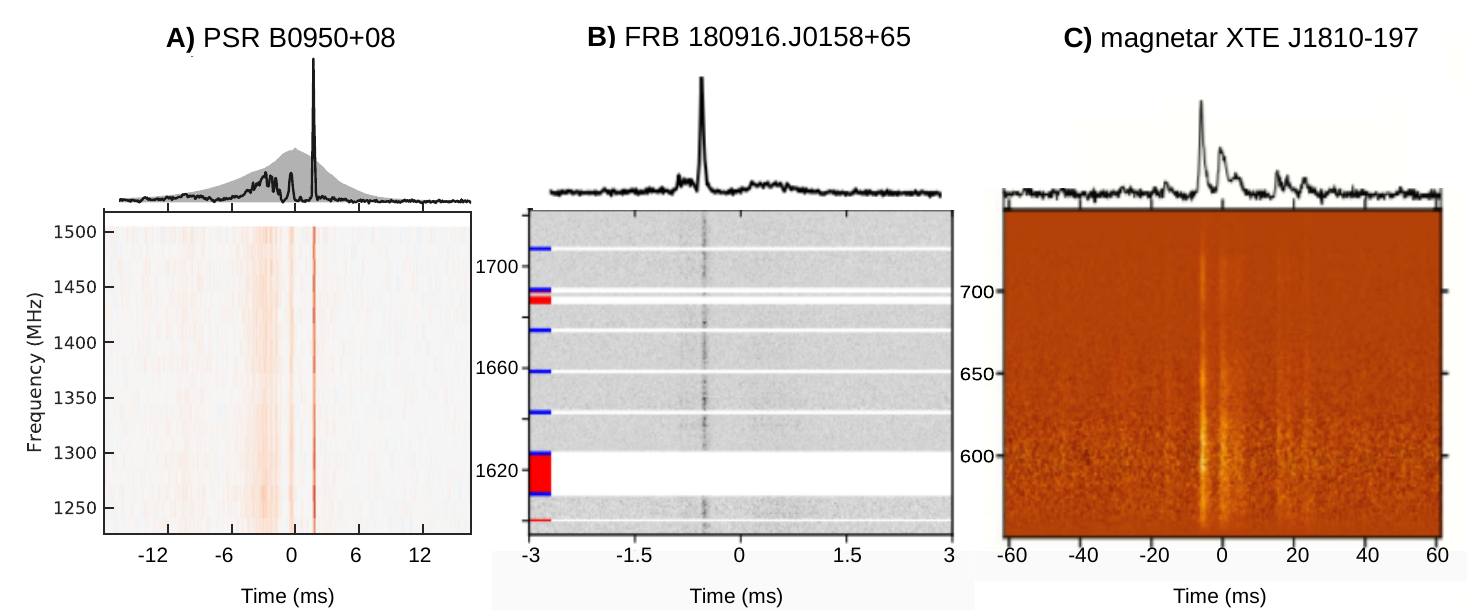}
  \caption{Apparent similarities in pulse morphology between radio pulses from \src, a repeating FRB source and a magnetar. \textit{Left:} one of single pulses from B0950+08 (this work). \textit{Middle:} 
  burst from repeating FRB~20180916B \citep{Marcote2020}. \textit{Right:} somewhat scattered single pulse from
  magnetar XTE~J1810$-$197 \citep[PSR J1809$-$1943;][]{Maan2019}. In all three cases there is present 
  the combination of amorphous structure and sharp peaks, with drops down to the  noise level in between.}
\label{fig:pulse_cf}
\end{figure*}

\begin{figure*}
\centering
 \includegraphics[width=0.4\textwidth, height=0.4\textwidth]{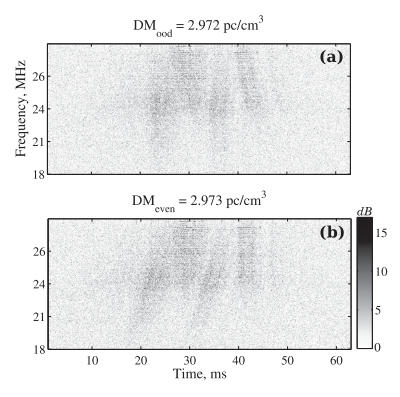}\includegraphics[width=0.6\textwidth, height=0.4\textwidth]{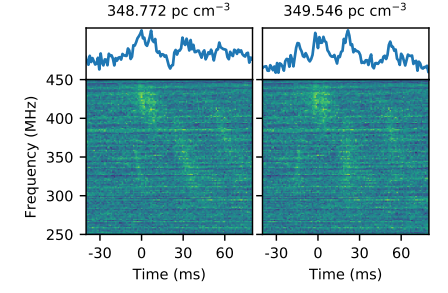}
 \caption{Interesting spectral structure of an individual pulse from \src\ observed at the very low radio frequencies by
   \citet[][left]{Ulyanov2016}  and of a burst from repeating  FRB~20180916B  observed at higher frequencies \citep[][right]{Pleunis2020}.}
\label{fig:SP_FRB}
\end{figure*}

Repeating FRBs often show so-called ``sad trombone'' behavior, downward drifting 
time-frequency structures (examples from Apertif are found in \citealt{Pastor2020}). 
Quite strikingly, the spectral features of \src\ strongly resemble these structures 
-- although the trombones are actually quite ``happy'' on the {leading} side of the MP 
(Fig.~\ref{fig:SP_trombones}). \citet{Wang2019} and \citet{Lyutikov2019} proposed 
a geometric explanation for such sad trombones in FRBs, by examining the radiation 
from groups of charged particles in a neutron star magnetosphere, in the classical 
framework of pulsar radio-to-frequency mapping (Fig.~\ref{fig:FRB_cartoon}).
In this framework, clumps of plasma emit radio waves tangentially to the field lines 
they propagate along. Higher radio frequencies are generated closer to the stellar 
surface (which is natural if emission is due to curvature radiation). In the model of 
\citet{Wang2019}, a sudden release of energy creates the required, separate, charged 
clumps, and these propagate along neighboring field lines. As the neutron star rotates, 
our line of sight (LOS) fortuitously first cuts through  the high-frequency emission 
from one of these clumps, at the lower altitude; then later cuts through the 
lower-frequency emission from another clump at higher altitude; thus creating apparent 
downward frequency drift (Fig.~\ref{fig:FRB_cartoon}A). This model would not create 
upward frequency drift if all clumps are created simultaneously, since clumps 
propagate away from the stellar surface (Fig.~\ref{fig:FRB_cartoon}B). On the other 
hand, one can imagine that for pulsars, clumps are created more regularly and that our 
LOS, passing through the leading half of the on-pulse window, can first pick up 
low-frequency emission from one batch of clumps and then high-frequency emission from 
another, resulting in apparent upward frequency drift (Fig.~\ref{fig:FRB_cartoon}C).
Interestingly, single pulses form \src's precursor show downward frequency drift, thereby 
complicating the interpretation of precursor emission within classical core/cone model.
We note that this model is purely geometrical and does not specify the exact mechanism 
for charged plasma or radio emission generation -- these may be different for 
rotation-powered pulsars and magnetars.

To our  knowledge, ``normal'' emission from radio pulsars was never considered as a 
potential explanation for FRB emission because of the vast disparity in luminosities 
and brightness temperatures. Figure~\ref{fig:lum_vs_w} shows that disparity on the 
duration versus luminosity diagram that is common to the field. However, we must note 
that there is much more variability to pulsar emission than is traditionally 
acknowledged. Single-pulse fluence distributions are not constrained well on a 
population level and single pulses exceeding 10$\times$ or 100$\times$ the mean 
fluence may not be as rare as previously thought. Also, very little is known about 
the fluence distribution of pulsar microstructure. Those scarce records of 
microstructure fluences from  ordinary nearby pulsars that do exist -- recorded 
within hours or even minutes-long sessions! -- show brightness temperatures which 
are quite different from the averages that are generally reported.

With the discovery of much fainter bursts from the repeating sources FRB~20180916B 
and FRB~20200120E, and the ``FRB-emitting'' galactic magnetar  SGR J1935+2154 
\citep[see][and references therein]{Nimmo2021}, the question of neutron-star radio 
emission variability attracts renewed attention. Perhaps the individual pulses from
neutron stars occupy a large fraction of the parameter space between GPs, average 
pulses and FRBs, and some of the latter are caused by extreme cases of the mechanism 
that operates in the magnetospheres of ordinary pulsars.

Finally, we note the curious but striking similarities in individual pulse morphology 
when we compare \src\ with a single burst from the repeating FRB~20180916B and 
with a pulse from  magnetar XTE~J1810$-$197 (Fig.~\ref{fig:pulse_cf}). The peculiar 
shape of \src's single pulses were best described by \citet{Hankins1981}: ``sometimes 
amorphous noisy pulses, sometimes very ``spiked'' micropulses interspersed with regions 
where the signal returns briefly to the system noise level, and occasional quasi-periodic 
micropulses''. The observations of \src\ performed at 18--26\,MHz by \citet{Ulyanov2016} 
revealed an interesting spectral behavior in one of the pulses recorded. A sequence 
of four ms-spaced subpulse components had visible dispersion difference between the 
pairs of components \#1/3 and \#2/4  (Fig.~\ref{fig:SP_FRB}). Similar behavior was 
recently noticed in one of the pulses from repeating FRB~20180916B, at frequencies an 
order of magnitude higher \citep{Pleunis2020}. \citet{Ulyanov2016} attributed the 
difference in visible dispersion measure to propagation through turbulent plasma in the 
neutron star magnetosphere and pulsar wind. It would be interesting to test this theory 
on FRB data. 

\section{Summary}

We have analyzed nonsimultaneous  single-pulse observations of \src\ in two widely 
separated frequency regions. Although coarse frequency resolution did not allow us 
to measure ISM parameters in our study, we argue that some of the high fluence 
variability that was  previously reported is due to the changing decorrelation 
bandwidth on our LOS. Similarly to \citet{Cairns2004}, we found a distinct pair of 
pulses pulse in the bridge region between MP and IP, with very low average level of 
emission. 

Comparing the fluence distributions and other properties of the \src's single pulses 
to those of ``classical'' GPs and ``normal'' pulses suggests that \src\ does not 
emit GPs, although the exact location of emission region remains elusive.  
We argue that the pulse population labeled by \citet{Kuiack2020} as separate and 
brighter may have been mischaracterized due to normalization with a noncontemporary 
average flux density from the literature, and we point out common mismatches in 
the GP energies reported by various  authors. 

Finally, we present upward and downward frequency drifts in single pulses below 
100\,MHz, and argue that the geometrical model of ``sad trombones'' from FRBs may 
be applicable here. We conclude that little is actually known about the $\mu$s-scale 
fluence variability of normal pulsars, and that its further study may bring important 
clues about the nature of FRBs.

\appendix

\section{Details on fluence distributions from the literature}

\subsection{\src}
\label{subsec:0950lit}

\citet{Kuiack2020} recorded bright emission from \src\ using AARTFAAC, an imaging 
transient detection instrument at LOFAR.  A sample of 275 one-second snapshot images with 
fluences ranging from 42 to 177\,kJy ms were collected during 96 hours of observations 
in 1.5-MHz bands centered at 58.3 and 61.8\,MHz. The time resolution was thus 4 
times larger than the pulse period. The normalizing AP fluence is obtained from 
interpolation over the average flux density values at nearby frequencies obtained in
earlier studies at other telescopes.

\citet{Singal2012} carried out 141 32-min observing sessions with the Rajkot telescope. 
Pulsed emission from \src\ was recorded over 1.6\,MHz of bandwidth centered at 103\,MHz. 
A single linear polarization was received and the sampling time was 48\,ms (thus, 
covering all of MP window). Over the time of observations, the mean pulse intensity 
was 3\,Jy, however the pulsar displayed a great deal of variability, between $<0.3$\,Jy 
and 26\,Jy. The authors used a 30-Jy criterion for bright pulse selection. The authors 
note that there is almost one-to-one correspondence between the rate of the strong 
pulses and the average pulse intensity of the session. The survival functions for 
three sessions are shown, with intensity of the pulses normalized by mean intensity 
(3\,Jy). We note that since the MP window is not resolved, these are essentially fluences.
Two of these sessions are from 1998, for which the average intensity per session is 
shown on their Fig.~3. For Feb 28 (59th day  of the year), the mean flux is 24\,Jy, 
for March 17 (day 76) it is 13\,Jy. 

\citet{Smirnova2012} observed \src\ with the Large Scanning Antenna of the Pushchino 
Radio Astronomy Observatory at 112\,MHz with 2.3-MHz bandwidth. One component of the 
polarized emission was received and the Faraday rotation was not removed. The fluence
distribution shown was compiled for pulses from three sessions with the highest S/N 
of the average profile. The duration of each session was about 765 pulse periods. In 
their Sec.~3, the authors note that 30-Jy peak flux density corresponds to the 2\,Jy 
average flux density. Using this relation, we can convert the average peak flux density 
of the three brightest sessions (130\,Jy) to the average fluence of 2.2\,Jy\,s. This 
fluence value was used for normalization on Fig.~\ref{fig:B0950_cf_lit}, however we note 
that there is a factor of two difference in peak fluxes between the most and the least 
bright sessions in the sample. Also, the fluence in \citet{Smirnova2012} was calculated 
in windows with edges at 0.3 of the peak flux of each given pulse, which may lead 
to some underestimation of single-pulse fluence as compared to using fixed-phase windows.

\citet{Tsai2016} recorded three 4-hr sessions on consecutive days with the Low Wavelength 
Array (LWA1). Observations were performed in the beamformed mode at 42 and 74\,MHz, each
with a bandwidth of 16\,MHz. Because of changing elevation of the source the System 
Equivalent Flux Density of the telescope changed by approximately 30\% over the course 
of each session. The authors fit Gaussians to single pulses and the average profile 
and report the ratio of the areas under the fit curves for individual and average 
pulse (thus, essentially, fluences). The authors searched for pulses using different 
DMs and smoothing time series with boxcars of different widths. They did not record 
fluences in each period. All pulses with fluences larger than ten of AP are shown in 
their work.

\subsection{Population studies}

\citet{Mickaliger2018} compiled single-pulse distributions for 264 pulsars and RRATs 
using 35-min pointings from the Parkes Multi-Beam Pulsar Survey \citep{Manchester2001}. 
The survey was conducted in 288-MHz bandwidth centered at 1374\,MHz. For each pointing, 
the authors identified an on- and off-pulse windows on the folded profile, subtracted 
the off-pulse mean in each spin period and integrated single-pulse fluences 
($F_\mathrm{SP}$) within the on-pulse window. These fluences were normalized by the 
fluence of the average profile of the pointing (hereafter, we denote them as 
$F \equiv F_\mathrm{SP}/F_\mathrm{AP}$). The influence of thermal noise was not explored.

Unfortunately, only the plots of fluence distributions were preserved (Mickaliger, 
priv. comm.). As input to our study, those plots (259 pulsars, RRATs omitted) were 
downloaded from the referenced web page\footnote{\url{http://astro.phys.wvu.edu/pmps/powerlaw.html}} 
and manually digitized using a web-based tool\footnote{\url{https://apps.automeris.io/wpd/}}. 
On the figures, fluence distributions were displayed as non-normalized histograms on 
the log-log scale. Logarithms of normalized fluences were binned with fixed-width bins,
where the bin width and fluence range varied from pointing to pointing.

Using the obtained values of $N(F_i)$ in bins with centers at $F_i$  and constant widths 
$\Delta F$, we constructed the survival functions as:
\begin{equation}
 \mathrm{SF}(F_i+\Delta F/2) = \frac{1}{N_\mathrm{periods}}\sum _{j\geq i} N(F_j),
 \end{equation}
where $N_\mathrm{periods}$ is number of spin periods per pointing.

A fraction of pulsars was observed more than once; the respective SFs were combined. 
As a sanity check we note that for all but one source, the SFs at the lowest fluence were 
smaller than one. That is expected, since the log-scale plots could not show periods where 
noise variations produced $F<0$. The only plot with $\mathrm{SF}(F_0)>1$, for PSR J1822-0848,  
potentially contained data from the wrong source.

The precision of this manual digitization process was assessed by comparing the retrieved 
values for the total numbers of pulses at the high-fluence end of the distribution, to 
the expected  number. For 1912 counts with $N<5$, the median of $N/\mathrm{round}(N)$ 
was 0.97 and 68\% range was [0.90, 1.03]. This we assume to be the fractional error for 
both $F$ and $N(F)$.

\subsection{Giant pulses}
\label{ap:flgp}

\citet{Abbate2020} explored the properties of GPs from PSR B1820$-$30A. The authors used 
the MeerKAT radio telescope in South Africa to record two 150-min ($\approx 3.1\times10^6 P$ 
total) sessions at 1284\,MHz with 642-MHz bandwidth. The authors measure the peak S/N of 
GPs in time series with 10\,$\mu$s resolution and normalize it by the total S/N of
the integrated profile by the square root of the number of rotations observed. GPs come 
from the phase regions of components C1 and C2 (there is also much weaker C3 component). 
The authors note that some of their GPs had multiple one-bin peaks. On Fig.~\ref{fig:B0950_cf_GP} 
we show pulses from main profile component (``C1'', 13522 GPs total). GPs from component 
C2 have distribution with similar slope, but are order of magnitude less frequent.

\citet{Bhat2008} recorded GPs from the Crab pulsar with the Australian Compact Telescope 
Array at 1300 MHz (32-MHz bandwidth) during one session.  Fluences were calculated in 
custom windows around each GP. The distribution for MP and IP giant pulses are combined. 
The authors estimate that uncertainties in fluence estimates depend on the pulse strength 
and width, ranging from 0.5\% for strong and narrow pulses, to 35\% for the weak and broad 
pulses. 

\citet{Bilous2015} studied the fluence distribution of GPs from PSR B1821$-$24A. The authors 
combined pulses from several observing sessions in UHF, L-band, and S-band. In $5.1\times10^7 P$, 
476 GPs were found on narrow phase window on the trailing edge of pulse component P1 and P2. 
Fluences of individual pulses (integrated within the boxcar that provided maximum S/N on 
the convolved signal) were normalized by \FAP\ of the corresponding session. PSR B1821$-$24A 
has a wide average profile that covers most of the spin phase. 
On Fig.~\ref{fig:B0950_cf_GP} only P1 is shown. 

\citet{Cognard1996} explored GPs from PSR B1937+21 while observing at 430 MHz with 
0.5-MHz bandwidth. About $1.7\times10^6$ pulse periods were accumulated. The authors 
calculated mean GP flux in fixed-size windows on the trailing edges of main pulse (MP) 
and interpulse (IP) and normalize it by the mean flux in MP and IP on-pulse windows 
($S_\mathrm{MP}=1.5$\,Jy and $S_\mathrm{IP}=0.8$\,Jy). Both MP and IP on-pulse windows 
span about 0.2 of spin period, thus the mean flux is about 0.46\,Jy, which agrees well 
with the broadband literature flux density compilation from \citet{Kondratiev2018}. 
Figure~\ref{fig:B0950_cf_GP} shows the distribution from MP GPs only. In order to 
convert the flux density ratios $S_m/S_\mathrm{MP}$ to the normalized fluence 
$F_m/F_\mathrm{AP}$ we set the size of MP on-pulse window to $0.2P$ and the 
size of MP GP window to $0.1P$ (Fig.~2 in \citealt{Cognard1996}). Then,
\begin{equation}
\dfrac{F_m}{F_\mathrm{AP}} = \dfrac{S_m\times 0.1P}{ S_\mathrm{MP}\times0.2P + S_\mathrm{IP}\times0.2P} \approx \dfrac{1}{3}\dfrac{S_m}{S_\mathrm{MP}}.
\end{equation}

\citet{Johnston2004} observed the extragalactic PSR B0540$-$69 at 1390\,MHz (bandwidth 
256\,MHz) for 80\,hrs or $5.7\times 10^6P$. Some of data were excised due to RFI. In 
total, 141 GPs were detected. In their Fig.~6, the maximum logarithm of survival 
probability is $-4.2$, whereas $\log(141/5.7/10^6)=-4.6$. In that same Fig., the 
maximum is at $-4.2$, corresponding to 31\,hr of observing time. The reason for that 
is unclear. The authors measured equivalent continuum flux densities of individual 
GPs and also the average flux density on folded profile (24\,$\mu$Jy). On 
Fig.~\ref{fig:B0950_cf_GP} we plot their ratio an renormalize SF to peak at $141\times 5.7^{-1}\times10^{-6}$.
GPs come from two regions coinciding with the on-pulse region on the average 
profile, although the S/N of the latter is quite low. More recent study of GPs 
from this pulsar probed SF on the lower-fluence end ($30<F/F_\mathrm{AP}<800$) 
and measured somewhat steeper PL index \citep{Geyer2021}.  

\citet{Karuppusamy2010} studied GPs from the classical GP source PSR B0531+21. 
The authors observed the Crab pulsar in 160-MHz band centered at 1400 MHz. GPs 
come from the on-pulse regions of MP and IP on the average profile (only MP GPs
are shown on Fig.~\ref{fig:B0950_cf_GP}). The authors measure equivalent widths 
of GPs and integrate fluences within those widths. The authors also report the
average pulsed flux density of 14\,mJy, which we use for 
$F_\mathrm{AP}=14\times33$\,Jy $\mu$s.

\citet{Knight2006} explored the properties of GPs from PSR J0218+4232. The 
observations were taken with the 100-m Green Bank Telescope at frequencies in the 
ranges of 793--921 and 1341--1469\,MHz (four sessions total). The authors searched 
for GPs by averaging time series with a set of boxcars of different widths. In 
total, 155 pulses were found, all of them in the minima of the very wide two-component 
average profile which stretched across all spin phases. It was noted that pulsar has
significant unpulsed components, which were not included in calculation of \FAP. On 
Fig.~\ref{fig:B0950_cf_GP} we show pulses from one phase range in one session of 
the 857-MHz observations, the most prolific one. 

\citet{Kinkhabwala2000} studied GPs from PSR B1937+21 using the Arecibo telescope 
at three frequencies. Several sessions were recorded, S/N varied strongly due to 
scintillation in 10-MHz bandwidth at a central frequency of 1420 MHz. The intensity 
of the average profile in each session was used to scale single-pulse amplitudes 
so the average flux matched the power-law spectral model from a previous study. 
GPs were searched by smoothing the signal with 2-sample rolling boxcars. Fluence was 
calculated in fixed 2-$\mu$s windows (larger than the typical duration of a scattered 
GP). The distribution at 1420\,MHz was made using all available data (only sessions 
with high S/N were kept for studying other pulse properties). Figure~\ref{fig:B0950_cf_GP} 
plots fluence distribution for both MP and IP GPs. 

\citet{Mahajan2018} observed the original Black Widow Pulsar B1957+20 in 
311.25--359.25\.MHz frequency range at the Arecibo Observatory. Over 9 hr 
($\approx 2.2\times10^7 P$) of data were accumulated in four daily sessions. 
Once in few seconds the pulsar switched rapidly between two discrete modes of 
radio emission (``high'' and ``low''). The authors searched for GPs in the time 
series convolved with an exponential decay filter with a scattering timescale 
obtained from analysing the average profile shape. Strong single pulses were 
found at the trailing sides of main pulse and interpulse components in both modes 
and the fluences were computed in fixed windows around each pulse. 
Figure~\ref{fig:B0950_cf_GP} shows fluence distribution for the main component 
pulses only, normalized by the integrated flux over the main pulse component, 
25 Jy\,$\mu$s. 

\citet{Mckee2019} collected the largest to-date sample of GPs from PSR B1937+21 
observed with the Large European Array for Pulsars. Twenty-one observing sessions 
spread over four years were conducted with different parts of the array in slightly 
varying frequency ranges, typically in 128-MHz bandwidth around the central frequency 
of 1396\,MHz. Fluences were integrated within half-max of each GP. No correlation 
was observed between integrated pulse profile S/N and the rate of GPs appeared to
fluctuate by the order of magnitude.

\citet{Soglasnov2004} analyzed observations of PSR B1937+21 made with 70-m DSS43 radio 
telescope at Tidbinbilla, Australia. The pulsar was observed in 32-MHz bandwidth 
centered at 1650\,MHz during one 39-min session. The authors employed complex
criteria based on the spin phase of high S/N event and its intensity in two subbands. 
The observations were affected by scintillation and the intensities of individual 
pulses in each subband were rescaled with $I(t)/\overline{I}$, where $I(t)$ is the 
local mean intensity of the normal main pulse (in each subband separately) and 
$\overline{I}$ is its average over the whole session. Fluences were calculated in 
custom-size windows around each GP. 
 
\begin{acknowledgements}
AB thanks M. Mickaliger and S. Burke-Spolaor for help in retrieving the single-pulse energy distributions reported in their works, K. Nimmo for comments on the time-luminosity phase-space plot, and A. Rohatgi for creating web based tool to extract data from plots.

This research was supported by 
the European Research Council under the European Union's Seventh Framework Programme
(FP/2007-2013)/ERC Grant Agreement No. 617199 `ALERT', 
and by NWO, the Dutch Research Council, under Vici research programme `ARGO' with project number
639.043.815 (PI:JvL).
JPWV acknowledges support by the Deutsche Forschungsgemeinschaft (DFG) through the Heisenberg programme (Project No. 433075039). 
GS acknowledges financial support provided under the European Union H2020 ERC Consolidator Grant ``Binary Massive Black Hole Astrophysics'' (B Massive, Grant Agreement: 818691).
DV acknowledges support from the Netherlands eScience Center (NLeSC) under grant ASDI.15.406.
EP acknowledges funding from an NWO Veni Fellowship.
JWK is a CITA Postdoctoral Fellow: This work was supported by the Natural Sciences and Engineering Research Council of Canada (NSERC), [funding reference \#CITA 490888-16].

This paper is partially based on data obtained using 
the NenuFAR radio-telescope. The development of NenuFAR has
been supported by personnel and funding from: Station de 
Radioastronomie de Nançay, CNRS-INSU, Observatoire de Paris-PSL, Universit\'{e}
d’Orl\'{e}ans, Observatoire des Sciences de l’Univers en R\'{e}gion Centre,
R\'{e}gion Centre-Val de Loire, DIM-ACAV and DIM-ACAV+ of R\'{e}gion
Ile-de-France, Agence Nationale de la Recherche.

We acknowledge the use of the Nan\c{c}ay Data Center computing facility (CDN
- Centre de Donn\'{e}es de Nan\c{c}ay). The CDN is hosted by the Station de
Radioastronomie de Nançay in partnership with Observatoire de Paris,
Universit\'{e} d’Orl\'{e}ans, OSUC and the CNRS. The CDN is supported by the
R\'{e}gion Centre-Val de Loire, d\'{e}partement du Cher.

The Nan\c{c}ay Radio Observatory is operated by the Paris Observatory,
associated with the French Centre National de la Recherche Scientifique
(CNRS).

This work makes use of data from the Apertif system installed at the Westerbork
Synthesis Radio Telescope owned by ASTRON. ASTRON, the Netherlands Institute for
Radio Astronomy, is an institute of NWO.
Apertif instrumentation development for this research
was supported  by NWO (grant 614.061.613 `ARTS') and the  
Netherlands Research School for Astronomy (`NOVA4-ARTS', `NOVA-NW3', and `NOVA5-NW3-10.3.5.14').
We thank Apertif Builders
E.~A.~K.	Adams,
B.	Adebahr,
W.~A.	van Cappellen,
W.~J.~G.	de Blok,
J.~P.~R.	de Reijer,
B.~S.	Frank,
J.~E.	Hargreaves,
K.~M.	Hess,
E.	Kooistra,
D.~M.	Lucero,
\'A.	Mika,
J.	Morawietz,
R.	Morganti,
V.~A.	Moss,
M.~J.	Norden,
T.~A.	Oosterloo,
E.	Orr\'u,
A.~A.	Ponomareva,
G.~W.	Schoonderbeek,
A.	Sclocco,
R.	van der Brink,
J.~M.	van der Hulst,
D.	van der Schuur,
G.~N.~J.	van Diepen,
M.~A.~W.	Verheijen, and
S.~J.	Wijnholds.

\end{acknowledgements}

\bibliographystyle{aa}
\bibliography{SPC_lit} 

\begin{thebibliography}{123}
\expandafter\ifx\csname natexlab\endcsname\relax\def\natexlab#1{#1}\fi

\bibitem[{{Abbate} {et~al.}(2020){Abbate}, {Bailes}, {Buchner}, {Camilo},
  {Freire}, {Geyer}, {Jameson}, {Kramer}, {Possenti}, {Ridolfi}, {Serylak},
  {Spiewak}, {Stappers}, \& {Venkatraman Krishnan}}]{Abbate2020}
{Abbate}, F., {Bailes}, M., {Buchner}, S.~J., {et~al.} 2020, \mnras, 498, 875

\bibitem[{{Abdollahi} {et~al.}(2020){Abdollahi}, {Acero}, {Ackermann},
  {Ajello}, {Atwood}, {Axelsson}, {Baldini}, {Ballet}, {Barbiellini},
  {Bastieri}, {Becerra Gonzalez}, {Bellazzini}, {Berretta}, {Bissaldi},
  {Blandford}, {Bloom}, {Bonino}, {Bottacini}, {Brandt}, {Bregeon}, {Bruel},
  {Buehler}, {Burnett}, {Buson}, {Cameron}, {Caputo}, {Caraveo}, {Casandjian},
  {Castro}, {Cavazzuti}, {Charles}, {Chaty}, {Chen}, {Cheung}, {Chiaro},
  {Ciprini}, {Cohen-Tanugi}, {Cominsky}, {Coronado-Bl{\'a}zquez}, {Costantin},
  {Cuoco}, {Cutini}, {D'Ammando}, {DeKlotz}, {de la Torre Luque}, {de Palma},
  {Desai}, {Digel}, {Di Lalla}, {Di Mauro}, {Di Venere}, {Dom{\'\i}nguez},
  {Dumora}, {Fana Dirirsa}, {Fegan}, {Ferrara}, {Franckowiak}, {Fukazawa},
  {Funk}, {Fusco}, {Gargano}, {Gasparrini}, {Giglietto}, {Giommi}, {Giordano},
  {Giroletti}, {Glanzman}, {Green}, {Grenier}, {Griffin}, {Grondin}, {Grove},
  {Guiriec}, {Harding}, {Hayashi}, {Hays}, {Hewitt}, {Horan},
  {J{\'o}hannesson}, {Johnson}, {Kamae}, {Kerr}, {Kocevski}, {Kovac'evic'},
  {Kuss}, {Landriu}, {Larsson}, {Latronico}, {Lemoine-Goumard}, {Li},
  {Liodakis}, {Longo}, {Loparco}, {Lott}, {Lovellette}, {Lubrano}, {Madejski},
  {Maldera}, {Malyshev}, {Manfreda}, {Marchesini}, {Marcotulli},
  {Mart{\'\i}-Devesa}, {Martin}, {Massaro}, {Mazziotta}, {McEnery}, {Mereu},
  {Meyer}, {Michelson}, {Mirabal}, {Mizuno}, {Monzani}, {Morselli},
  {Moskalenko}, {Negro}, {Nuss}, {Ojha}, {Omodei}, {Orienti}, {Orlando},
  {Ormes}, {Palatiello}, {Paliya}, {Paneque}, {Pei}, {Pe{\~n}a-Herazo},
  {Perkins}, {Persic}, {Pesce-Rollins}, {Petrosian}, {Petrov}, {Piron}, {Poon},
  {Porter}, {Principe}, {Rain{\`o}}, {Rando}, {Razzano}, {Razzaque}, {Reimer},
  {Reimer}, {Remy}, {Reposeur}, {Romani}, {Saz Parkinson}, {Schinzel},
  {Serini}, {Sgr{\`o}}, {Siskind}, {Smith}, {Spandre}, {Spinelli}, {Strong},
  {Suson}, {Tajima}, {Takahashi}, {Tak}, {Thayer}, {Thompson}, {Tibaldo},
  {Torres}, {Torresi}, {Valverde}, {Van Klaveren}, {van Zyl}, {Wood},
  {Yassine}, \& {Zaharijas}}]{Abdollahi2020}
{Abdollahi}, S., {Acero}, F., {Ackermann}, M., {et~al.} 2020, \apjs, 247, 33

\bibitem[{Adams \& van Leeuwen(2019)}]{al19}
Adams, E. A.~K. \& van Leeuwen, J. 2019, Nature Astronomy, 3, 188

\bibitem[{{Bartel}(1978)}]{Bartel1978}
{Bartel}, N. 1978, \aap, 62, 393

\bibitem[{{Bartel} \& {Hankins}(1982)}]{Bartel1982}
{Bartel}, N. \& {Hankins}, T.~H. 1982, \apjl, 254, L35

\bibitem[{{Becker} {et~al.}(2004){Becker}, {Weisskopf}, {Tennant}, {Jessner},
  {Dyks}, {Harding}, \& {Zhang}}]{Becker2004}
{Becker}, W., {Weisskopf}, M.~C., {Tennant}, A.~F., {et~al.} 2004, \apj, 615,
  908

\bibitem[{{Bell} {et~al.}(2016){Bell}, {Murphy}, {Johnston}, {Kaplan}, {Croft},
  {Hancock}, {Callingham}, {Zic}, {Dobie}, {Swiggum}, {Rowlinson},
  {Hurley-Walker}, {Offringa}, {Bernardi}, {Bowman}, {Briggs}, {Cappallo},
  {Deshpand e}, {Gaensler}, {Greenhill}, {Hazelton}, {Johnston-Hollitt},
  {Lonsdale}, {McWhirter}, {Mitchell}, {Morales}, {Morgan}, {Oberoi}, {Ord},
  {Prabu}, {Shankar}, {Srivani}, {Subrahmanyan}, {Tingay}, {Wayth}, {Webster},
  {Williams}, \& {Williams}}]{Bell2016}
{Bell}, M.~E., {Murphy}, T., {Johnston}, S., {et~al.} 2016, \mnras, 461, 908

\bibitem[{{Bera} \& {Chengalur}(2019)}]{Bera2019}
{Bera}, A. \& {Chengalur}, J.~N. 2019, \mnras, 490, L12

\bibitem[{{Bhat} {et~al.}(2008){Bhat}, {Tingay}, \& {Knight}}]{Bhat2008}
{Bhat}, N.~D.~R., {Tingay}, S.~J., \& {Knight}, H.~S. 2008, \apj, 676, 1200

\bibitem[{{Bilous} {et~al.}(2016){Bilous}, {Kondratiev}, {Kramer}, {Keane},
  {Hessels}, {Stappers}, {Malofeev}, {Sobey}, {Breton}, {Cooper}, {Falcke},
  {Karastergiou}, {Michilli}, {Os{\l}owski}, {Sanidas}, {ter Veen}, {van
  Leeuwen}, {Verbiest}, {Weltevrede}, {Zarka}, {Grie{\ss}meier}, {Serylak},
  {Bell}, {Broderick}, {Eisl{\"o}ffel}, {Markoff}, \& {Rowlinson}}]{Bilous2016}
{Bilous}, A.~V., {Kondratiev}, V.~I., {Kramer}, M., {et~al.} 2016, \aap, 591,
  A134

\bibitem[{{Bilous} {et~al.}(2015){Bilous}, {Pennucci}, {Demorest}, \&
  {Ransom}}]{Bilous2015}
{Bilous}, A.~V., {Pennucci}, T.~T., {Demorest}, P., \& {Ransom}, S.~M. 2015,
  \apj, 803, 83

\bibitem[{{Bochenek} {et~al.}(2020){Bochenek}, {Ravi}, {Belov}, {Hallinan},
  {Kocz}, {Kulkarni}, \& {McKenna}}]{Bochenek2020}
{Bochenek}, C.~D., {Ravi}, V., {Belov}, K.~V., {et~al.} 2020, \nat, 587, 59

\bibitem[{{Bondonneau} {et~al.}(2020){Bondonneau}, {Grie{\ss}meier},
  {Theureau}, {Bilous}, {Kondratiev}, {Serylak}, {Keith}, \&
  {Lyne}}]{Bondonneau2020}
{Bondonneau}, L., {Grie{\ss}meier}, J.~M., {Theureau}, G., {et~al.} 2020, \aap,
  635, A76

\bibitem[{{Bondonneau} {et~al.}(2021){Bondonneau}, {Grie{\ss}meier},
  {Theureau}, {Cognard}, {Brionne}, {Kondratiev}, {Bilous}, {McKee}, {Zarka},
  {Viou}, {Guillemot}, {Chen}, {Main}, {Pilia}, {Possenti}, {Serylak},
  {Shaifullah}, {Tiburzi}, {Verbiest}, {Wu}, {Wucknitz}, {Yerin}, {Briand},
  {Cecconi}, {Corbel}, {Dallier}, {Girard}, {Loh}, {Martin}, {Tagger}, \&
  {Tasse}}]{Bondonneau2021}
{Bondonneau}, L., {Grie{\ss}meier}, J.~M., {Theureau}, G., {et~al.} 2021, \aap,
  652, A34

\bibitem[{{Burke-Spolaor} {et~al.}(2012){Burke-Spolaor}, {Johnston}, {Bailes},
  {Bates}, {Bhat}, {Burgay}, {Champion}, {D'Amico}, {Keith}, {Kramer}, {Levin},
  {Milia}, {Possenti}, {Stappers}, \& {van Straten}}]{BurkeSpolaor2012}
{Burke-Spolaor}, S., {Johnston}, S., {Bailes}, M., {et~al.} 2012, \mnras, 423,
  1351

\bibitem[{{Cairns} {et~al.}(2004){Cairns}, {Johnston}, \& {Das}}]{Cairns2004}
{Cairns}, I.~H., {Johnston}, S., \& {Das}, P. 2004, \mnras, 353, 270

\bibitem[{{Cognard} {et~al.}(1996){Cognard}, {Shrauner}, {Taylor}, \&
  {Thorsett}}]{Cognard1996}
{Cognard}, I., {Shrauner}, J.~A., {Taylor}, J.~H., \& {Thorsett}, S.~E. 1996,
  \apjl, 457, L81

\bibitem[{{Connor} {et~al.}(2016){Connor}, {Sievers}, \& {Pen}}]{Connor2016}
{Connor}, L., {Sievers}, J., \& {Pen}, U.-L. 2016, \mnras, 458, L19

\bibitem[{{Connor} \& {van Leeuwen}(2018)}]{Connor2018}
{Connor}, L. \& {van Leeuwen}, J. 2018, \aj, 156, 256

\bibitem[{{Cordes} {et~al.}(2004){Cordes}, {Bhat}, {Hankins}, {McLaughlin}, \&
  {Kern}}]{Cordes2004}
{Cordes}, J.~M., {Bhat}, N.~D.~R., {Hankins}, T.~H., {McLaughlin}, M.~A., \&
  {Kern}, J. 2004, \apj, 612, 375

\bibitem[{{Cordes} \& {Lazio}(1991)}]{Cordes1991}
{Cordes}, J.~M. \& {Lazio}, T.~J. 1991, \apj, 376, 123

\bibitem[{{Cordes} \& {Wasserman}(2016)}]{Cordes2016}
{Cordes}, J.~M. \& {Wasserman}, I. 2016, \mnras, 457, 232

\bibitem[{{Cordes} {et~al.}(1985){Cordes}, {Weisberg}, \&
  {Boriakoff}}]{Cordes1985}
{Cordes}, J.~M., {Weisberg}, J.~M., \& {Boriakoff}, V. 1985, \apj, 288, 221

\bibitem[{{Cordes} {et~al.}(1990){Cordes}, {Weisberg}, \&
  {Hankins}}]{Cordes1990}
{Cordes}, J.~M., {Weisberg}, J.~M., \& {Hankins}, T.~H. 1990, \aj, 100, 1882

\bibitem[{{De} {et~al.}(2016){De}, {Gupta}, \& {Sharma}}]{De2016}
{De}, K., {Gupta}, Y., \& {Sharma}, P. 2016, \apjl, 833, L10

\bibitem[{{Eftekhari} {et~al.}(2016){Eftekhari}, {Stovall}, {Dowell},
  {Schinzel}, \& {Taylor}}]{Eftekhari2016}
{Eftekhari}, T., {Stovall}, K., {Dowell}, J., {Schinzel}, F.~K., \& {Taylor},
  G.~B. 2016, \apj, 829, 62

\bibitem[{{Esamdin} {et~al.}(2012){Esamdin}, {Abdurixit}, {Manchester}, \&
  {Niu}}]{Esamdin2012}
{Esamdin}, A., {Abdurixit}, D., {Manchester}, R.~N., \& {Niu}, H.~B. 2012,
  \apjl, 759, L3

\bibitem[{{Everett} \& {Weisberg}(2001)}]{Everett2001}
{Everett}, J.~E. \& {Weisberg}, J.~M. 2001, \apj, 553, 341

\bibitem[{{Fallows} {et~al.}(2016){Fallows}, {Bisi}, {Forte}, {Ulich},
  {Konovalenko}, {Mann}, \& {Vocks}}]{Fallows2016}
{Fallows}, R.~A., {Bisi}, M.~M., {Forte}, B., {et~al.} 2016, \apjl, 828, L7

\bibitem[{{Gajjar} {et~al.}(2014){Gajjar}, {Joshi}, \& {Wright}}]{Gajjar2014}
{Gajjar}, V., {Joshi}, B.~C., \& {Wright}, G. 2014, \mnras, 439, 221

\bibitem[{{Geyer} {et~al.}(2021){Geyer}, {Serylak}, {Abbate}, {Bailes},
  {Buchner}, {Chilufya}, {Johnston}, {Karastergiou}, {Main}, {van Straten}, \&
  {Shamohammadi}}]{Geyer2021}
{Geyer}, M., {Serylak}, M., {Abbate}, F., {et~al.} 2021, \mnras, 505, 4468

\bibitem[{{Gil}(1983)}]{Gil1983}
{Gil}, J. 1983, \aap, 127, 267

\bibitem[{{Goldman}(2021)}]{Goldman2021}
{Goldman}, I. 2021, \mnras, 504, 4493

\bibitem[{{Gupta} {et~al.}(1993){Gupta}, {Rickett}, \& {Coles}}]{Gupta1993}
{Gupta}, Y., {Rickett}, B.~J., \& {Coles}, W.~A. 1993, \apj, 403, 183

\bibitem[{{Hankins}(1971)}]{Hankins1971}
{Hankins}, T.~H. 1971, \apj, 169, 487

\bibitem[{{Hankins} \& {Boriakoff}(1978)}]{Hankins1978}
{Hankins}, T.~H. \& {Boriakoff}, V. 1978, \nat, 276, 45

\bibitem[{{Hankins} \& {Cordes}(1981)}]{Hankins1981}
{Hankins}, T.~H. \& {Cordes}, J.~M. 1981, \apj, 249, 241

\bibitem[{{Hankins} \& {Eilek}(2007)}]{Hankins2007}
{Hankins}, T.~H. \& {Eilek}, J.~A. 2007, \apj, 670, 693

\bibitem[{{Hankins} {et~al.}(2003){Hankins}, {Kern}, {Weatherall}, \&
  {Eilek}}]{Hankins2003}
{Hankins}, T.~H., {Kern}, J.~S., {Weatherall}, J.~C., \& {Eilek}, J.~A. 2003,
  \nat, 422, 141

\bibitem[{{Hassall} {et~al.}(2012){Hassall}, {Stappers}, {Hessels}, {Kramer},
  {Alexov}, {Anderson}, {Coenen}, {Karastergiou}, {Keane}, {Kondratiev},
  {Lazaridis}, {van Leeuwen}, {Noutsos}, {Serylak}, {Sobey}, {Verbiest},
  {Weltevrede}, {Zagkouris}, {Fender}, {Wijers}, {B{\"a}hren}, {Bell},
  {Broderick}, {Corbel}, {Daw}, {Dhillon}, {Eisl{\"o}ffel}, {Falcke},
  {Grie{\ss}meier}, {Jonker}, {Law}, {Markoff}, {Miller-Jones}, {Osten}, {Rol},
  {Scaife}, {Scheers}, {Schellart}, {Spreeuw}, {Swinbank}, {ter Veen}, {Wise},
  {Wijnands}, {Wucknitz}, {Zarka}, {Asgekar}, {Bell}, {Bentum}, {Bernardi},
  {Best}, {Bonafede}, {Boonstra}, {Brentjens}, {Brouw}, {Br{\"u}ggen},
  {Butcher}, {Ciardi}, {Garrett}, {Gerbers}, {Gunst}, {van Haarlem}, {Heald},
  {Hoeft}, {Holties}, {de Jong}, {Koopmans}, {Kuniyoshi}, {Kuper}, {Loose},
  {Maat}, {Masters}, {McKean}, {Meulman}, {Mevius}, {Munk}, {Noordam},
  {Orr{\'u}}, {Paas}, {Pandey-Pommier}, {Pandey}, {Pizzo}, {Polatidis},
  {Reich}, {R{\"o}ttgering}, {Sluman}, {Steinmetz}, {Sterks}, {Tagger}, {Tang},
  {Tasse}, {Vermeulen}, {van Weeren}, {Wijnholds}, \&
  {Yatawatta}}]{Hassall2012}
{Hassall}, T.~E., {Stappers}, B.~W., {Hessels}, J.~W.~T., {et~al.} 2012, \aap,
  543, A66

\bibitem[{{Hobbs} {et~al.}(2004){Hobbs}, {Lyne}, {Kramer}, {Martin}, \&
  {Jordan}}]{Hobbs2004}
{Hobbs}, G., {Lyne}, A.~G., {Kramer}, M., {Martin}, C.~E., \& {Jordan}, C.
  2004, \mnras, 353, 1311

\bibitem[{{Hotan} {et~al.}(2004){Hotan}, {van Straten}, \&
  {Manchester}}]{Hotan2004}
{Hotan}, A.~W., {van Straten}, W., \& {Manchester}, R.~N. 2004, Proc. Astron.
  Soc., 21, 302

\bibitem[{{Huguenin} {et~al.}(1969){Huguenin}, {Taylor}, \&
  {Jura}}]{Huguenin1969}
{Huguenin}, G.~R., {Taylor}, J.~H., \& {Jura}, M. 1969, \aplett, 4, 71

\bibitem[{{Jankowski} {et~al.}(2018){Jankowski}, {van Straten}, {Keane},
  {Bailes}, {Barr}, {Johnston}, \& {Kerr}}]{Jankowski2018}
{Jankowski}, F., {van Straten}, W., {Keane}, E.~F., {et~al.} 2018, \mnras, 473,
  4436

\bibitem[{{Johnston} {et~al.}(1998){Johnston}, {Nicastro}, \&
  {Koribalski}}]{Johnston1998}
{Johnston}, S., {Nicastro}, L., \& {Koribalski}, B. 1998, \mnras, 297, 108

\bibitem[{{Johnston} \& {Romani}(2002)}]{Johnston2002}
{Johnston}, S. \& {Romani}, R.~W. 2002, \mnras, 332, 109

\bibitem[{{Johnston} {et~al.}(2004){Johnston}, {Romani}, {Marshall}, \&
  {Zhang}}]{Johnston2004}
{Johnston}, S., {Romani}, R.~W., {Marshall}, F.~E., \& {Zhang}, W. 2004,
  \mnras, 355, 31

\bibitem[{{Karuppusamy} {et~al.}(2012){Karuppusamy}, {Stappers}, \&
  {Lee}}]{Karuppusamy2012}
{Karuppusamy}, R., {Stappers}, B.~W., \& {Lee}, K.~J. 2012, \aap, 538, A7

\bibitem[{{Karuppusamy} {et~al.}(2010){Karuppusamy}, {Stappers}, \& {van
  Straten}}]{Karuppusamy2010}
{Karuppusamy}, R., {Stappers}, B.~W., \& {van Straten}, W. 2010, \aap, 515, A36

\bibitem[{{Kazantsev} \& {Basalaeva}(2020)}]{Kazantsev2020}
{Kazantsev}, A.~N. \& {Basalaeva}, M.~Y. 2020, arXiv e-prints, arXiv:2005.07244

\bibitem[{{Keane} {et~al.}(2010){Keane}, {Ludovici}, {Eatough}, {Kramer},
  {Lyne}, {McLaughlin}, \& {Stappers}}]{Keane2010}
{Keane}, E.~F., {Ludovici}, D.~A., {Eatough}, R.~P., {et~al.} 2010, \mnras,
  401, 1057

\bibitem[{{Kijak} \& {Gil}(2003)}]{Kijak2003}
{Kijak}, J. \& {Gil}, J. 2003, \aap, 397, 969

\bibitem[{{Kinkhabwala} \& {Thorsett}(2000)}]{Kinkhabwala2000}
{Kinkhabwala}, A. \& {Thorsett}, S.~E. 2000, \apj, 535, 365

\bibitem[{{Kirsten} {et~al.}(2020){Kirsten}, {Snelders}, {Jenkins}, {Nimmo},
  {van den Eijnden}, {Hessels}, {Gawro{\'n}ski}, \& {Yang}}]{Kirsten2020}
{Kirsten}, F., {Snelders}, M.~P., {Jenkins}, M., {et~al.} 2020, Nature
  Astronomy

\bibitem[{{Knight}(2006)}]{Knight2006a}
{Knight}, H.~S. 2006, Chinese Journal of Astronomy and Astrophysics Supplement,
  6, 41

\bibitem[{{Knight} {et~al.}(2006){Knight}, {Bailes}, {Manchester}, {Ord}, \&
  {Jacoby}}]{Knight2006}
{Knight}, H.~S., {Bailes}, M., {Manchester}, R.~N., {Ord}, S.~M., \& {Jacoby},
  B.~A. 2006, \apj, 640, 941

\bibitem[{{Kondratiev} {et~al.}(2018){Kondratiev}, {Bilous}, \& {LOFAR
  PWG}}]{Kondratiev2018}
{Kondratiev}, V., {Bilous}, A., \& {LOFAR PWG}. 2018, in Pulsar Astrophysics
  the Next Fifty Years, ed. P.~{Weltevrede}, B.~B.~P. {Perera}, L.~L.
  {Preston}, \& S.~{Sanidas}, Vol. 337, 358--359

\bibitem[{{Kramer} {et~al.}(2002){Kramer}, {Johnston}, \& {van
  Straten}}]{Kramer2002}
{Kramer}, M., {Johnston}, S., \& {van Straten}, W. 2002, \mnras, 334, 523

\bibitem[{{Kuiack} {et~al.}(2020){Kuiack}, {Wijers}, {Rowlinson}, {Shulevski},
  {Huizinga}, {Molenaar}, \& {Prasad}}]{Kuiack2020}
{Kuiack}, M., {Wijers}, R. A.~M.~J., {Rowlinson}, A., {et~al.} 2020, \mnras,
  497, 846

\bibitem[{{Kuzmin}(2007)}]{Kuzmin2007}
{Kuzmin}, A.~D. 2007, in WE-Heraeus Seminar on Neutron Stars and Pulsars 40
  years after the Discovery, ed. W.~{Becker} \& H.~H. {Huang}, 72

\bibitem[{{Kuzmin} {et~al.}(2003){Kuzmin}, {Hamilton}, {Shitov}, {McCulloch},
  {McConnell}, \& {Pugatchev}}]{Kuzmin2003}
{Kuzmin}, A.~D., {Hamilton}, P.~A., {Shitov}, Y.~P., {et~al.} 2003, \mnras,
  344, 1187

\bibitem[{{Lange} {et~al.}(1998){Lange}, {Kramer}, {Wielebinski}, \&
  {Jessner}}]{Lange1998}
{Lange}, C., {Kramer}, M., {Wielebinski}, R., \& {Jessner}, A. 1998, \aap, 332,
  111

\bibitem[{{Lorimer} {et~al.}(2007){Lorimer}, {Bailes}, {McLaughlin},
  {Narkevic}, \& {Crawford}}]{Lorimer2007}
{Lorimer}, D.~R., {Bailes}, M., {McLaughlin}, M.~A., {Narkevic}, D.~J., \&
  {Crawford}, F. 2007, Science, 318, 777

\bibitem[{{Lorimer} \& {Kramer}(2005)}]{Lorimer2005}
{Lorimer}, D.~R. \& {Kramer}, M. 2005, {Handbook of Pulsar Astronomy (Cambridge
  University Press)}

\bibitem[{{Lyubarsky}(2019)}]{Lyubarsky2019}
{Lyubarsky}, Y. 2019, \mnras, 483, 1731

\bibitem[{{Lyutikov}(2019)}]{Lyutikov2019}
{Lyutikov}, M. 2019, arXiv e-prints, arXiv:1908.07313

\bibitem[{{Maan} {et~al.}(2019){Maan}, {Joshi}, {Surnis}, {Bagchi}, \&
  {Manoharan}}]{Maan2019}
{Maan}, Y., {Joshi}, B.~C., {Surnis}, M.~P., {Bagchi}, M., \& {Manoharan},
  P.~K. 2019, \apjl, 882, L9

\bibitem[{{Maan} \& {van Leeuwen}(2017)}]{Maan2017}
{Maan}, Y. \& {van Leeuwen}, J. 2017, in 2017 XXXIInd General Assembly and
  Scientific Symposium of the International Union of Radio Science (URSI GASS,
  2

\bibitem[{{Mahajan} {et~al.}(2018){Mahajan}, {van Kerkwijk}, {Main}, \&
  {Pen}}]{Mahajan2018}
{Mahajan}, N., {van Kerkwijk}, M.~H., {Main}, R., \& {Pen}, U.-L. 2018, \apjl,
  867, L2

\bibitem[{{Main} {et~al.}(2017){Main}, {van Kerkwijk}, {Pen}, {Mahajan}, \&
  {Vanderlinde}}]{Main2017}
{Main}, R., {van Kerkwijk}, M., {Pen}, U.-L., {Mahajan}, N., \& {Vanderlinde},
  K. 2017, \apjl, 840, L15

\bibitem[{{Manchester} {et~al.}(2005){Manchester}, {Hobbs}, {Teoh}, \&
  {Hobbs}}]{Manchester2005}
{Manchester}, R.~N., {Hobbs}, G.~B., {Teoh}, A., \& {Hobbs}, M. 2005, \aj, 129,
  1993

\bibitem[{{Manchester} {et~al.}(2001){Manchester}, {Lyne}, {Camilo}, {Bell},
  {Kaspi}, {D'Amico}, {McKay}, {Crawford}, {Stairs}, {Possenti}, {Kramer}, \&
  {Sheppard}}]{Manchester2001}
{Manchester}, R.~N., {Lyne}, A.~G., {Camilo}, F., {et~al.} 2001, \mnras, 328,
  17

\bibitem[{{Marcote} {et~al.}(2020){Marcote}, {Nimmo}, {Hessels}, {Tendulkar},
  {Bassa}, {Paragi}, {Keimpema}, {Bhardwaj}, {Karuppusamy}, {Kaspi}, {Law},
  {Michilli}, {Aggarwal}, {Andersen}, {Archibald}, {Bandura}, {Bower}, {Boyle},
  {Brar}, {Burke-Spolaor}, {Butler}, {Cassanelli}, {Chawla}, {Demorest},
  {Dobbs}, {Fonseca}, {Giri}, {Good}, {Gourdji}, {Josephy}, {Kirichenko},
  {Kirsten}, {Landecker}, {Lang}, {Lazio}, {Li}, {Lin}, {Linford}, {Masui},
  {Mena-Parra}, {Naidu}, {Ng}, {Patel}, {Pen}, {Pleunis}, {Rafiei-Ravandi},
  {Rahman}, {Renard}, {Scholz}, {Siegel}, {Smith}, {Stairs}, {Vanderlinde}, \&
  {Zwaniga}}]{Marcote2020}
{Marcote}, B., {Nimmo}, K., {Hessels}, J.~W.~T., {et~al.} 2020, \nat, 577, 190

\bibitem[{{McKee} {et~al.}(2019){McKee}, {Stappers}, {Bassa}, {Chen},
  {Cognard}, {Gaikwad}, {Janssen}, {Karuppusamy}, {Kramer}, {Lee}, {Liu},
  {Perrodin}, {Sanidas}, {Smits}, {Wang}, \& {Zhu}}]{Mckee2019}
{McKee}, J.~W., {Stappers}, B.~W., {Bassa}, C.~G., {et~al.} 2019, \mnras, 483,
  4784

\bibitem[{{McLaughlin} {et~al.}(2006){McLaughlin}, {Lyne}, {Lorimer}, {Kramer},
  {Faulkner}, {Manchester}, {Cordes}, {Camilo}, {Possenti}, {Stairs}, {Hobbs},
  {D'Amico}, {Burgay}, \& {O'Brien}}]{McLaughlin2006}
{McLaughlin}, M.~A., {Lyne}, A.~G., {Lorimer}, D.~R., {et~al.} 2006, \nat, 439,
  817

\bibitem[{{Mickaliger} {et~al.}(2018){Mickaliger}, {McEwen}, {McLaughlin}, \&
  {Lorimer}}]{Mickaliger2018}
{Mickaliger}, M.~B., {McEwen}, A.~E., {McLaughlin}, M.~A., \& {Lorimer}, D.~R.
  2018, \mnras, 479, 5413

\bibitem[{{Mickaliger} {et~al.}(2012){Mickaliger}, {McLaughlin}, {Lorimer},
  {Langston}, {Bilous}, {Kondratiev}, {Lyutikov}, {Ransom}, \&
  {Palliyaguru}}]{Mickaliger2012}
{Mickaliger}, M.~B., {McLaughlin}, M.~A., {Lorimer}, D.~R., {et~al.} 2012,
  \apj, 760, 64

\bibitem[{{Narayan} \& {Vivekanand}(1983)}]{Narayan1983}
{Narayan}, R. \& {Vivekanand}, M. 1983, \apj, 274, 771

\bibitem[{{Nicastro} {et~al.}(2001){Nicastro}, {Nigro}, {D'Amico}, {Lumiella},
  \& {Johnston}}]{Nicastro2001}
{Nicastro}, L., {Nigro}, F., {D'Amico}, N., {Lumiella}, V., \& {Johnston}, S.
  2001, \aap, 368, 1055

\bibitem[{{Nimmo} {et~al.}(2021){Nimmo}, {Hessels}, {Kirsten}, {Keimpema},
  {Cordes}, {Snelders}, {Hewitt}, {Karuppusamy}, {Archibald}, {Bezukovs},
  {Bhardwaj}, {Blaauw}, {Buttaccio}, {Cassanelli}, {Conway}, {Corongiu},
  {Feiler}, {Fonseca}, {Forssen}, {Gawronski}, {Giroletti}, {Kharinov},
  {Leung}, {Lindqvist}, {Maccaferri}, {Marcote}, {Masui}, {Mckinven},
  {Melnikov}, {Michilli}, {Mikhailov}, {Ng}, {Orbidans}, {Ould-Boukattine},
  {Paragi}, {Pearlman}, {Petroff}, {Rahman}, {Scholz}, {Shin}, {Smith},
  {Stairs}, {Surcis}, {Tendulkar}, {Vlemmings}, {Wang}, {Yang}, \&
  {Yuan}}]{Nimmo2021}
{Nimmo}, K., {Hessels}, J.~W.~T., {Kirsten}, F., {et~al.} 2021, arXiv e-prints,
  arXiv:2105.11446

\bibitem[{{Oostrum} {et~al.}(2020){Oostrum}, {Maan}, {van Leeuwen}, {Connor},
  {Petroff}, {Attema}, {Bast}, {Gardenier}, {Hargreaves}, {Kooistra}, {van der
  Schuur}, {Sclocco}, {Smits}, {Straal}, {ter Veen}, {Vohl}, {Adams},
  {Adebahr}, {de Blok}, {van den Brink}, {van Cappellen}, {Coolen}, {Damstra},
  {van Diepen}, {Frank}, {Hess}, {van der Hulst}, {Hut}, {Ivashina}, {Loose},
  {Lucero}, {Mika}, {Morganti}, {Moss}, {Mulder}, {Norden}, {Oosterloo},
  {Orr{\'u}}, {de Reijer}, {Ruiter}, {Vermaas}, {Wijnholds}, \&
  {Ziemke}}]{Oostrum2020}
{Oostrum}, L.~C., {Maan}, Y., {van Leeuwen}, J., {et~al.} 2020, \aap, 635, A61

\bibitem[{{Pastor-Marazuela} {et~al.}(2021){Pastor-Marazuela}, {Connor}, {van
  Leeuwen}, {Maan}, {ter Veen}, {Bilous}, {Oostrum}, {Petroff}, {Straal},
  {Vohl}, {Attema}, {Boersma}, {Kooistra}, {van der Schuur}, {Sclocco},
  {Smits}, {Adams}, {Adebahr}, {de Blok}, {Coolen}, {Damstra}, {D{\'e}nes},
  {Hess}, {van der Hulst}, {Hut}, {Ivashina}, {Kutkin}, {Loose}, {Lucero},
  {Mika}, {Moss}, {Mulder}, {Norden}, {Oosterloo}, {Orr{\'u}}, {Ruiter}, \&
  {Wijnholds}}]{Pastor2020}
{Pastor-Marazuela}, I., {Connor}, L., {van Leeuwen}, J., {et~al.} 2021, \nat,
  596, 505

\bibitem[{{Pennucci}(2015)}]{Pennucci2015}
{Pennucci}, T.~T. 2015, PhD thesis, University of Virginia

\bibitem[{{Pennucci}(2019)}]{Pennucci2019}
{Pennucci}, T.~T. 2019, \apj, 871, 34

\bibitem[{{Pennucci} {et~al.}(2014){Pennucci}, {Demorest}, \&
  {Ransom}}]{Pennucci2014}
{Pennucci}, T.~T., {Demorest}, P.~B., \& {Ransom}, S.~M. 2014, \apj, 790, 93

\bibitem[{{Petroff} {et~al.}(2019){Petroff}, {Hessels}, \&
  {Lorimer}}]{Petroff2019}
{Petroff}, E., {Hessels}, J.~W.~T., \& {Lorimer}, D.~R. 2019, \aapr, 27, 4

\bibitem[{{Petroff} {et~al.}(2021){Petroff}, {Hessels}, \&
  {Lorimer}}]{Petroff2021}
{Petroff}, E., {Hessels}, J.~W.~T., \& {Lorimer}, D.~R. 2021, arXiv e-prints,
  arXiv:2107.10113

\bibitem[{{Petrova}(2004{\natexlab{a}})}]{Petrova2004b}
{Petrova}, S.~A. 2004{\natexlab{a}}, \aap, 424, 227

\bibitem[{{Petrova}(2004{\natexlab{b}})}]{Petrova2004a}
{Petrova}, S.~A. 2004{\natexlab{b}}, \aap, 417, L29

\bibitem[{{Petrova}(2008)}]{Petrova2008}
{Petrova}, S.~A. 2008, \mnras, 384, L1

\bibitem[{{Pilkington} {et~al.}(1968){Pilkington}, {Hewish}, {Bell}, \&
  {Cole}}]{Pilkington1968}
{Pilkington}, J.~D.~H., {Hewish}, A., {Bell}, S.~J., \& {Cole}, T.~W. 1968,
  \nat, 218, 126

\bibitem[{{Platts} {et~al.}(2019){Platts}, {Weltman}, {Walters}, {Tendulkar},
  {Gordin}, \& {Kandhai}}]{Platts2019}
{Platts}, E., {Weltman}, A., {Walters}, A., {et~al.} 2019, \physrep, 821, 1

\bibitem[{{Pleunis} {et~al.}(2021){Pleunis}, {Michilli}, {Bassa}, {Hessels},
  {Naidu}, {Andersen}, {Chawla}, {Fonseca}, {Gopinath}, {Kaspi}, {Kondratiev},
  {Li}, {Bhardwaj}, {Boyle}, {Brar}, {Cassanelli}, {Gupta}, {Josephy},
  {Karuppusamy}, {Keimpema}, {Kirsten}, {Leung}, {Marcote}, {Masui},
  {Mckinven}, {Meyers}, {Ng}, {Nimmo}, {Paragi}, {Rahman}, {Scholz}, {Shin},
  {Smith}, {Stairs}, \& {Tendulkar}}]{Pleunis2020}
{Pleunis}, Z., {Michilli}, D., {Bassa}, C.~G., {et~al.} 2021, \apjl, 911, L3

\bibitem[{{Popov} {et~al.}(2009){Popov}, {Soglasnov}, {Kondratiev}, {Bilous},
  {Moshkina}, {Oreshko}, {Ilyasov}, {Sekido}, \& {Kondo}}]{Popov2009}
{Popov}, M., {Soglasnov}, V., {Kondratiev}, V., {et~al.} 2009, \pasj, 61, 1197

\bibitem[{{Popov} {et~al.}(2002){Popov}, {Bartel}, {Cannon}, {Novikov},
  {Kondratiev}, \& {Altunin}}]{Popov2002}
{Popov}, M.~V., {Bartel}, N., {Cannon}, W.~H., {et~al.} 2002, \aap, 396, 171

\bibitem[{{Popov} {et~al.}(2006){Popov}, {Kuz'min}, {Ul'yanov}, {Deshpand e},
  {Ershov}, {Zakharenko}, {Kondrat'ev}, {Kostyuk}, {Losovski{\"a}­}, \&
  {Soglasnov}}]{Popov2006}
{Popov}, M.~V., {Kuz'min}, A.~D., {Ul'yanov}, O.~M., {et~al.} 2006, Astronomy
  Reports, 50, 562

\bibitem[{{Radhakrishnan} \& {Cooke}(1969)}]{Radhakrishnan1969}
{Radhakrishnan}, V. \& {Cooke}, D.~J. 1969, \aplett, 3, 225

\bibitem[{{Raithel} {et~al.}(2015){Raithel}, {Shannon}, {Johnston}, \&
  {Kerr}}]{Raithel2015}
{Raithel}, C.~A., {Shannon}, R.~M., {Johnston}, S., \& {Kerr}, M. 2015, \apjl,
  804, L18

\bibitem[{{Rankin}(1990)}]{Rankin1990}
{Rankin}, J.~M. 1990, \apj, 352, 247

\bibitem[{{Rankin}(1993)}]{Rankin1993}
{Rankin}, J.~M. 1993, \apj, 405, 285

\bibitem[{{Rickett}(1990)}]{Rickett1990}
{Rickett}, B.~J. 1990, \araa, 28, 561

\bibitem[{{Rickett} \& {Cordes}(1981)}]{Rickett1981}
{Rickett}, B.~J. \& {Cordes}, J.~M. 1981, in Pulsars: 13 Years of Research on
  Neutron Stars, ed. W.~{Sieber} \& R.~{Wielebinski}, Vol.~95, 107--109

\bibitem[{{Rickett} {et~al.}(1975){Rickett}, {Hankins}, \&
  {Cordes}}]{Rickett1975}
{Rickett}, B.~J., {Hankins}, T.~H., \& {Cordes}, J.~M. 1975, \apj, 201, 425

\bibitem[{{Ruan} {et~al.}(2020){Ruan}, {Taylor}, {Dowell}, {Stovall},
  {Schinzel}, \& {Demorest}}]{Ruan2020}
{Ruan}, D., {Taylor}, G.~B., {Dowell}, J., {et~al.} 2020, \mnras, 495, 2125

\bibitem[{{Sclocco} {et~al.}(2016){Sclocco}, {van Leeuwen}, {Bal}, \& {van
  Nieuwpoort}}]{Sclocco2016}
{Sclocco}, A., {van Leeuwen}, J., {Bal}, H.~E., \& {van Nieuwpoort}, R.~V.
  2016, Astronomy and Computing, 14, 1

\bibitem[{{Seymour} {et~al.}(2014){Seymour}, {Lorimer}, \&
  {Ridley}}]{Seymour2014}
{Seymour}, A.~D., {Lorimer}, D.~R., \& {Ridley}, J.~P. 2014, \mnras, 439, 3951

\bibitem[{{Shabanova} \& {Shitov}(2004)}]{Shabanova2004}
{Shabanova}, T.~V. \& {Shitov}, Y.~P. 2004, \aap, 418, 203

\bibitem[{{Singal} \& {Vats}(2012)}]{Singal2012}
{Singal}, A.~K. \& {Vats}, H.~O. 2012, \aj, 144, 155

\bibitem[{{Smirnova}(2012)}]{Smirnova2012}
{Smirnova}, T.~V. 2012, Astronomy Reports, 56, 430

\bibitem[{{Smirnova} \& {Shishov}(2008)}]{Smirnova2008}
{Smirnova}, T.~V. \& {Shishov}, V.~I. 2008, Astronomy Reports, 52, 736

\bibitem[{{Smith}(1973)}]{Smith1973}
{Smith}, F.~G. 1973, \mnras, 161, 9P

\bibitem[{{Soglasnov} {et~al.}(2004){Soglasnov}, {Popov}, {Bartel}, {Cannon},
  {Novikov}, {Kondratiev}, \& {Altunin}}]{Soglasnov2004}
{Soglasnov}, V.~A., {Popov}, M.~V., {Bartel}, N., {et~al.} 2004, \apj, 616, 439

\bibitem[{{Song} {et~al.}(2018){Song}, {Kondratiev}, \& {Bilous}}]{Song2018}
{Song}, X., {Kondratiev}, V., \& {Bilous}, A. 2018, in IAU Symposium, Vol. 337,
  Pulsar Astrophysics the Next Fifty Years, ed. P.~{Weltevrede}, B.~B.~P.
  {Perera}, L.~L. {Preston}, \& S.~{Sanidas}, 412--413

\bibitem[{{Tsai} {et~al.}(2016){Tsai}, {Simonetti}, {Akukwe}, {Bear}, {Gough},
  {Shawhan}, \& {Kavic}}]{Tsai2016}
{Tsai}, Wei, J., {Simonetti}, J.~H., {Akukwe}, B., {et~al.} 2016, \aj, 151, 28

\bibitem[{{Tsai} {et~al.}(2015){Tsai}, {Simonetti}, {Akukwe}, {Bear},
  {Cutchin}, {Dowell}, {Gough}, {Kanner}, {Kassim}, {Schinzel}, {Shawhan},
  {Taylor}, {Yancey}, {Quezada}, \& {Kavic}}]{Tsai2015}
{Tsai}, J.-W., {Simonetti}, J.~H., {Akukwe}, B., {et~al.} 2015, \aj, 149, 65

\bibitem[{{Ulyanov} {et~al.}(2016){Ulyanov}, {Skoryk}, {Shevtsova}, {Plakhov},
  \& {Ulyanova}}]{Ulyanov2016}
{Ulyanov}, O.~M., {Skoryk}, A.~O., {Shevtsova}, A.~I., {Plakhov}, M.~S., \&
  {Ulyanova}, O.~O. 2016, \mnras, 455, 150

\bibitem[{{Ulyanov} {et~al.}(2006){Ulyanov}, {Zakharenko}, {Konovalenko},
  {Lecacheux}, {Rosolen}, \& {Rucker}}]{Ulyanov2006}
{Ulyanov}, O.~M., {Zakharenko}, V.~V., {Konovalenko}, O.~O., {et~al.} 2006,
  Russian Radio Physics and Radio Astronomy, 11, 113

\bibitem[{{van Leeuwen} {et~al.}(2020){van Leeuwen}, {Mikhailov}, {Keane},
  {Coenen}, {Connor}, {Kondratiev}, {Michilli}, \& {Sanidas}}]{vanLeeuwen2020}
{van Leeuwen}, J., {Mikhailov}, K., {Keane}, E., {et~al.} 2020, \aap, 634, A3

\bibitem[{{van Straten} \& {Bailes}(2011)}]{vanStraten2011}
{van Straten}, W. \& {Bailes}, M. 2011, \pasa, 28, 1

\bibitem[{{Wang} {et~al.}(2020){Wang}, {Hobbs}, {Wang}, {Manchester}, {Wang},
  {Zhang}, {Feng}, {Wang}, {Li}, {Dai}, {Lee}, {Dang}, \& {Zhang}}]{Wang2020}
{Wang}, S.~Q., {Hobbs}, G., {Wang}, J.~B., {et~al.} 2020, \apjl, 902, L13

\bibitem[{{Wang} {et~al.}(2019){Wang}, {Zhang}, {Chen}, \& {Xu}}]{Wang2019}
{Wang}, W., {Zhang}, B., {Chen}, X., \& {Xu}, R. 2019, \apjl, 876, L15

\bibitem[{{Zakharenko} {et~al.}(2013){Zakharenko}, {Vasylieva}, {Konovalenko},
  {Ulyanov}, {Serylak}, {Zarka}, {Grie{\ss}meier}, {Cognard}, \&
  {Nikolaenko}}]{Zakharenko2013}
{Zakharenko}, V.~V., {Vasylieva}, I.~Y., {Konovalenko}, A.~A., {et~al.} 2013,
  \mnras, 431, 3624

\bibitem[{{Zarka} {et~al.}(2020){Zarka}, {Denis}, {Tagger}, {Girard}, {Coffre},
  {Dumez-Viou}, {Taffoureau}, {Charrier}, {Bondonneau}, {Briand}, \&
  {Casoli}}]{Zarka2020}
{Zarka}, P., {Denis}, L., {Tagger}, M., {et~al.} 2020, in URSI GASS 2020,
  Session J01 New Telescopes on the Frontier

\end{thebibliography}

\end{document}